\documentclass[pdflatex,sn-mathphys,Numbered]{sn-jnl}

\usepackage[markup=underlined]{changes}
\definechangesauthor[name={Jos\'{e} Rodal}, color=red]{author}

\usepackage{tcolorbox}

\usepackage{amsmath,amssymb,amsfonts}%
\usepackage{fix-cm}%
\usepackage{amsthm}%
\usepackage[cal=rsfs]{mathalfa}
\usepackage{graphicx}%
\usepackage{multirow}%
\usepackage[title]{appendix}%
\usepackage{xcolor}%
\usepackage{textcomp}%
\usepackage{manyfoot}%
\usepackage{booktabs}%
\usepackage{algorithm}%
\usepackage{algorithmicx}%
\usepackage{algpseudocode}%
\usepackage{listings}%
\usepackage{bm}%
\usepackage{mathtools}

\usepackage{silence}
\usepackage{caption}
\usepackage{subcaption}

\usepackage{todonotes}

\DeclarePairedDelimiter\abs{|}{|}
\newcommand{\sign}{\text{sign}}

\begin{document}

\title[Article Title]{A Closer Look at  Natário’s Zero-Expansion Warp Drive}


\author*[1]{\fnm{Jos\'{e}} \sur{Rodal}}\email{jrodal@alum.mit.edu}



\affil*[1]{\orgname{Rodal Consulting}, \orgaddress{\street{205 Firetree Ln.}, \city{Cary}, \state{NC}, \postcode{27513}, \country{USA}}}




\abstract{We conduct a detailed analysis of Natário’s ``zero-expansion'' warp drive spacetime, focusing on scalar curvature invariants within the 3+1 formalism. This paper has four primary objectives: First, we establish the Petrov type classification of Natário’s spacetime, which has not been previously determined in the literature. We prove that Natário’s spacetime is Petrov type I, not fitting the Class $B$ warped product spacetime definition. Second, we assess the relative magnitude of the Weyl scalar curvature invariant and compare it with the amplitudes of Einstein’s scalar and the Ricci quadratic and cubic invariants within the warp-bubble zone. Previous studies have focused on Ricci curvature and the energy-momentum tensor, neglecting the Weyl curvature, which we demonstrate plays a significant role due to the sharp localization of the form function near the warp-bubble radius. Third, we visualize several curvature invariants for Natário’s warp drive, as well as momentum density, which we show as the critical physical quantity governing the orientation of the warp drive trajectory, overshadowing space volume changes. Fourth, we critically examine claims that Natário’s warp drive is more realistic than Alcubierre’s. We demonstrate that Natário’s spacetime exhibits curvature invariant amplitudes 35 times greater than Alcubierre’s, given identical warp-bubble parameters, making Natário’s concept even less viable. Additionally, we address Mattingly et al.'s analysis, highlighting their underestimation of curvature invariant amplitudes by 21 orders of magnitude.}




\maketitle

\clearpage 

\tableofcontents

\section{Introduction}\label{sec1}

\makeatletter
\newcommand\incircbin
{%
  \mathpalette\@incircbin
}
\newcommand\@incircbin[2]
{%
  \mathbin%
  {%
    \ooalign{\hidewidth$#1#2$\hidewidth\crcr$#1\bigcirc$}%
  }%
}
\newcommand{\oland}{\incircbin{\land}}
\makeatother

The purpose of this paper is to further investigate Nat\'{a}rio’s zero-expansion warp drive spacetime with four primary objectives. First, establish its Petrov type classification and examine whether it is a Class $B_1$ spacetime, as claimed in  \cite{Mattingly}. To date, the Petrov type of Nat\'{a}rio's spacetime has not been explicitly determined in the literature. Second, assess the relative magnitude of Weyl’s scalar curvature invariant and compare it with the amplitude of Einstein’s scalar as well as the Ricci quadratic and cubic invariants within the warp-bubble zone. Previous studies of warp drives have focused on the Ricci curvature and the associated energy-momentum tensor, neglecting the Weyl curvature. Typically, Weyl curvature is associated with lower magnitude far zone effects, such as tidal effects, but we hypothesize that for Nat\'{a}rio's warp drive, due to the sharp localization of the form function $f(r)$ near the warp-bubble radius, Weyl curvature plays a significant role. Third, visualize several curvature invariants for Nat\'{a}rio's warp drive.  We also visualize the momentum density, which we show is the critical physical quantity governing the orientation of the trajectory of a constant velocity warp drive. Fourth, critically examine claims in the literature that the Nat\'{a}rio warp drive is ``more realistic'' than Alcubierre's.

It has been 30 years since Alcubierre’s seminal paper on the hyperdrive in 1994 \cite{Alcubierre_1994}, which sparked a surge of interest and led to the publication of hundreds of articles on warp drives. The Alcubierre metric, which introduced the concept of faster-than-light travel within a warp-bubble, captivated researchers due to its implications for superluminal space travel. This groundbreaking idea has fueled theoretical explorations and inspired discussions on the physical constraints of achieving such travel within the framework of General Relativity. Historically and currently, a major focus has been the investigation of energy conditions in such spacetimes, similar to the approach taken in other spacetime geometries where the metric is predefined, such as in the study of wormholes. This interest is driven by the challenge of satisfying the metric’s requirements for `exotic' negative matter, as the practical realization of the warp drive is constrained by the need for exorbitant amounts of negative energy densities \cite{Hiscock_1997, GonzalezDiaz, HiscockClark, LoboVisser, SantiagoVisser, Olum, Pfenning}.  As remarked by Synge (p. 213 of \cite{synge1966}) three decades before Alcubierre’s paper on warp drives, it has long been known in General Relativity that when a metric $( g_{\mu\nu})$ is pre-assigned and the stress-energy-momentum tensor $(T^{\mu\nu})$ is calculated by mere differentiation of the metric, one is very likely to encounter negative energy densities and tensile stress, whereas one would physically expect only positive energy densities and compressive stress (pressure) to occur in nature. 

Since the literature on warp drives is extensive and it is not the purpose of this article to provide a comprehensive bibliography, we present here a selection of papers published since 2022 \cite{marquet2022, SantiagoVisser, ShoshanyAndStober2023, ShoshanyAndSnodgrass2023, Helmerich2024, Fuchs2024, garattini2024, Eroshenko2023, AbellánVasilevGRG, Abellán2023lapse}. Additionally, several other pertinent articles are cited throughout this paper where relevant.

As noted in \cite{SchusterSantiagoVisser2023}: ``\dots two distinct modifications (zero-expansion and generic) were subsequently developed by Natário \cite{Natário} in 2001.'' And as pointed out in \cite{SantiagoVisser}: ``Note that the second half of Nat\'{a}rio's 2001 paper \cite{Natário} focuses on this zero-expansion warp field, whereas the first half of that paper deals with the generic Nat\'{a}rio warp field spacetime\dots.  When mentioning Nat\'{a}rio's warp field, it is therefore important to make explicit which of these two specific warp drives one is referring to.'' ``The generic Natário warp drive line element\dots is sufficiently general to cover well over 99\% of the relevant literature.'' For example, the generic Natário warp drive line element includes Alcubierre's \cite{Alcubierre_1994}, Natário's \cite{Natário} ``zero-expansion'' and Lentz \cite{Lentz_2021}/Fell-Heisenberg's \cite{Fell} zero-vorticity warp drive models.
In this article, we focus exclusively on what Nat\'{a}rio \cite{Natário} and Santiago et al. \cite{SantiagoVisser} call Nat\'{a}rio’s zero-expansion warp drive, which we refer to as Nat\'{a}rio’s warp drive or Nat\'{a}rio’s spacetime for short. We also often use the term `zero differential volume change' as it is more descriptive than the usual term `zero expansion,' indicating that the differential volume can expand or shrink in the warp-bubble. Nat\'{a}rio constructed a warp drive vector field that effectively set the scalar extrinsic curvature to zero ($K = 0$), a condition necessary but not sufficient for maximal slicing within the 3+1 formalism, where additional constraints on the lapse function must be satisfied \cite{Gourgoulhon, baumgarte_shapiro_2010, AlcubierreBook}. The condition of zero extrinsic curvature is also analogous to isochoric (constant volume) flow. Previously, a warp drive solution was commonly believed to require contracting space in front of the warp-bubble and expanding it behind.

This article is organized as follows. Section \ref{sec2} introduces the notation and units used throughout this paper and presents the coordinate system of the zero-differential-volume-change Natário warp drive to ensure clarity in subsequent discussions. This section also includes definitions of the tensors and curvature invariants utilized in our analysis and computational visualizations. 

Section \ref{sec3} addresses the tensorial solution of Natário's spacetime, starting with the curvilinear coordinate tensor components of the 3+1 ADM shift vector field. To date, Natário's solution in curvilinear tensor components has not been presented in the literature; Natário instead employed the tetrad formalism, which represents `physical components' in a local Minkowski frame that do not transform according to tensor transformation laws (instead, they transform under Lorentz transformations). Subsequent subsections discuss the spacetime metric, matter energy density, and momentum density. Here we address White's question \cite{WhiteGRG}: ``\dots how does the ship know which way to go? The energy density curves local spacetime, but since it has no bias along the $x$-axis\dots.'' We show that the momentum density, determined by the spatial gradient of the extrinsic curvature tensor, is the key physical quantity that governs the orientation of the constant-velocity warp drive trajectory in both the Natário and Alcubierre models.

Subsections \ref{Newman Penrose} to \ref{Clarifying} focus on establishing the Petrov type classification of the zero-differential-volume-change Natário warp drive and examining the claim in \cite{Mattingly} that it is a Class $B_1$ spacetime. We prove, using the Newman-Penrose (double-null, complex) formalism, that Natário's spacetime is actually Petrov type I, characterized by the lack of algebraic symmetries and the existence of four unique and real principal null directions. We establish that Natário’s spacetime does not fit the definition of a Class $B$ warped product spacetime, as these are restricted to Petrov types $D$ or $O$. The final subsection of section \ref{sec3} discusses the radially-dependent warp-bubble form function $f(r)$ and its approximations, which are essential for calculations and visualizations since Natário \cite{Natário} did not suggest a particular $f(r)$ function.

Section \ref{Visualizing curvature invariants} covers the computation and visualization of curvature invariants in Natário’s spacetime, examining their radial and circumferential distributions, relative amplitudes, and comparisons with those in Alcubierre's spacetime for different warp drive velocities and bubble dimensions. Subsection \ref{Invariant comparison with Mattingly} compares our computational results with those presented by Mattingly et al. \cite{Mattingly}, who assert that Natário’s model is more realistic than Alcubierre’s. Similarly, Loup \cite{Loup, Loup6}, referenced by \cite{Mattingly}, claims greater feasibility for Natário’s isochoric flow model. However, our investigation shows the opposite; Natário’s enforcement of zero-differential-volume-change $K=0$ renders his model even more impractical than Alcubierre’s. Our study contrasts sharply with that of \cite{Mattingly}, who underestimated the amplitudes of the curvature invariants by 21 orders of magnitude, potentially misleading the understanding of Natário’s warp drive viability.

Finally, section \ref{SummaryResults} presents a summary of results, highlighting that the Weyl scalar is the curvature invariant with the highest amplitude (subsubsection \ref{sec:amplitude}). Previous warp drive studies have examined the Ricci curvature and the corresponding energy-momentum tensor, ignoring the crucial role of the Weyl curvature. We attribute the Weyl curvature's high amplitude to the form function's sharp localization $f(r)$ near the warp-bubble radius (subsection \ref{form function}). As discussed in subsubsection \ref{sec:amplitude}, the second derivatives of the Weyl curvature tensor are connected to the second derivatives of the stress-energy-momentum tensor (with nonlinear coupling terms) through an equation derived from a Bianchi identity and Einstein’s equations \cite{padmanabhan, deFelice}. The form function approximates a top-hat function near the warp-bubble radius, leading to pronounced amplitudes of higher-order derivatives $\partial^n f / \partial r^n$ for orders $n \geq 2$. The high amplitude of these higher-order derivatives of $f(r)$ near the warp-bubble radius elucidates the critical, yet previously under-emphasized, local, and significant role of the Weyl curvature tensor in warp-bubble spacetimes.

\section{Preliminaries}\label{sec2}

 This section introduces the notation and units used throughout this paper and presents the coordinate system of the zero-differential-volume-change Natário warp drive. We then define the tensors and curvature invariants used in our analysis and computational visualizations. These definitions are crucial to ensure that the analysis of Natário's warp drive in the remainder of the paper is clear and unambiguous.

\subsection{Notation, Units, and Coordinate System}\label{Notation}

\begin{figure}[htbp]
    \centering
    \begin{subfigure}[b]{0.7\textwidth}
        \includegraphics[width=\textwidth]{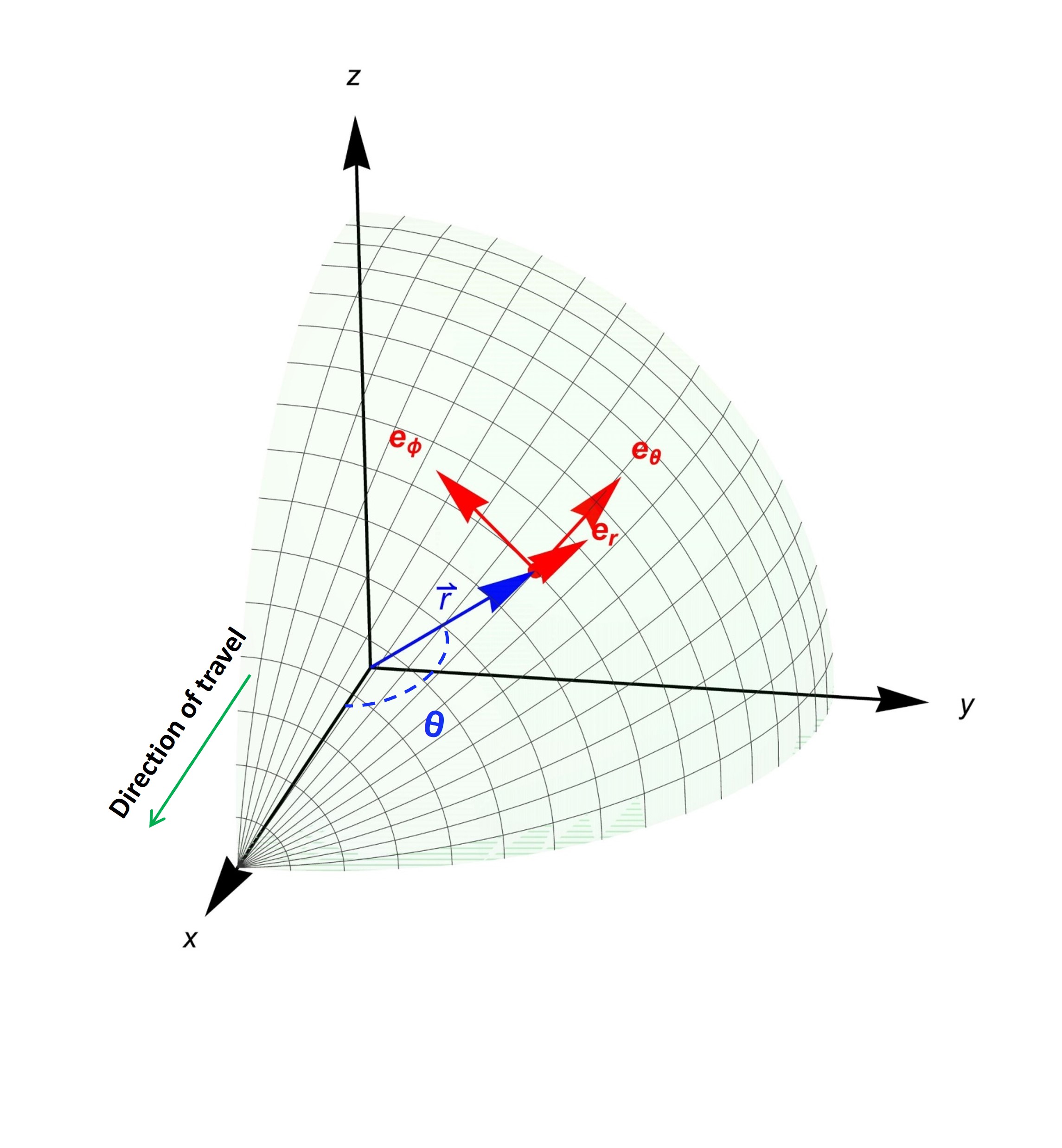}
       \caption{$\mathit{P}$ at $(r=\rho,\;0<\theta<\pi/2,\;0<\phi<\pi/2)$, with $\theta$ measured from $+x$ and $\phi$ in the $yz$ (equatorial) plane from $+y$ toward $+z$}

        \label{fig:1a}
    \end{subfigure}
    \hfill
    \begin{subfigure}[b]{0.4\textwidth}
        \includegraphics[width=\textwidth]{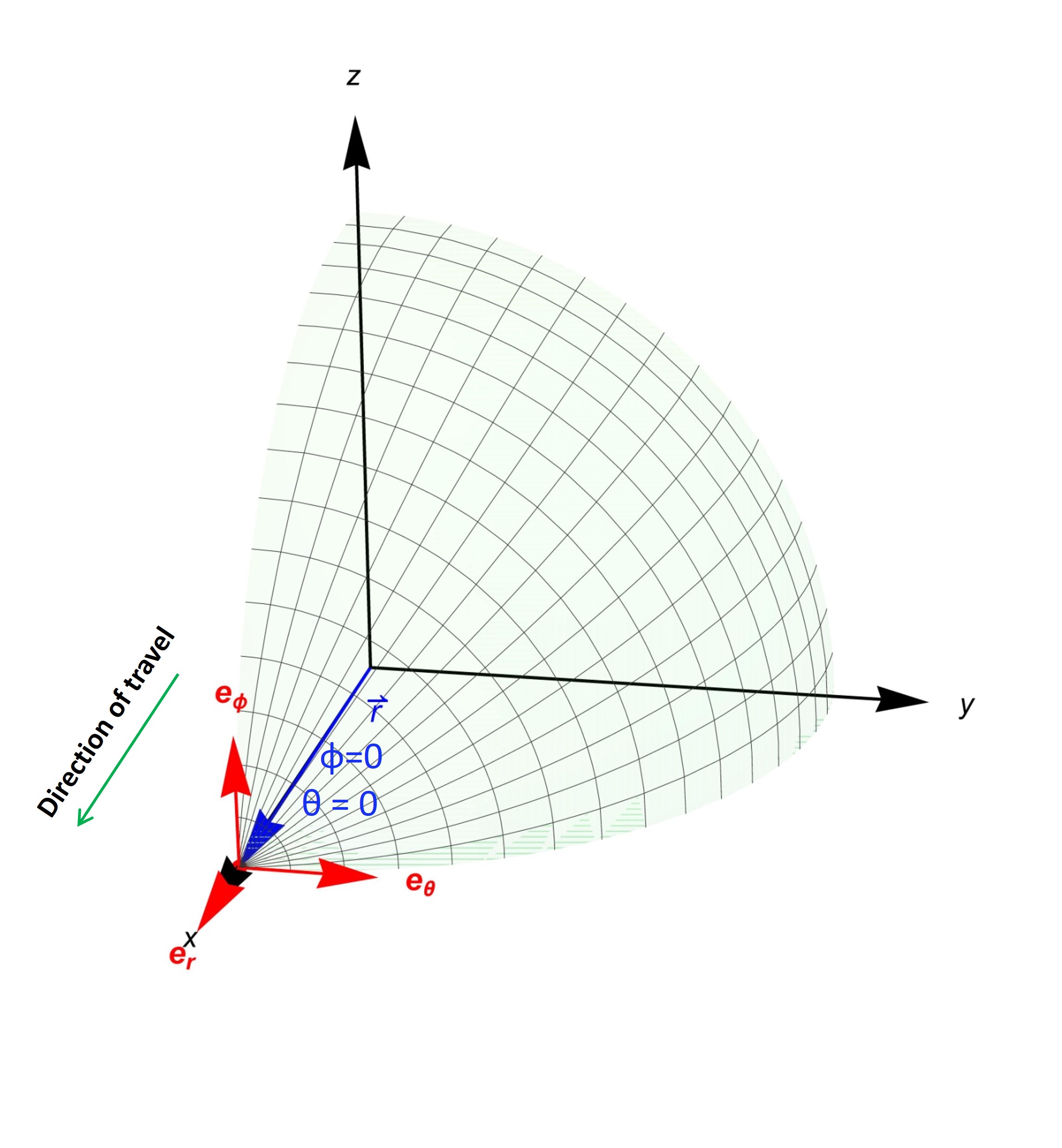}
        \caption{$\mathit{P}$ at $(r=\rho,\,\theta=0,\,\phi=0)$}
        \label{fig:1b}
    \end{subfigure}
    \hfill
    \begin{subfigure}[b]{0.4\textwidth}
        \includegraphics[width=\textwidth]{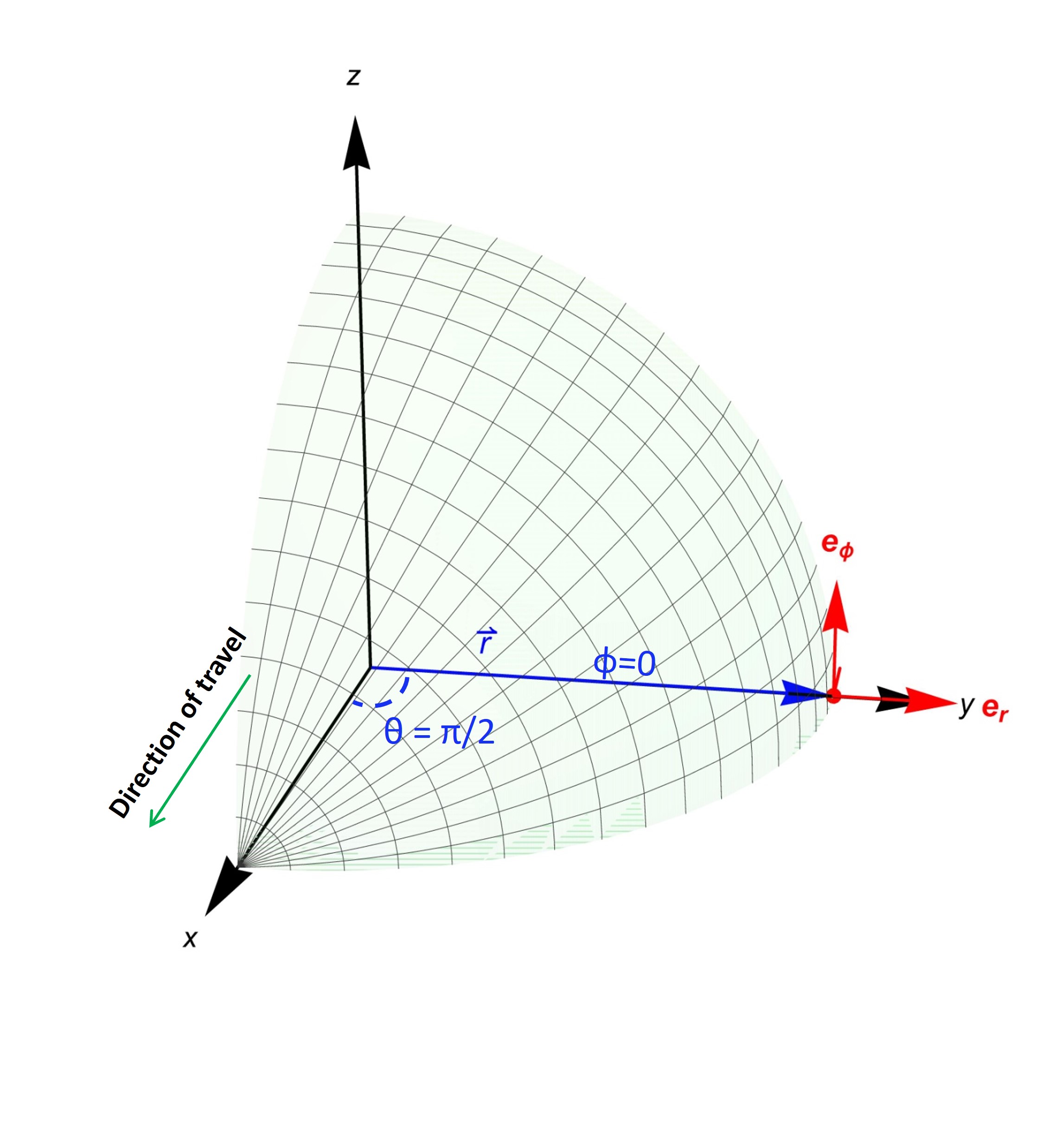}
        \caption{$\mathit{P}$ at $(r=\rho,\,\theta=\pi/2,\,\phi=0)$}
        \label{fig:1c}
    \end{subfigure}
    \caption{Natário's spherical coordinate system $(r, \theta, \phi)$ for the warp drive moving in the $x$ direction.}
    \label{Fig1_SpherCoord}
\end{figure}

\begin{itemize}
    \item Spacetime 4-tensors are denoted with lowercase Greek indices $\alpha, \beta, \ldots = 0, 1, 2, 3$.
    \item Spatial 3-tensors are denoted with lowercase Latin indices $a, b, \ldots = 1, 2, 3$.
    \item Spacetime 4-tensors (that have spatial counterparts) are denoted by the index $(4)$ on the top left of the tensor: $^{(4)}R_{\mu \nu}$.
    \item Spatial 3-tensors (that have spacetime counterparts) are denoted by the index $(3)$ on the top left of the tensor: $^{(3)}R_{i j}$.  
    \item The spacetime metric is denoted by $g_{\mu \nu} := \, ^{(4)}g_{\mu \nu}$, and has the Lorentzian signature $(-, +, +, +)$.
    \item The spatial metric is denoted by $\gamma_{ij} := \, ^{(3)}g_{ij}$, and has a positive definite signature $(+, +, +)$.
   \item ``3+1 formalism'' refers to an approach for solving Einstein's field equations that involves decomposing 4-dimensional spacetime into 3-dimensional spatial hypersurfaces evolving in time \cite{AlcubierreBook, Gourgoulhon, baumgarte_shapiro_2010, Misner1973}. The acronym ADM is derived from the initials of Arnowitt, Deser, and Misner, who developed a Hamiltonian formulation of General Relativity using this formalism \cite{ADM}.
    \item Covariant differentiation in space (in 3+1 decomposition) is denoted by $D_a$ or $D^a$.
    \item The components of the Riemann curvature tensor $^{(4)}R^{\kappa}_{\; \mu \kappa \nu}$ and of the Ricci curvature tensor $^{(4)}R_{\mu \nu}$ are defined following the conventions in Misner et al. \cite{Misner1973}.
    \item Tetrad components of tensors are denoted with carets (hats) above the indices: $K_{\hat{i}\hat{j} }$, as in \cite{Misner1973}.
    \item Units are denoted by square brackets, e.g., $[\text{m}]$. Geometrized (natural) units are employed in the analysis. SI units are used in a few physical examples. Specifically:

    \begin{itemize}
       \item $G, G^{00}, G_{i}^{\ 0}, G^{0}_{\ i}$: $[\text{m}^{-2}]$ in both geometrized and in SI.
       \item $I, r_{1}$: $[\text{m}^{-4}]$ in both geometrized and in SI.
       \item $r_{2}$: $[\text{m}^{-6}]$ in both geometrized and in SI.
       \item $x, y, z, r, \rho$: $[\text{m}]$ in both geometrized and in SI.
       \item $\sigma$: $[\text{m}^{-1}]$ in both geometrized and in SI.
       \item $f, \theta$: $[\text{1}]$ (dimensionless) in both geometrized and in SI.
       \item $v, c$: $[\text{1}]$ (dimensionless) in geometrized and $[\text{m}\, \text{s}^{-1}]$ in SI.
       \item $t$: $[\text{m}]$ in geometrized and $[\text{s}]$ in SI.
       \item $\kappa$: $[\text{1}]$ (dimensionless) in geometrized and $[\text{N}^{-1}]$ in SI.
       \item $T, T^{00}=\varrho, T_{i}^{\ 0}, T^{0}_{\ i}$: $[\text{m}^{-2}]$ in geometrized and $[\text{N}\, \text{m}^{-2}] = [\text{J} \, \text{m}^{-3}]$ in SI.
\end{itemize}

     \item We adopt Nat\'{a}rio's \cite{Natário} spherical coordinate system convention, in which the direction of travel is designated as the polar $x$-axis and the angle $\theta$ is defined as the polar angle, measured from the polar $x$-axis towards the position vector $\mathbf{r}$ in the spherical coordinate system $(r, \theta, \phi)$ (Fig.~\ref{Fig1_SpherCoord}). Nat\'{a}rio also defines orthonormal basis vectors $\mathbf{e_{r}}$, $\mathbf{e_{\theta}}$ and $\mathbf{e_{\phi}}$ at a point $\mathit{P}$. Nat\'{a}rio's application of the Hodge star operator implies a right-handed coordinate system, as evidenced by his definitions of the basis vectors in relation to the spherical coordinates $(r, \theta, \phi)$.

\end{itemize}

\subsection{Tensor definitions}\label{Tensor definitions subsec2.3}  

We introduce several trace-adjusted tensors rendered traceless in four-dimensional spacetime ($n=g_{\alpha \beta} \: g^{\alpha \beta}=\delta_{\alpha}^{\alpha}=4$) \cite{Rodal}. Such tensors are denoted by a hat, e.g., $\widehat{T}^{\alpha \beta}$, distinguishing them from their unadjusted counterparts, e.g., $T^{\alpha \beta}$.\par 
A traceless Ricci curvature tensor $\widehat{R}_{\alpha \beta}$ is defined as follows: in terms of the 2nd-rank Ricci curvature tensor $R_{\alpha \beta}$, Ricci curvature scalar $R \equiv R_{\alpha}\,^{\alpha}$, and metric tensor $g_{\alpha \beta}$:

\begin{equation}
\widehat{R}_{\alpha \beta} \equiv R_{\alpha \beta}-\frac{1}{4} R \: g_{\alpha \beta}.
\end{equation}

Similarly a trace-adjusted Einstein curvature tensor $\widehat{G}_{\alpha \beta}$ is defined in terms of the 2nd-rank Einstein curvature tensor $G_{\alpha \beta}$, and Einstein curvature scalar $G \equiv G_{\alpha}\,^{\alpha}$:

\begin{subequations}
\begin{align}
\widehat{G}_{\alpha \beta}&\equiv G_{\alpha \beta} -\frac{1}{4} \: G \:g_{\alpha \beta} \quad \text{for } n = 4 \label{eq:HatEins_a} \\
&= \widehat{R}_{\alpha \beta} \quad \text{for } n = 4 \label{eq:HatEins_b}
\end{align}
\end{subequations}

In four-dimensional spacetime ($n=4$), the trace-adjusted Einstein curvature tensor $\widehat{G}_{\alpha \beta}$ happens to be exactly equal to the trace-adjusted Ricci curvature tensor $\widehat{R}_{\alpha \beta}$, while the traces of the unadjusted tensors are equal in magnitude with opposite signs \cite{Rodal}:

\begin{equation}
\begin{aligned}
G &= -R
\end{aligned}
\quad \text{for } n = 4.
\label{EinsRiccN4}
\end{equation}

The Weyl curvature tensor $C_{\alpha\beta\gamma\delta}$ is a traceless, 4th-rank tensor invariant under conformal rescaling of the metric. It is defined as what remains of the 4th-rank Riemann curvature tensor $R_{\alpha\beta\gamma\delta}$ after subtraction of the Kulkarni–Nomizu product \cite{andrews2010ricci} (denoted by $\oland$ ) of the 2nd-rank trace adjusted Ricci tensor $\widehat{R}_{\alpha \beta}$ with the metric tensor $g_{\alpha \beta}$, and subtraction of the $\oland$ product of the metric tensor with itself:

\begin{equation}
\begin{aligned}
C_{\alpha\beta\gamma\delta} &\equiv R_{\alpha\beta\gamma\delta} -\frac{1}{2} \widehat{R}_{\alpha \beta} \oland g_{\alpha \beta}-\frac{1}{24} g_{\alpha \beta} \oland g_{\alpha \beta}\\
\label{weylDef}
\end{aligned}
\quad \text{for } n = 4.
\end{equation}

Eqs.~(\ref{weylDef}) and (\ref{eq:HatEins_b}) can be combined to decompose the Riemann curvature tensor $R_{\alpha\beta\gamma\delta}$ into the traceless tensor $C_{\alpha\beta\gamma\delta}$, the trace adjusted Einstein tensor $\widehat{G}_{\alpha \beta} $ and the metric tensor.

We also define a trace-adjusted stress-energy-momentum tensor ($\widehat{T}_{\alpha \beta}$) as follows, in terms of the 2nd-rank stress-energy-momentum tensor $T_{\alpha \beta}$, its contracted scalar $T\equiv T_{\alpha}\,^{\alpha}$:

\begin{equation}
\widehat{T}_{\alpha \beta} \equiv T_{\alpha \beta}-\frac{1}{4} T \: g_{\alpha \beta} \quad \text{for } n = 4.\label{HatT}
\end{equation}

With the cosmological constant omitted due to its minimal impact on local scales, Einstein's field equations are expressed as $G_{\alpha \beta} = \kappa \, T_{\alpha \beta}$. Here, $\kappa$ represents the Einstein gravitational coupling parameter. Einstein’s field equations can be decomposed into an invariant scalar equation in terms of the traces $G$  and $T$, and a tensor equation in terms of traceless tensors (known in the continuum mechanics literature as `deviatoric' tensors) $\widehat{G}$  and $\widehat{T}$, with $1/\kappa$ acting as a spacetime stiffness:

\begin{subequations}
\begin{align}
    T &=\frac{1}{\kappa} G \quad \text{for } n = 4 \label{eq:FieldEq_a} \\
    \widehat{T}_{\alpha \beta} &= \frac{1}{\kappa}\widehat{G}_{\alpha \beta} \quad \text{for } n = 4 \label{eq:FieldEq_b}
\end{align}
\end{subequations}

\subsection{Curvature invariants in General Relativity}\label{StdCurvatureInvariants subsec2.3}

Einstein \cite{EinsteinCollected} expressed his preference for the term ``theory of invariance'' over ``theory of relativity'' in a letter to Eberhard Zschimmer dated Nov. 30, 1921. Einstein wrote that ``invariance theory'' would describe the research method of his theory more accurately but noted that it would be confusing to change the commonly accepted name ``relativity theory.'' 

In General Relativity, invariants are crucial for analyzing spacetime properties, as they remain unchanged under coordinate transformations \cite{Carminati, Harvey, Zakhary, Cherubini, Santosuosso, MacCallum}. They help distinguish between true physical singularities (such as the one at the center of a black hole) and coordinate artifacts (such as the apparent singularity at a black hole's event horizon), which can be resolved through appropriate coordinate choices.

Their ability to encapsulate the geometric essence of spacetime in scalar quantities makes them a fundamental aspect of theoretical investigations in General Relativity.
Curvature invariants can be used to classify different types of spacetimes. By examining specific invariants, we can determine whether a spacetime is conformally flat, or if it possesses certain symmetries or characteristics, such as being of Petrov type $D$ or exhibiting specific gravitational wave patterns. This classification helps in understanding the possible behaviors and stability of spacetimes under various physical conditions.
Additionally, curvature invariants play a role in the study of exact solutions to Einstein's field equations. They provide a powerful tool for verifying the correctness of these solutions and for comparing different solutions. For example, the quadratic Kretschmann scalar is often used to compare the curvature of different spacetimes and to study the effects of tidal forces near massive objects like black holes.

Curvature invariants are scalar quantities derived from curvature tensors like the Riemann, Weyl, Einstein, and Ricci tensors. They are formulated using contraction, covariant differentiation, trace adjustment, and dualization operations. Polynomial methods are commonly used to construct these invariants, resulting in linear, quadratic, cubic, quartic, quintic, and higher-order polynomial invariants.

Despite their broad utility, polynomial invariants have limitations. Notably, they fail to differentiate certain Kundt spacetimes – Lorentzian manifolds with geodesic null congruences and vanishing optical scalars (expansion, twist, and shear) \cite{Coley_2009}. This limitation is particularly evident in spacetimes with vanishing scalar invariants, such as certain Petrov type $N$ spacetimes, characterized by a quadruple principal null direction and relevant to transverse gravitational waves, and Petrov type III spacetimes, with one triple and one simple principal null direction, typically associated with longitudinal gravitational radiation. \par

One of the most important invariants of the Riemann tensor of a four-dimensional Einsteinian manifold is the quadratic scalar invariant, formed from the contraction of the Riemann tensor with itself, known as the Kretschmann scalar, defined as:

\begin{equation}
    K \equiv R_{\alpha\beta\gamma\delta} R^{\alpha\beta\gamma\delta},  
    \label{Kretschmann}
\end{equation}
where $R_{\alpha\beta\gamma\delta}$ are the fully covariant and $R^{\alpha\beta\gamma\delta}$ are the fully contravariant components of the Riemann curvature tensor.  If the Kretschmann scalar is finite at a specific point in a spacetime manifold, it indicates that the squared components of the Riemann curvature tensor, summed over all indices, are finite at that point.\par

Another quadratic scalar invariant (which appears as a gravitational term in the Lagrangian in conformal gravity theories), formed from the contraction of the traceless Weyl curvature tensor with itself, is known as the Weyl scalar $I$ defined as:

\begin{equation}
\begin{aligned}
I &\equiv C_{\alpha\beta\gamma\delta} C^{\alpha\beta\gamma\delta}, \\
\end{aligned}
\label{WeylScalar}
\end{equation}
where $C_{\alpha\beta\gamma\delta}$ and $C^{\alpha\beta\gamma\delta}$ are the fully covariant and the fully contravariant components, respectively, of the Weyl curvature tensor.\par

Similarly, we can define a quadratic scalar invariant (which also appears as a gravitational term in the Lagrangian in conformal gravity theories) in terms of the trace-adjusted Ricci curvature tensor (which for 4-dimensional spacetime can also be expressed in terms of the trace-adjusted Einstein curvature tensor using Eq.~(\ref{EinsRiccN4})) as:
    
\begin{subequations}
\begin{align}
r_{1} &\equiv \widehat{R}_{\alpha}^{\:\:\:{\beta}}\widehat{R}_{\beta}^{\:\:\:{\alpha}} \label{eq:r1_a} \\
&= \widehat{G}_{\alpha}^{\:\:\:{\beta}}\widehat{G}_{\beta}^{\:\:\:{\alpha}} 
\quad \text{for } n = 4.\label{eq:r1_b} 
\end{align}
\end{subequations}

This definition differs from Carminatti et al. \cite{Carminati} who define this quadratic traceless Ricci invariant  with an arbitrary prefactor of $1/4$ in the 4-dimensional Einsteinian manifold.\par

In the Einsteinian four-dimensional manifold with $n=4$, the Kretschmann scalar can be decomposed into the traceless Weyl scalar $I$, the quadratic invariant $r_{1}= \widehat{G}_{\alpha}^{\:\:\:{\beta}}\widehat{G}_{\beta}^{\:\:\:{\alpha}}$ (formed from the traceless Einstein curvature tensor), and the square of the Einstein curvature scalar $G$ as follows (from Cherubini et al.\cite{Cherubini} and Eq.~\ref{EinsRiccN4}):

\begin{equation}
\begin{aligned}
K &= I+2 \:{r}_{1} +\frac{1}{6} {G}^2  \quad \text{for } n = 4.\\
\end{aligned}
\label{KIG}
\end{equation}\par

This quadratic invariant decomposition of the Kretschmann scalar can be interpreted (considering free fall geodesic motion) as decomposing the deformation into volumetric and shear modes of deformation. \par
\par
For a fluid under isotropic pressure, the square of the trace of the Einstein curvature tensor, $G^2$, or equivalently, the square of the trace of the Ricci curvature tensor, $R^2$, provides information about spacetime curvature change related to volume changes—for instance, due to the uniform radial deformation of a perfect sphere. In the case of anisotropic pressure, $G^2$, or equivalently $R^2$, encapsulates information about spacetime curvature due to pure-shear strain (shear without spatial rotation of the principal axes of stress); in other words, unequal curvature change in different directions. For example, the curvature change due to the deformation of an initially perfect sphere into an oblate or prolate spheroid. 

The quadratic traceless invariants, $r_{1}$ and $I$, contain information about the curvature due to a simple-shear change of spacetime (shear due to planar spatial rotation of the principal axes of stress). \par

The invariant $r_{1}$ represents the curvature due to simple-shear change arising from local sources, from the self-contraction of the traceless curvature tensor $\widehat{G}_{\alpha}^{\:\:\:{\beta}}\widehat{G}_{\beta}^{\:\:\:{\alpha}}$. This is locally sourced, by virtue of the Einstein field equations, by the traceless stress-energy-momentum tensor $\widehat{T}_{\alpha \beta}$. \par

We will consider one cubic invariant formed from the trace-adjusted Ricci curvature tensor (or, equivalently, in four-dimensional spacetime, from the traceless Einstein curvature tensor) as follows:

\begin{subequations}
\begin{align}
r_{2} &\equiv \widehat{R}_{\alpha}^{\:\:\:{\beta}}\widehat{R}_{\beta}^{\:\:\:{\gamma}} \widehat{R}_{\gamma}^{\:\:\:{\alpha}} \label{eq:r2_a} \\
&= \widehat{G}_{\alpha}^{\:\:\:{\beta}}\widehat{G}_{\beta}^{\:\:\:{\gamma}} \widehat{G}_{\gamma}^{\:\:\:{\alpha}}\quad \text{for } n = 4. \label{eq:r2_b}
\end{align}
\end{subequations}

Carminati et al. \cite{Carminati} define the cubic invariant $r_{2}$ with an arbitrary prefactor of $-1/8$. This invariant encapsulates curvature information attributable to the three-dimensional shear deformation of spatial components. While $r_{1}$ pertains to curvature induced by in-plane simple-shear deformation, characterized by rotation of principal stress axes, $r_{2}$ captures the curvature resulting from a more complex shear deformation involving spatial, three-dimensional rotation of principal axes.\par

\section{Tensorial and tetrad analyses of Nat\'{a}rio's spacetime}\label{sec3}

This section analyzes Natário's zero-differential-volume-change spacetime, utilizing both the curvilinear coordinate tensor formalism and the tetrad (Newman-Penrose) formalism. Both the Alcubierre and Natário warp drive models are based on the 3+1 ADM formalism, therefore the analysis focuses on the 3+1 decomposition of spacetime.

Unlike previous works which relied on Natário's orthonormal tetrad (vierbein) formalism, presenting `physical components' in a local Minkowski frame, our analysis in subsections \ref{Tensor comp. Vector field} to \ref{Momentum density}  employs the tensorial solution of zero-differential-volume-change Natário's spacetime using the curvilinear coordinate tensor components to provide a detailed curvilinear tensor component solution. This approach fills a gap in the literature where Natário's `zero-expansion' solution in curvilinear tensor components has not been previously presented. The subsections delve into the spacetime metric, matter energy density, and momentum density.

Subsections \ref{Newman Penrose} to \ref{Clarifying} focus on the Petrov type classification of the zero-differential-volume-change Natário warp drive and also examine the literature claim \cite{Mattingly} that it is a Class $B_1$ spacetime. Using the Newman-Penrose (double-null, complex) tetrad formalism, we prove that Natário's spacetime is actually Petrov type I, characterized by the lack of algebraic symmetries and the presence of four unique and real principal null directions. We further establish that Natário’s spacetime does not fit the definition of a Class $B$ warped product spacetime, as these are restricted to Petrov types $D$ or $O$.

Subsection \ref{form function} discusses the radially-dependent warp-bubble form function $f(r)$ and its approximations. This discussion is necessary for performing calculations and visualizations since Natário \cite{Natário} did not suggest a specific $f(r)$ function.

\subsection{Tensor components of the 3+1 ADM shift vector field}\label{Tensor comp. Vector field}

Nat\'{a}rio \cite{Natário} characterized a warp drive spacetime, generated by a velocity vector field $\mathbf{X}$, as a spacetime foliated by Euclidean 3-spaces within the framework of the 3+1 Arnowitt-Deser-Misner (ADM) formalism \cite{Misner1973,Gourgoulhon,AlcubierreBook,baumgarte_shapiro_2010}. This formalism decomposes spacetime into a series of evolving three-dimensional spatial slices, each representing a distinct instant in time.  In Nat\'{a}rio's model, the integral curves (the paths traced out by this vector field) of $\mathbf{X}$ correspond to time-like geodesics. These geodesics are associated with Eulerian observers, whose 4-velocity is the future-pointing covariant component of the unit normal vector to the Cauchy surfaces.  The 3+1 foliation's feasibility necessitates that initial conditions on the Cauchy surfaces uniquely determine the spacetime’s evolution. Alcubierre \cite{Alcubierre_1994} first wrote ``Notice that as long as the metric $\gamma_{ij}$ is positive definite for all values of t (as it should in order for it to be a spatial metric), the spacetime is guaranteed to be globally hyperbolic.''  Later, Ralph et al. \cite{SpinAlcubierre}, presented a metric for an Alcubierre warp drive on a rotating platform. The rotation in their model, leading to the formation of closed timelike curves within the spacetime, constitutes a violation of global hyperbolicity, however, their 3+1 foliation remains valid \textit{locally} within the horizons. Nat\'{a}rio \cite{Natário} wrote ``This is a globally hyperbolic spacetime foliated by Euclidean 3-spaces such that the integral lines of $\mathbf{X}$ represent time-like geodesics (corresponding to the so-called \textit{Eulerian observers}).''  

A (Nat\'{a}rio or Alcubierre) warp drive satisfies the global hyperbolicity condition \textit{only at subliminal speeds}. In contrast, a superluminal warp drive will exhibit Cauchy horizons, as demonstrated, for instance, in the Carter–Penrose diagrams depicted in Fig.2 of Finazzi et al.\cite{Finazzi}. Consequently, for a superluminal warp drive, the 3+1 decomposition decomposition remains valid solely \textit{locally} within these Cauchy horizons.

Nat\'{a}rio \cite{Natário} first specified the line element with signature $(-,+,+,+)$ with time coordinate $t$ and orthogonal Cartesian Euclidean coordinates $(x,y,z)$ for a generalized Alcubierre spacetime. Denoting the associated components of the vector field $(\mathbf{X})_{\text{Alcubierre}}$ as $(X,Y,Z)$, this Alcubierre warp drive line element is given by:

\begin{equation}
    ds^2 = -dt^2 + \left(dx - X\, dt\right)^2 + \left(dy - Y\, dt\right)^2 + \left(dz - Z\,dt\right)^2 .
    \label{eq:lineelementnatario}
\end{equation}

Nat\'{a}rio \cite{Natário} then introduced spherical coordinates $(r,\theta,\phi)$ (Fig.~\ref{Fig1_SpherCoord})  to characterize the foliated Euclidean three-dimensional space, and designated the direction of travel, the $x$-axis, as the polar axis. In this coordinate system, Nat\'{a}rio \cite{Natário} defined the vector field $\mathbf{X}$ and expressed its components relative to the unit orthonormal basis vectors $(\mathbf{e_r},\mathbf{e_{\theta}},\mathbf{e_{\phi}})$ instead of the curvilinear coordinate basis vectors  $(\mathbf{g_{r}},\mathbf{g_{\theta}},\mathbf{g_{\phi}})$.  Consequently, the components defined by Nat\'{a}rio are tetrad components (``physical components'') \cite{Misner1973, Wald, deFelice, PlebanskiKrasinski, Chandrasekhar} by definition, by virtue of their association with the unit orthonormal basis.  In contrast, curvilinear coordinate tensor components are resolved in terms of the coordinate basis vectors  $(\mathbf{g_{r}},\mathbf{g_{\theta}},\mathbf{g_{\phi}})$ characterizing the change of the position vector $\mathbf{r}$.  Notably, the spherical coordinate basis vectors $(\mathbf{g_{\theta}},\mathbf{g_{\phi}})$ are not unit vectors. Tetrad components generally do not transform according to the curvilinear coordinate tensor transformation laws and are not components of tensors. Instead, tetrad components transform from one tetrad basis to another through Lorentz transformations, which are not the general coordinate transformations of curvilinear coordinates. Furthermore, the tetrad formalism utilizes Ricci rotation coefficients, which encode the connection between the tetrad fields and the spacetime geometry. These coefficients provide information on how the tetrad must rotate in a curved spacetime, as a curved spacetime is not compatible with a Minkowski metric for the entire spacetime. They describe how the local inertial frames (tetrads) change from point to point in a curved manifold, ensuring the correct representation of gravitational effects. 

We adopt the notation for tetrad components as presented by Misner et al. \cite{Misner1973} wherein carets (hats) are employed above the indices of the tetrad components. This is to distinguish tetrad components from the curvilinear coordinate components, which are denoted without hats. This leads to the following representation of the vector field $\mathbf{X}$  in terms of its tetrad components with respect to the unit orthonormal basis vectors$(\mathbf{e_r},\mathbf{e_{\theta}},\mathbf{e_{\phi}})$:

\begin{equation}
    \mathbf{X}=X^{\hat{r}} \mathbf{e_r}+X^{\hat{\theta}} \mathbf{e_{\theta}}+X^{\hat{\phi}} \mathbf{e_{\phi}}
    \label{eq:Xphysicoordnatario}
\end{equation}

Tetrad orthonormal basis components reflect measurements an observer in free fall (not subjected to non-gravitational forces) would record.  In General Relativity, a single inertial frame cannot globally encompass spacetime curvature. However, in a sufficiently small spacetime region, one can establish a frame with negligible spacetime curvature effects, and Special Relativity applies. This is known as a local Lorentz frame.

Nat\'{a}rio constructed a divergenceless (zero expansion-contraction of the Eulerian observers’ volume element) vector field $\mathbf{X}$ in his specific spherical coordinate system (Fig.~\ref{Fig1_SpherCoord}). Adjusted for caret notation to distinguish tetrad components, Nat\'{a}rio's definition of the tetrad components of the vector field $\mathbf{X}$, in terms of the polar angle $\theta$, a scalar function $f(r)$ of the radial distance $r$, and a constant velocity parameter $v$, is as follows:

\begin{equation}
\begin{pmatrix}
X^{\hat{r}} \\
X^{\hat{\theta}} \\
X^{\hat{\phi}}
\end{pmatrix}
=
\begin{pmatrix}
-2 \, v \, f \, \cos(\theta) \\
v \, (2 f + r \, \frac{\partial f}{\partial r} ) \sin(\theta) \\
0
\end{pmatrix} .
\label{eq:Xvectorphysic}
\end{equation}

From hereon, we denote the three-dimensional Euclidean spatial metric as $\gamma_{ij} =\, ^{(3)}g_{ij}$ to distinguish it from the four-dimensional pseudo-Riemannian metric of General Relativity, denoted as $g_{\alpha \beta} =\, ^{(4)}g_{\alpha \beta}$.

The covariant components $\gamma_{ij}$ of the Euclidean spatial metric in the spherical coordinate system $(r,\theta,\phi)$, as defined by Nat\'{a}rio (see subsection \ref{Notation}), are represented by the following matrix:

\begin{equation}
\gamma_{ij}=
\begin{bmatrix}
1 & 0 & 0 \\
0 & r^2 & 0 \\
0 & 0 & (r \sin(\theta))^2
\end{bmatrix} .
\label{eq:gmatrix3}
\end{equation}

Given that this matrix has zero off-diagonal entries, its inverse (the contravariant components $\gamma^{ij} =\, ^{(3)}g^{ij}$) is simply the inverse of the diagonal entries of $\gamma_{ij}$.

Consequently, the contravariant tensor components of the vector field $\mathbf{X}$ are:

\begin{equation}
\begin{pmatrix}
X^{r} \\
X^{\theta}\\
X^{\phi}
\end{pmatrix}
=
\begin{pmatrix}
X^{\hat{r}}\,/ \sqrt{\gamma_{rr}}\\
X^{\hat{\theta}}\,/ \sqrt{\gamma_{\theta \theta}} \\
X^{\hat{\phi}}\,/ \sqrt{\gamma_{\phi \phi}}
\end{pmatrix}
=
\begin{pmatrix}
-2 \, v \, f \, \cos(\theta) \\
v \, \left(\frac{2 f}{r} + \frac{\partial f}{\partial r} \right) \sin(\theta) \\
0
\end{pmatrix} .
\label{eq:Xcontra}
\end{equation}

And the covariant tensor components of the vector field $\mathbf{X}$ are:

\begin{equation}
\begin{pmatrix}
X_{r} \\
X_{\theta}\\
X_{\phi}
\end{pmatrix}
=
\begin{pmatrix}
-2 \, v \, f \, \cos(\theta) \\
v \, \left(2 f \, r + r^2 \frac{\partial f}{\partial r} \right) \sin(\theta) \\
0
\end{pmatrix} .
\label{eq:Xcov}
\end{equation}

We adopt the notation $\alpha$ for the lapse function and $\boldsymbol{\beta}$ for the shift vector as used by Gourgoulhon \cite{Gourgoulhon}, Alcubierre \cite{AlcubierreBook}, and Baumgarte et al. \cite{baumgarte_shapiro_2010}, aligning with the current standard convention in the field.  In contrast, Misner et al. \cite{Misner1973} used N with indices for the shift vector and without indices for the lapse function.

The 3+1 ADM formalism lapse function $\alpha = d\tau / dt$ measures how much proper time elapses between neighboring time slices along the normal vector. It determines the rate at which time flows as one moves from one slice to another in the spacetime manifold. In Nat\'{a}rio \cite{Natário}'s spacetime, $\alpha=1$.

The shift vector $\boldsymbol{\beta}$ in the 3+1 ADM formalism is defined on each spatial hypersurface. It measures the amount by which the spatial coordinates are shifted within each spatial slice relative to the normal vector to the hypersurface. This shift is essential for analyzing how the spatial coordinates evolve from one time slice to the next. 
The interpretation of the shift vector $\boldsymbol{\beta}$ is influenced by the perspective of the Eulerian observer, who is stationary relative to the spatial coordinates. In the Alcubierre warp drive model, the Eulerian observer is chosen to be at a sufficiently large distance,  $r \to \infty$, from the warp-bubble, such that this observer is stationary relative to the warp-bubble which is observed to be moving at a relative velocity $v$. Conversely, in Nat\'{a}rio's warp drive model, the Eulerian observer is selected to be at rest relative to the center of the warp-bubble $r=0$, and from this perspective, distant stars appear to move at a velocity of $-v$. Both of these perspectives are valid within the framework of the 3+1 ADM decomposition in General Relativity. The freedom to choose a shift vector reflects the freedom to relabel spatial coordinates on each slice and is a key aspect of the coordinate freedom inherent in General Relativity. It can be demonstrated that the selection of the sign of the shift vector arbitrarily determines the orientation (positive or negative) of the warp drive's velocity vector. As noted in \cite{SantiagoVisser} ``a  flow vector $v^{i}=-$ (shift vector)\dots The sign-flip on the shift vector is traditional in a warp drive context, and inspired by the notion of ``flow'' rather than ``shift.''.'' Accordingly, we adopt Nat\'{a}rio's perspective of the distant stars moving at velocity $-v$ with respect to the Eulerian observer (at rest relative to the center of the warp-bubble $r=0$), and therefore the shift vector $\boldsymbol{\beta}$ is defined as:

\begin{equation}
\boldsymbol{\beta}=\mathbf{X} .
\label{eq:shift vector}
\end{equation}

Consequently (from Eqs.~\ref{eq:Xcontra} and \ref{eq:shift vector}), the contravariant tensor components of the shift vector $\boldsymbol{\beta}$ are:

\begin{equation}
\begin{pmatrix}
\beta^{1} \\
\beta^{2}\\
\beta^{3}
\end{pmatrix}
=
\begin{pmatrix}
\beta^{r} \\
\beta^{\theta}\\
\beta^{\phi}
\end{pmatrix}
=
\begin{pmatrix}
-2 \, v \, f \, \cos(\theta) \\
v \, \left(\frac{2 f}{r} + \frac{\partial f}{\partial r} \right) \sin(\theta) \\
0
\end{pmatrix} .
\label{eq:BetaCtv}
\end{equation}

And (from Eqs.~\ref{eq:Xcov} and \ref{eq:shift vector}) the covariant tensor components of the shift vector $\boldsymbol{\beta}$ are:

\begin{equation}
\begin{pmatrix}
\beta_{1} \\
\beta_{2}\\
\beta_{3}
\end{pmatrix}
=
\begin{pmatrix}
\beta_{r} \\
\beta_{\theta}\\
\beta_{\phi}
\end{pmatrix}
=
\begin{pmatrix}
-2 \, v \, f \, \cos(\theta) \\
v \, \left(2 f \, r + r^2 \frac{\partial f}{\partial r} \right) \sin(\theta) \\
0
\end{pmatrix} .
\label{eq:BetaCov}
\end{equation}

\subsection{Spacetime metric}\label{Spacetime metric}

Understandably, Nat\'{a}rio \cite{Natário} did not present the curved spacetime metric for his zero-expansion warp drive because Nat\'{a}rio used instead an orthonormal tetrad with a local Minkowski metric. While a curved manifold cannot be globally represented by a Minkowski metric, the tetrad formalism provides a locally flat (Minkowski) representation at each point of the curved spacetime. Tetrad components generally do not transform according to the curvilinear coordinate tensor transformation laws and are not components of tensors. Instead, tetrad components transform from one tetrad basis to another through Lorentz transformations. The tetrad formalism utilizes Ricci rotation coefficients \cite{deFelice, Chandrasekhar, PlebanskiKrasinski}, which encode the connection between the local inertial frames (tetrads) and the curved spacetime geometry. The Ricci rotation coefficients provide the information on how the tetrad must rotate from point to point in a curved manifold to correctly represent gravitational effects in the curved spacetime.

Mattingly et al. \cite{Mattingly} construct a timelike comoving orthonormal tetrad field for the Nat\'{a}rio spacetime:

\begin{equation}
e^{\hat{\alpha}}\,_{\mu} =
\begin{bmatrix}
E^{\hat{1}} \\
E^{\hat{2}} \\
E^{\hat{3}} \\
E^{\hat{4}}
\end{bmatrix}
=
\begin{bmatrix}
1 & 0 & 0 & 0 \\
- 2 \,v_s\, f\, cos(\theta) & -1 & 0 & 0 \\
v_s \, \left(2 f  + r \frac{\partial f}{\partial r} \right) \sin(\theta) & 0 & -r & 0 \\
0 & 0 & 0 & r \, \sin(\theta)
\end{bmatrix}
.
\label{eq: MattinglyTetrad}
\end{equation}

Mattingly et al. \cite{Mattingly} then write that ``it can be verified that $g_{ij} = E_{i} \cdot E_{j}$.'' However, this equation is incorrect since it mixes the metric coordinate frame (lower) indices with the tetrad vectors' frame (which should be upper) indices (in this case). Most importantly, the resulting dot product is not Lorentzian: the time-time component does not have the opposite sign to the space-space components of the resulting, incorrect result. The correct equation \cite{PlebanskiKrasinski, Chandrasekhar, deFelice} to reconstruct the curvilinear coordinate metric should employ the Minkowski metric $\eta_{\hat{\alpha} \hat{\beta}}$ of the tetrad as follows:

\begin{equation}
g_{\mu \nu} = e_{\mu}\,^{\hat{\alpha}} \;\eta_{\hat{\alpha} \hat{\beta}}\;e^{\hat{\beta}}\,_{\nu}
\label{eq: Metric reconstruction}
\end{equation}

or, in matrix notation:

\begin{equation}
\begin{bmatrix}
g 
\end{bmatrix} =
\begin{bmatrix}
E^{\hat{1}} \\
E^{\hat{2}} \\
E^{\hat{3}} \\
E^{\hat{4}}
\end{bmatrix}^{\mathrm{T}}
\,
\begin{bmatrix}
-1 & 0 & 0 & 0 \\
0 & 1 & 0 & 0 \\
0 & 0 & 1 & 0 \\
0 & 0 & 0 & 1 
\end{bmatrix}
\,
\begin{bmatrix}
E^{\hat{1}} \\
E^{\hat{2}} \\
E^{\hat{3}} \\
E^{\hat{4}}
\end{bmatrix}
\label{eq: Matrix Metric reconstruction}
\end{equation}

Eq.~(\ref{eq: Metric reconstruction}) is fundamental in the tetrad formalism; it is used to translate between the curved spacetime metric $g_{\mu \nu}$ and the flat spacetime metric $\eta_{\hat{\alpha} \hat{\beta}}$  of local inertial tetrad frames. The azimuthal angle ($\phi$) coordinate component of the $E^{\hat{4}}$ tetrad vector defined in \cite{Mattingly} contributes the last positive component (for $r\, \sin(\theta)>0$) to the signature $(+,-,-,+)$ of the diagonal of the tetrad matrix $e^{\hat{\beta}}\,_{\nu}$ in Eq.~(\ref{eq: MattinglyTetrad}). This tetrad matrix signature is inconsequential for metric $g_{\mu \nu}$ reconstruction because all the diagonal components of the tetrad matrix $e^{\hat{\beta}}\,_{\nu}$ get squared in the reconstruction operation, thereby canceling the effects of any negative signs. The reconstructed metric's $g_{\mu \nu}$ signature is solely governed by the Minkowski metric's $\eta_{\hat{\alpha} \hat{\beta}}$ signature used in the Eq.~(\ref{eq: Metric reconstruction}) reconstruction operation.

We can also obtain the curved spacetime metric for Nat\'{a}rio's spacetime using the $3+1$ ADM decomposition formalism \cite{Misner1973, Gourgoulhon, AlcubierreBook, baumgarte_shapiro_2010}.

The line element, when expressed in terms of the 3-dimensional spatial metric $\gamma_{ij}$, the shift vector $\boldsymbol{\beta}$, and the lapse function $\alpha$ must be:

\begin{equation}
\mathrm{d}s^2 = -\alpha^2 \mathrm{d}t^2 + \gamma_{ij}(\mathrm{d}x^i + \beta^i \mathrm{d}t)(\mathrm{d}x^j + \beta^j \mathrm{d}t) .
\label{eq: Natario line element}
\end{equation}

Therefore the covariant components of the 4-dimensional spacetime metric $g_{\mu \nu}$, are:

\begin{equation}
g_{\mu \nu} =
\begin{pmatrix}
- \alpha^2 + \beta_{k} \beta^{k} & \beta_{1} & \beta_{2} & \beta_{3} \\
\beta_{1} & \gamma_{11} & \gamma_{12} & \gamma_{13} \\
\beta_{2} & \gamma_{21} & \gamma_{22} & \gamma_{23} \\
\beta_{3} & \gamma_{31} & \gamma_{32} & \gamma_{33}
\end{pmatrix} .
\label{eq: spacetime metric}
\end{equation}

In this matrix, the time-space and space-time components are given by the following expression (from Eq.~(\ref{eq:BetaCov})):
\begin{equation}
\left(
\begin{array}{c}
 g_{01} \\
 g_{02} \\
 g_{03} \\
\end{array}
\right)
=
\left(
\begin{array}{c}
 g_{10} \\
 g_{20} \\
 g_{30} \\
\end{array}
\right)
=
\left(
\begin{array}{c}
 \beta_{1} \\
 \beta_{2} \\
 \beta_{3} \\
\end{array}
\right) =
\left(
\begin{array}{c}
 -2\; v\; f \cos(\theta) \\
 v\; r\; \left(r \frac{\partial f}{\partial r} + 2 f\right)  \sin(\theta)\\
 0 \\
\end{array}
\right).
\label{eq: timespace metric}
\end{equation}
And the time-time component of this metric is:
\begin{equation}
g_{00} = - \alpha^2 + \beta_{k} \beta^{k} = -1 + 4 v^2 f^2 \cos^2(\theta) + v^2 \sin^2(\theta) \left(r \frac{\partial f}{\partial r} + 2 f\right)^2.
\label{eq: timetime metric}
\end{equation}

It is easy to verify that Eq.~\ref{eq: Matrix Metric reconstruction} is identical to Eq.~\ref{eq: spacetime metric} if one accommodates the fact that our velocity is opposite in sign, i.e., $ v = -v_s $.

Finally, the $3 \times 3$ submatrix of components $\gamma_{ij}$ in the lower right-hand corner of the $4 \times 4$ spacetime metric $g_{\mu \nu}$ is the same spatial metric previously defined in Eq.~(\ref{eq:gmatrix3}).

Notice that although, in the matrix of covariant components $g_{\mu \nu}$, the lower right-hand corner $3 \times 3$ submatrix is independent of the shift vector $\boldsymbol{\beta}$; this is not the case for the matrix of contravariant metric components $g^{\mu \nu}$:

\begin{equation}
g^{\mu \nu} = \frac {1}{\alpha^2} 
\begin{pmatrix}
-1 & \beta^{1} & \beta^{2} & \beta^{3} \\
\beta^{1} & \left(\alpha^2 \gamma^{11} - \beta^1 \beta^1 \right) & \left(\alpha^2 \gamma^{12} - \beta^1 \beta^2 \right) & \left(\alpha^2 \gamma^{13} - \beta^1 \beta^3 \right) \\
\beta^{2} & \left(\alpha^2 \gamma^{21} - \beta^2 \beta^1 \right) & \left(\alpha^2 \gamma^{22} - \beta^2 \beta^2 \right) & \left(\alpha^2 \gamma^{23} - \beta^2 \beta^3 \right) \\
\beta^{3} & \left(\alpha^2 \gamma^{31} - \beta^3 \beta^1 \right) & \left(\alpha^2 \gamma^{32} - \beta^3 \beta^2 \right) & \left(\alpha^2 \gamma^{33} - \beta^3 \beta^3 \right)
\end{pmatrix} .
\label{eq: spacetime contra metric}
\end{equation}

Also, Mattingly et al. \cite{Mattingly} define a line element with a signature $(+,-,-,-)$ which is contrary to the $(-,+,+,+)$ signature utilized for the Alcubierre line element within the same article. This choice is also contrary to the conventions used in the $3+1$ ADM references \cite{Misner1973, AlcubierreBook, Gourgoulhon, baumgarte_shapiro_2010}, and the conventions adopted by Nat\'{a}rio \cite{Natário} and Alcubierre \cite{Alcubierre_1994, AlcubierreLobo}. This choice of signature results in the opposite sign for $R$, $G$, and $r_{2}$. However, the sign of $I$ (Weyl invariant), $r_1$, $R_{ij}$, $R^{ij}$, $G_{ij}$, and $G^{ij}$ (including $G^{00} = \varrho$ energy density) do not flip with signature change.  These conventions are arbitrary and should not impact the accuracy of calculations. This distinction is highlighted for clarity as it affects the sign comparison of curvature invariants between Alcubierre and Nat\'{a}rio warp drives.

\subsection{Tensor components of extrinsic curvature}\label{Extrinsic Curvature}

The extrinsic curvature $\mathbf{K}$ in the 3+1 ADM decomposition of spacetime is a second-order, purely spatial tensor. It quantifies the rate of change of the normal vectors to a hypersurface. Both the fully covariant components $K_{ij}$ and the fully contravariant components $K^{ij}$ of this tensor are symmetric matrices. It provides a mathematical characterization of the deviation of the normal vectors at neighboring points on the three-dimensional hypersurface, which is embedded in a four-dimensional spacetime. The extrinsic curvature effectively measures the rate at which the curvature of the hypersurface itself changes when it is `transported' along its normal vector.

The extrinsic curvature tensor components, derived from the Lie derivative of  the spatial metric $\boldsymbol{\gamma}$ (the evolution equation), expressed in terms of the curvilinear coordinate tensor components of the shift vector  $\boldsymbol{\beta}$, and the lapse function $\alpha$ are:

\begin{subequations}
\begin{align}
K_{ij} &= \frac{1}{2 \alpha} \left( D_{i}\left( \beta_{j} \right) + D_{j}\left( \beta_{i} \right) - \frac{\partial \gamma_{ij}}{\partial t}\right) ,\\
K_{i}^{\; j} &= \frac{1}{2 \alpha} \left( D_{i}\left( \beta^{j} \right) + D^{j}\left( \beta_{i} \right) - \gamma^{jk} \frac{\partial \gamma_{ik}}{\partial t} \right) ,\\
K^{i}_{\; j} &= \frac{1}{2 \alpha}\left( D^{i}\left( \beta_{j} \right) + D_{j}\left( \beta^{i} \right) - \gamma^{ik} \frac{\partial \gamma_{kj}}{\partial t}\right) ,\\
K^{ij} &= \frac{1}{2 \alpha}\left( D^{i}\left( \beta^{j} \right) + D^{j}\left( \beta^{i} \right) - \frac{\partial \gamma^{ij}}{\partial t} \right) ,
\end{align}
\label{eq:ExtrCurvDefinition}
\end{subequations}
where 

\begin{equation}
D^{i} (\beta^{j})=\gamma^{im} \, D_{m} (\beta^{j})
\end{equation}

denotes spatial covariant differentiation using the Christoffel symbols $^{(3)}\Gamma^i_{jk}$  computed from the three-dimensional metric $\gamma_{ij}$.
 
From Eq.~(\ref{eq:gmatrix3}), it is evident that the spatial metric \( \boldsymbol{\gamma} \) is independent of time. Additionally, $\alpha = 1$. Consequently, the extrinsic curvature depends only on the gradients of the shift vector $\boldsymbol{\beta}$. 
The non-zero components of the Christoffel symbols $^{(3)}\Gamma^i_{jk}$ in this coordinate system are:

\begin{subequations}
\begin{align}
^{(3)}\Gamma^1_{\ 2 2}=^{(3)}\Gamma^r_{\ \theta \theta} &= -r, & ^{(3)}\Gamma^1_{\ 33}=^{(3)}\Gamma^r_{\ \phi \phi} &= -r \sin^2 \theta, \\
^{(3)}\Gamma^{2}_{\ 33}=^{(3)}\Gamma^{\theta}_{\ \phi \phi} &= -\cos \theta \sin \theta, \\
^{(3)}\Gamma^{2}_{\ 12}=^{(3)}\Gamma^{\theta}_{\ r \theta} &= \frac{1}{r}, & ^{(3)}\Gamma^{2}_{\ 21}=^{(3)}\Gamma^{\theta}_{\ \theta r} &= \frac{1}{r}, \\
^{(3)}\Gamma^{3}_{\ 13}=^{(3)}\Gamma^{\phi}_{\ r \phi} &= \frac{1}{r}, &
^{(3)}\Gamma^{3}_{\ 31}=^{(3)}\Gamma^{\phi}_{\ \phi r} &= \frac{1}{r},\\
^{(3)}\Gamma^{3}_{\ 23}=^{(3)}\Gamma^{\phi}_{\ \theta \phi} &= \cot \theta, &
^{(3)}\Gamma^{3}_{\ 32}=^{(3)}\Gamma^{\phi}_{\ \phi \theta} &= \cot \theta. & &
\end{align}
\label{eq:Christoffel3D}
\end{subequations}

Armed with the Christoffel symbols, we can calculate the extrinsic curvature curvilinear tensor components from Eq.~(\ref{eq:ExtrCurvDefinition}); for example, the mixed covariant-contravariant components are:

\begin{equation}
K_{i}^{\; j}=
\begin{bmatrix}
-2 v \frac{\partial f}{\partial r} \cos(\theta)  & \frac{1}{r} v \left(\frac{\partial f}{\partial r}+\frac{r}{2} \, \frac{\partial^2 f}{\partial r^2} \right) \sin(\theta)  & 0 \\
 r\; v  \left(\frac{\partial f}{\partial r}+\frac{r}{2} \, \frac{\partial^2 f}{\partial r^2} \right) \sin(\theta)  & v \frac{\partial f}{\partial r} \cos(\theta) & 0 \\
0 & 0 &  v \frac{\partial f}{\partial r} \cos(\theta) 
\end{bmatrix} .
\label{eq:Ksubiexpj}
\end{equation}
The diagonal entries of the matrix of the mixed tensor components in curvilinear coordinates are identical to the diagonal components of the tetrad (``physical'') component matrix used by Nat\'{a}rio. The off-diagonal curvilinear coordinate tensor components differ from the tetrad components. $K_{2}^{\; 1} =\gamma_{22}\; K_{1}^{\; 2}=r^2 K_{1}^{\; 2}$ due to the fact that the metric diagonal component $\gamma_{22}$ is not unity. We also verify that the trace of the extrinsic curvature matrix is indeed zero, $K := \text{Tr}(K_{i}^{\; j}) = K_{i}^{\; i} = 0 $, by Nat\'{a}rio's construction aimed at achieving zero expansion-contraction of volume elements as observed by an Eulerian observer.

The ``zero change in differential volume'' discussed in the second half of Natário's paper is due to the specific choice of the flow vector field (the shift vector field in the 3+1 ADM formulation) constructed by Natário, which purposely satisfied the condition $K=0$. Natário achieved this ``zero change in differential volume'' by deliberately constructing the vector field to include a flow component in the plane perpendicular to the direction of motion (with continuous rotational symmetry around the axis of motion) in addition to the usual flow vector in the direction of motion, thus enforcing ``zero expansion'' of differential volume.

To prove that Natário's zero-expansion flow field satisfies continuous rotational symmetry around the axis of travel, consider that Natário's flow field components depend only on the radial distance $r$ and the polar angle $\theta$, but they are not functions of the azimuthal angle $\phi$ (Fig. \ref{Fig1_SpherCoord}). The azimuthal angle 

\begin{equation}
\phi = \mathrm{sgn}(z) \arccos\left(\frac{y}{\sqrt{y^2 + z^2}}\right)
\end{equation}

is defined within the plane ($yz$) orthogonal to the axis of travel (the $x$-axis). It is measured from the positive $y$-axis to the projection of the position vector $\mathbf{r}$ onto this plane. Its range is $\phi \in [-\pi, \pi]$. The surfaces $\phi = \text{constant}$ are planes containing the $x$-axis. Varying the angle $\phi$ effectively rotates these planes around the $x$-axis. Each of these planes intersects the spherical warp bubble along a circle of radius $\rho$, analogous to how meridians intersect the Earth's surface.

In these spherical coordinates, $r$ and $\theta$ are invariant under rotations around the axis of travel (the $x$-axis). Specifically, under a rotation by an angle $\phi$ around the $x$-axis, the coordinates $(y, z)$ transform as follows:
\begin{align}
y' &= y \cos(\phi) - z \sin(\phi), \\
z' &= y \sin(\phi) + z \cos(\phi).
\end{align}
Since the radial distance $r = \sqrt{x^2 + y^2 + z^2}$ and the the polar angle $\theta = \arccos\left(\frac{x}{r}\right)$ do not depend on the azimuthal angle $\phi$, any function $g(r, \theta)$ remains unchanged under such a rotation. Thus, Natário's flow field being of the form $g(r, \theta)$ exhibits continuous rotational (`cylindrical') symmetry around the axis of travel.

Constructing Natário's zero-expansion flow field in Cartesian coordinates is straightforward. From Eq.~(\ref{eq:lineelementnatario}) and Eq.~(\ref{eq:ExtrCurvDefinition}), it follows that for Natário's generic line element in Cartesian coordinates, setting the trace of the extrinsic curvature to zero:

\begin{align}
\text{Tr}(K_{ij}) = 0 = \frac{1}{\alpha} \left( \frac{\partial X}{\partial x} + \frac{\partial Y}{\partial y} + \frac{\partial Z}{\partial z} - \frac{\partial \gamma_{ij}}{\partial t} \right)
\end{align}

If the spatial metric $\gamma_{ij}$ is time-independent, i.e., $\frac{\partial \gamma_{ij}}{\partial t} = 0$, it follows that for a warp drive to satisfy zero expansion ($\text{Tr}(K_{ij}) = 0$), the flow field must be multiaxial. This is because if it is a uniaxial flow field, as in Alcubierre's original warp drive \cite{Alcubierre_1994, AlcubierreLobo} that had $Y = Z = 0$ and $X = v f(r)$, then the equation becomes simply

\begin{align}
0 = v \left( \frac{\partial f(r)}{\partial r} \, \frac{\partial r}{\partial x} \right)
\end{align}

with trivial solution $v = 0$ (showing that isochoric uniaxial flow for $v>0$ is impossible for the Alcubierre warp drive). Since spatial volume is 3-dimensional, the flow field must be multiaxial to satisfy the zero expansion constraint for a generic warp drive to be able to move at a non-zero velocity (as observed by Eulerian observers). For time-independent $\gamma_{ij}$, it  must satisfy:

\begin{align}
\text{Tr}(K_{ij}) = 0 = \frac{\partial X}{\partial x} + \frac{\partial Y}{\partial y} + \frac{\partial Z}{\partial z}
\end{align}

Which shows that for time-independent $\gamma_{ij}$, zero expansion is equivalent to zero divergence of the flow field.

The simplest case of zero expansion is, therefore, a flow field satisfying continuous rotational symmetry around the axis of travel, in which case, a zero-expansion warp drive must satisfy:

\begin{align}
\frac{\partial Y}{\partial y} &= \frac{\partial Z}{\partial z} = -\frac{1}{2} \left( \frac{\partial X}{\partial x} \right)
\end{align}

This shows that isochoric flow of an isotropic fluid necessitates that if the line differential element stretches (or, conversely, shrinks) in the direction of motion, it simultaneously must contract (or expand) by half as much in the directions orthogonal to this axis, for the differential volume to remain constant.

\subsection{Energy density and Hamiltonian constraint}\label{Energy density}

Given the extrinsic curvature, we can proceed to calculate the energy density. Start with the Gauss equation \cite{Misner1973, Gourgoulhon, AlcubierreBook}:

\begin{equation}
^{(4)}R + 2 n_{\mu} n_{\nu} \; ^{(4)}R^{\mu\nu} =\, ^{(3)}R + K^2 - K_{i}^{\; j} K_{j}^{\; i} \; ,
\label{eq:gausscodazzi}
\end{equation}
where the covariant components $n_{\mu}$ of the future-pointing normal vector to the hypersurface are:

\begin{equation}
\begin{pmatrix}
n_{0} \\
n_{1} \\
n_{2} \\
n_{3}
\end{pmatrix}
=
\begin{pmatrix}
-\alpha \\
0 \\
0 \\
0
\end{pmatrix}
=
\begin{pmatrix}
-1 \\
0 \\
0 \\
0
\end{pmatrix}.
\label{eq:covnormal}
\end{equation}

Furthermore, from the definition of the 4-dimensional spacetime Einstein's tensor $G^{\mu \nu}$ (where we drop the $(4)$ index because we invoke Einstein's tensor only in a 4-dimensional context), we have the identity:

\begin{equation}
^{(4)}R + 2 \; ^{(4)}R^{00} = 2 \; G^{00}.
\label{eq:RicciandEinstein}
\end{equation}

Therefore, from Eqs.~\eqref{eq:RicciandEinstein}, \eqref{eq:covnormal} and \eqref{eq:gausscodazzi}, it follows that:

\begin{equation}
2 \; G^{00}= \, ^{(3)}R + K^2 -K_{i}^{\; j} K_{j}^{\; i} \; .
\label{eq:Einsteinenergy}
\end{equation}

In General Relativity, the matter energy density $\varrho$ is an invariant quantity (referred to by Synge \cite{synge1966} as the `proper energy density'), given by the scalar relation (which therefore holds in any coordinate system) $\varrho=T^{\alpha\beta} u_{\alpha} u_{\beta}$ where $u_{\alpha}$ are the covariant components of the 4-velocity vector of the Eulerian observer and $T^{\alpha\beta}$ are the fully contravariant components of the stress-energy-momentum tensor. This matter energy density $\varrho$ encompasses internal forms of energy such as rest mass energy, thermal energy, and chemical energy while excluding macroscopic kinetic energy and macroscopic potential energy from external fields, which are frame-dependent \cite{PlebanskiKrasinski}.  In the standard 3+1 ADM formalism, the covariant components $n_{\mu}$ of the vector normal to the hypersurface (Eq.~\eqref{eq:covnormal}) coincides with the 4-velocity covariant components $(u_{\mu} = {-1, 0, 0, 0})$ of the Eulerian observer who remains at fixed spatial coordinates on each slice of the foliation.  Therefore $\varrho = T^{\alpha\beta} n_{\alpha} n_{\beta}$ simplifies to $\varrho = T^{00}$ in the standard 3+1 ADM foliation. Hence, $\varrho$ remains invariant under coordinate transformations that respect the standard 3+1 ADM formalism's structure.  Einstein's field equations Eqs.~\eqref{eq:FieldEq_a} and \eqref{eq:FieldEq_b} can be expressed in contravariant components as $G^{\alpha \beta} = \kappa \, T^{\alpha \beta}$, therefore the time-time component is $G^{0 0} = \kappa \, T^{0 0}= \kappa  \; \varrho$ with $\varrho$ being the matter energy density and $\kappa$ Einstein's coupling constant.  Therefore, from Einstein's equation, it follows that the matter energy density $\varrho$, observed by Eulerian observers in the standard 3+1 ADM foliation with 4-velocity Eq.~\eqref{eq:covnormal} is:

\begin{equation}
 \varrho = \frac{1}{2 \; \kappa} \;(\, ^{(3)}R + K^2 -K_{i}^{\; j} K_{j}^{\; i}) \; .
\label{eq:energy density}
\end{equation}

It is important to note that the general validity of Eq.~\eqref{eq:energy density} depends on a foliation that conforms to the standard 3+1 ADM decomposition, where the vector components $n_{\alpha}$ normal to the hypersurface (Eq.~\eqref{eq:covnormal}) coincide with the 4-velocity covariant components $u_{\alpha}$ of the Eulerian observer.

  Other foliations are generally possible, in which case Eq.~\eqref{eq:energy density} would need to be modified accordingly. For example, more general foliations can be constructed where the hypersurfaces are not orthogonal to the 4-velocity of the matter. This has been found useful in numerical relativity, where gauge choices (slicing conditions and coordinate conditions) are employed to simplify the evolution equations. Also, tilted foliations can be used where the hypersurface normal is not aligned with the matter 4-velocity. These are used for example in studies involving accretion disks around black holes or other astrophysical phenomena where the flow is not orthogonal to the slicing. Also, in a Lagrangian formulation of relativistic fluid flow, the 4-velocity of comoving observers does not necessarily coincide with the normal vector of the spatial hypersurfaces unless the matter flow is orthogonal to the hypersurfaces.

The spatial Ricci scalar $^{(3)}R$ is computed solely from the spatial metric $\gamma_{ij}$ and represents curvature in the spatial dimensions without involving time derivatives or temporal components in the computation. On the other hand, the spacetime Ricci scalar $^{(4)}R$ involves time and spatial derivatives of the full spacetime metric $g_{\mu\nu}$ and encompasses both spatial and temporal components of curvature.

This Eq.~(\ref{eq:Einsteinenergy}) is the `Hamiltonian constraint' in the 3+1 AMD formalism.
Since the hypersurfaces' three-dimensional intrinsic geometry is Euclidean (flat) in Nat\'{a}rio's spacetime, it follows that the spatial Ricci scalar vanishes, $^{(3)}R=0$. Furthermore, the trace of the extrinsic curvature also vanishes by construction, $K=0$. Therefore, the energy density $\varrho$ in Nat\'{a}rio's spacetime is proportional to the quadratic scalar $- K_{i}^{\; j} K_{j}^{\; i}$ (the trace of the cofactor of the matrix) of the extrinsic curvature mixed components:

\begin{subequations}
\label{eq:Einsteingausscodazzi}
\begin{align}
\varrho =& \;\frac{1}{\kappa} G^{00} \\
=& - \frac{1}{2 \, \kappa} K_{i}^{\; j} K_{j}^{\; i} \label{eq:Einsteingausscodazzi-a}\\
=& \; \frac{1}{\kappa} \; \text{Trace}\left(\text{cofactor}(K_{i}^{\; j})\right) \label{eq:Einsteingausscodazzi-b}\\
=& -\frac{1}{\kappa} v^2 \left( 3 \left(\frac{\partial f}{\partial r}\right)^2 \,  \cos^2(\theta)+  \left( \frac{\partial f}{\partial r} + \frac{r}{2} \frac{\partial^2 f}{\partial r^2}\right)^2 \, \sin^2(\theta) \right) .\label{eq:Einsteingausscodazzi-c}
\end{align}
\end{subequations}

This result, obtained here using curvilinear tensor components, agrees exactly with Nat\'{a}rio's expression obtained using tetrad vector components \cite{Natário}. The matter energy density $\varrho$ is a negative definite quadratic function of the variables and exhibits a symmetric distribution fore and aft of the spacecraft traveling at a constant velocity $v$.

This strictly negative expression Eq.~\eqref{eq:Einsteingausscodazzi-c} for the energy density ($\varrho < 0$) observed by Eulerian observers in the standard 3+1 ADM foliation follows from the lapse function $\alpha = 1$ (therefore, the proper time equals coordinate time) and the vector flow field Eq.~\eqref{eq:Xvectorphysic} imposed by Nat\'{a}rio. This vector field satisfies zero change in differential volume elements $K = 0$ and zero spatial Ricci scalar $^{(3)}R = 0$. Due to these particular choices by Nat\'{a}rio, the energy density $\varrho$ is proportional to $- K_{i}^{\; j} K_{j}^{\; i}$. In this specific context, $\varrho$ behaves as an invariant as per Eq.~\eqref{eq:Einsteingausscodazzi-a}. This invariance holds within the restricted context of the standard 3+1 ADM foliated spatial hypersurface using Nat\'{a}rio's vector field Eq.~\eqref{eq:Xvectorphysic} and lapse function $\alpha=1$. The spatial Ricci scalar $^{(3)}R$ of the hypersurface depends only on the intrinsic curvature of the hypersurface, and not on how it is embedded or foliated in the higher-dimensional spacetime. However, the trace $K$ and the quadratic scalar $K_{i}^{\; j} K_{j}^{\; i}$ of the extrinsic curvature are not generally invariant under arbitrary changes in the foliation of spacetime.  In Eq.~\eqref{eq:energy density}, the scalars $K$ and $K_{i}^{\; j} K_{j}^{\; i}$ are invariant under smooth coordinate transformations that map the spatial hypersurface onto itself or onto a hypersurface with an identical embedding. Different choices of foliations can lead to different extrinsic curvatures (a measure of how much the normal vectors to a hypersurface differ at neighboring points) and, therefore, different values of $K$ and $K_{i}^{\; j} K_{j}^{\; i}$. Consequently, from Eq.~\eqref{eq:energy density}, this can lead to different values of $\varrho$.  For example, an initially flat 2-dimensional surface embedded in 3-dimensional Euclidean space has zero extrinsic principal curvatures $k_1 = 0$ and $k_2 = 0$, but when rolled into a cylindrical shape, the extrinsic principal curvatures become $k_1 > 0$, $k_2 = 0$; therefore, the extrinsic mean curvature becomes positive $(k_1 + k_2)/2 > 0$, despite the intrinsic measures of curvature remaining flat, namely the Gaussian curvature $\mathcal{K} = k_1 \times k_2 = 0$ and therefore the Ricci scalar $^{(2)}R = 2\mathcal{K} = 0$.

The continuous rotational symmetry around the axis of travel of Nat\'{a}rio's vector flow field (Eq.~\eqref{eq:Xvectorphysic}) (which is a function only of the radial distance $r$ and the polar angle $\theta$, but not a function of the azimuthal angle $\phi$) ensures that the spacetime geometry remains invariant under rotations around the axis of travel (the polar axis $x$ in Fig. \ref{Fig1_SpherCoord}), simplifying the equations that satisfy the $K=0$ constraint. 

Furthermore, from Eq.~\eqref{eq:energy density} it is clear that the imposition of zero change in differential volume $K=0$ constrains the energy density in Natário’s drive to be more negative than in Alcubierre’s drive, since a non-zero change in volume $K\neq 0$ enters as a positive term in this energy density expression, and therefore with the opposite sign to the term $-K_{i}^{\; j} K_{j}^{\; i}$.

Theoretically \cite{marquet2009}, the total energy density observed by Eulerian observers can become positive ($\varrho > 0$) if the spatial Ricci scalar curvature $^{(3)}R$ becomes sufficiently large and positive such that the positive contribution from $^{(3)}R + K^2$ outweighs the negative term $- K_{i}^{\; j} K_{j}^{\; i}$ in Eq.~\eqref{eq:energy density}. According to Einstein's field equations Eqs.~\eqref{eq:FieldEq_a} and \eqref{eq:FieldEq_b}, changing Ricci's curvature from flat ($^{(3)}R = 0$) to positive ($^{(3)}R > 0$) can only occur within the positive matter distribution of a non-vacuum spacetime. In our prior example, deforming the initially flat surface $^{(2)}R = 2\mathcal{K} = 0$ into a spherical shape with positive extrinsic principal curvatures $k_1 = k_2 > 0$ introduces positive intrinsic curvature, $^{(2)}R = 2\mathcal{K} = 2 (k_1)^2 > 0$.  This spherical deformation is impossible without allowing in-plane membrane stretching or shrinking of the initially flat surface sheet and would lead to buckling, fractures, or other damage to a physical sheet if the in-plane membrane stiffness is much greater than the out-of-plane bending stiffness of the sheet. In General Relativity, because the (effective `spacetime stiffness') inverse $1/\kappa$ of the coupling constant in Einstein's field equations Eqs.~\eqref{eq:FieldEq_a} and \eqref{eq:FieldEq_b} is so large, significantly changing the intrinsic spatial curvature $^{(3)}R$ requires an extremely large energy density.

Consider Eq.~\ref{eq:Einsteinenergy} for Natário’s zero-expansion drive $K=0$:

\begin{equation}
\begin{aligned}
\varrho &= \frac{1}{2 \; \kappa} \;(\, ^{(3)}R - K_{i}^{\; j} K_{j}^{\; i}) \\
&= \varrho_{\text{int}} + \varrho_{\text{ext}} \; .
\end{aligned}
\label{eq:energy density Natario}
\end{equation}

where the energy density due to intrinsic curvature is $\varrho_{\text{int}}:=^{(3)}R/{2 \; \kappa}$ and the one due to extrinsic curvature is  $\varrho_{\text{ext}}:= -K_{i}^{\; j} K_{j}^{\; i}/{2 \; \kappa}$. Natário \cite{Natário} assumed a flat spatial metric, hence $^{(3)}R = 0$, and $\varrho_{\text{int}} = 0$. Hence, what remains is a negative energy density purely due to the contribution from extrinsic curvature $\varrho_{\text{ext}}$, as per Eq.~\ref{eq:Einsteingausscodazzi-c}.

Taking as an example the case from Fig.~\ref{fig:pi2 v1 rho5}, for the zero-expansion Natário warp drive traveling with $v=c$, a warp bubble radius $\rho = 5 \, [\text{m}]$, and warp bubble inverse thickness $\sigma = 4 \, [\text{m}^{-1}]$, the largest negative amplitude of the Einstein scalar $G$ is approximately $-130 \, [\text{m}^{-2}]$. From Eq.~(\ref{eq: G appendix}) in Appendix~\ref{secA1}, Einstein's scalar $G = 2 \, \varrho \, \kappa$ for Natário's zero-expansion warp drive. Therefore, in SI units, the contribution from extrinsic curvature is $\varrho_{\text{ext}} \approx -130 \, [\text{m}^{-2}] / (2 \, \kappa \, c^2) = -3.48 \times 10^{27} \, \text{kg/m}^3$. The contribution from the energy density due to positive intrinsic curvature $\varrho_{\text{int}}$ would need to be $\varrho_{\text{int}} = -\varrho_{\text{ext}} = 3.48 \times 10^{27} \, \text{kg/m}^3$ to cancel out the contribution from the negative energy density $\varrho_{\text{ext}}$. This is a fantastically large density. For comparison, the energy density of the densest neutron star is approximately $10^{18} \, \text{kg/m}^3$ \cite{Steiner2016}, which is about 9 orders of magnitude smaller. 
Black holes have a singularity approaching infinite density at their center; therefore, we consider them in the following calculations. The average density decreases with increasing mass of the black hole. The average density of supermassive black holes can be lower than the density of water \cite{Celotti1999}. The average density of the smallest observed black hole, with a mass of about 3.8 solar masses \cite{Shaposhnikov2009}, is approximately $1.3 \times 10^{18} \, \text{kg/m}^3$, which is still 9 orders of magnitude smaller than required for cancellation.  Primordial black holes could theoretically have formed in the early universe with a wide range of masses. The minimum average density required for a primordial black hole to survive Hawking evaporation until the present day corresponds to a mass of approximately $10^{11} \, \text{kg}$ or slightly higher \cite{Barco2021}.  Therefore, considering a surviving primordial black hole with mass of $10^{12}  \, \text{kg}$ we calculate a Schwarzschild radius of $10^{-15} \, \text{m}$ and therefore an average density of $10^{55}  \, \text{kg/m}^3$.  Therefore, a surviving primordial-sized black hole can cancel out this example's negative energy density. We calculate that a primordial black hole with the required average density of $3.48 \times 10^{27} \, \text{kg/m}^3$ has a Schwarzschild radius of $0.215 \, \text{m}$ (which approximately matches the warp bubble thickness $1/\sigma = 0.25 \, [\text{m}]$) and a mass of $1.45 \times 10^{26} \, \text{kg}$, which is about $25\%$ the mass of Saturn.

Canceling the total energy density does not eliminate the warp drive requirement for an extremely large exotic negative energy density.  A distant Eulerian observer observes a warp drive with $v=c$ and zero total energy density, but this zero energy density results from the algebraic addition of oppositely signed extremely large energy densities, each having an absolute magnitude of $3.48 \times 10^{27} \, \text{kg/m}^3$. 

One inconsistency in warp drive literature, with a few exceptions such as \cite{SchusterSantiagoVisser2023} (which only considers, for simplicity, generic Natário warp drives with zero-vorticity flow field) is the often unstated assumption that the warp drive (including the warp bubble) is a test mass. In General Relativity, a `test mass' is an idealized mass that is so small that it does not influence the curvature of spacetime. It moves solely under the influence of the gravitational field, following geodesics, but it does not contribute to the gravitational field itself. This idealization allows for the study of the properties of spacetime and the effects of gravity without the complexities introduced by the test mass's own gravitational influence.  This assumption is severely inconsistent with the exorbitant amount of exotic negative energy density required by warp drives in order to satisfy Einstein's field equations.

As discussed by \cite{SchusterSantiagoVisser2023}, it is challenging to disentangle the `warp field' from the `payload' from the `rest of the universe' in the warp drive spacetime formulation. \cite{SchusterSantiagoVisser2023} stated that their future work would study how the (non-compact) horizon of a superluminal warp drive might interact with the compact horizon of a black hole, either moving as a warp drive, inside a warp drive as payload, or present as an external immovable background geometry. When one considers the cancellation of the total energy density (by algebraic addition of oppositely signed energy densities, including the positive energy density of primordial black holes, as discussed here), the formulation of the warp drive becomes self-consistent with the assumption of being a test mass.

Returning now to Eq.~\eqref{eq:energy density}, where the extrinsic curvature was expressed in terms of the covariant-contravariant mixed components $K_{i}^{\; j}$, this energy density can be equally obtained from the $K^{i}_{\; j} $ contravariant-covariant mixed components of the extrinsic curvature, since 
\begin{equation}
\text{Trace}\left(\text{cofactor}(K_{i}^{\; j})\right) = \text{Trace}\left(\text{cofactor}(K^{i}_{\; j})\right).
\label{eq:traceCofactorEquivalence}
\end{equation}
The matrix of the $K^{i}_{\; j} $ contravariant-covariant mixed components of the extrinsic curvature is the transpose of the matrix of the $K_{i}^{\; j}$ covariant-contravariant mixed components, given in Eq.~(\ref{eq:Ksubiexpj}).

We can also calculate the extrinsic curvature fully covariant tensor components as follows:

\begin{equation}
K_{ij}=\begin{bmatrix}
-2 v \frac{\partial f}{\partial r} \cos(\theta)  & r v \left(\frac{\partial f}{\partial r}+\frac{r}{2}  \frac{\partial^2 f}{\partial r^2} \right) \sin(\theta)  & 0 \\
r v  \left(\frac{\partial f}{\partial r}+\frac{r}{2}  \frac{\partial^2 f}{\partial r^2} \right) \sin(\theta)  &r^2 v \frac{\partial f}{\partial r} \cos(\theta) & 0 \\
0 & 0 & r^2 \sin^2 (\theta) v \frac{\partial f}{\partial r} \cos(\theta) 
\end{bmatrix}. 
\label{eq:Kcovsubiexpj}
\end{equation}

The matrix representing the fully covariant components of the extrinsic curvature tensor, $K_{ij}$, is symmetric. However, unlike the mixed components $K_{i}^{\; j}$ and $K^{i}_{\; j}$, the trace of the matrix $K_{ij}$ is not zero. As expected, only the trace of the matrix of the mixed tensor components vanishes. Furthermore, the trace of the cofactor matrix derived from the fully covariant extrinsic curvature tensor $K_{ij}$ does not correspond to the energy density.  Of course, when contracting the contravariant components $\gamma^{ij}$ of the spatial metric with the covariant components $K_{ij}$ one will indeed obtain a null result for the trace of the extrinsic curvature tensor.

\subsection{Momentum density and momentum constraint}\label{Momentum density}

\begin{figure}[htbp]
    \centering
    \includegraphics[width=\textwidth]{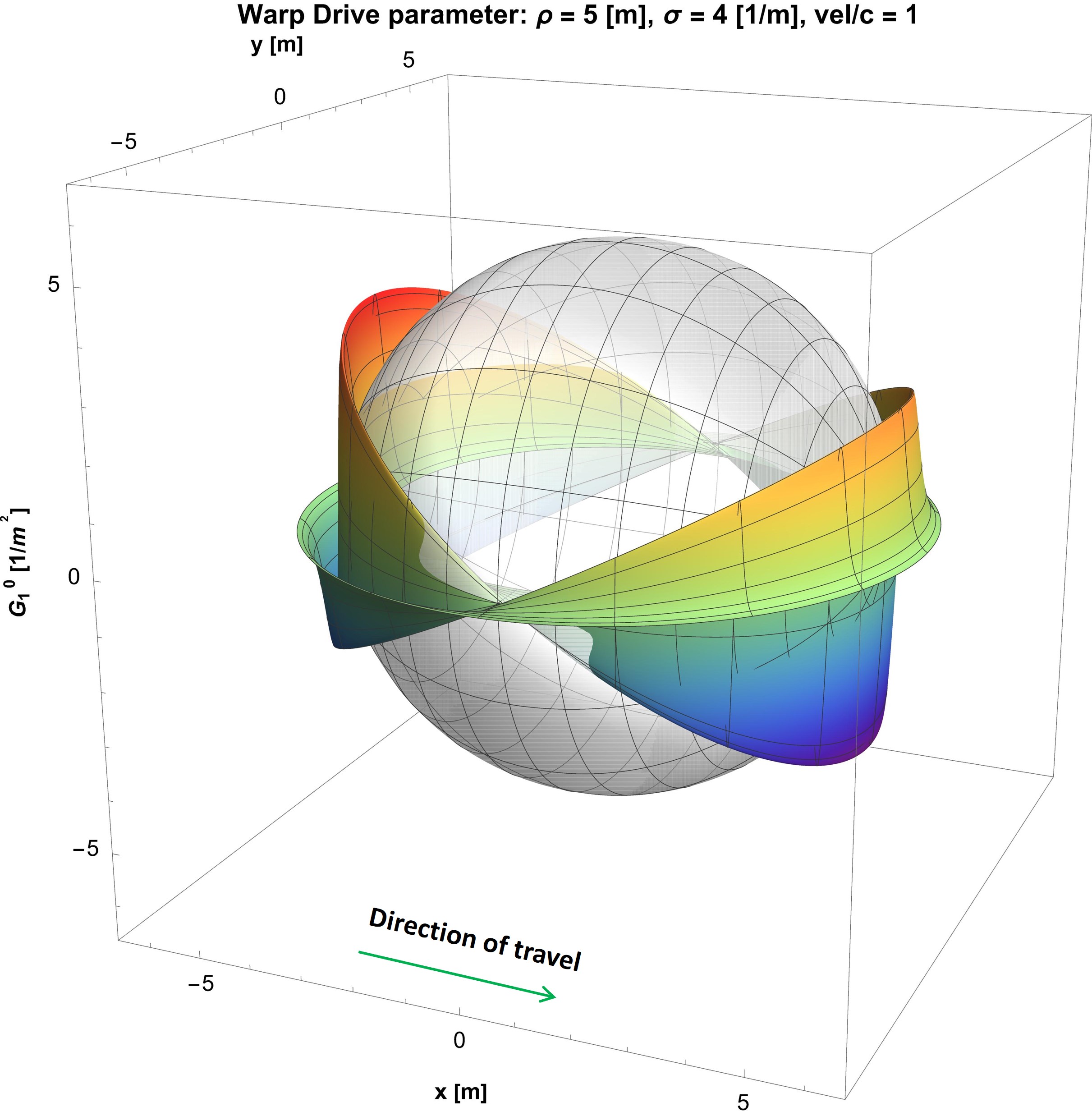}
    \caption{3-D plot of $G_{1}^{\; 0}$ vs. the $x,y$ coordinates, for $\frac{v}{c}=1,\rho=5[\text{m}],\sigma=4[\text{m}^{-1}]$ and also illustrating a sphere (overlaid with circles of longitude and latitude) matching Nat\'{a}rio's warp-bubble radius $\rho$.}
    \label{fig:G_1^0 v1.5 rho5}
\end{figure}

The previously discussed `Hamiltonian energy constraint' arises from the Gauss equation, which relates the full spatial projection of the Riemann tensor in four-dimensional spacetime, $^{(4)}R_{\mu\nu\rho\sigma}$, to the Riemann tensor in the three-dimensional hypersurfaces, $^{(3)}R_{ijkl}$. There is also a momentum constraint that results from the projection of the Riemann tensor in four-dimensional spacetime with one index projected in the future-pointing normal direction to the hypersurface \cite{Gourgoulhon, baumgarte_shapiro_2010, AlcubierreBook, Misner1973}. 
Both the (full-spatial projection) Gauss and the (mixed spatial-normal projection) Codazzi equations depend on the spatial metric, $\boldsymbol{\gamma}$, and the extrinsic curvature, $\mathbf{K}$. Additionally, the Codazzi equation depends on the spatial derivatives of the extrinsic curvature tensor $\mathbf{K}$.
The Codazzi equation can be expressed as the `momentum constraint':
\begin{equation}
G_{i}^{\; 0}=D^{j} (K_{ij}- \gamma_{ij}K)=G^{0}_{\; i}=D^{j}(K_{ji}-\gamma_{ji}K) \; ,
\end{equation}
where 

\begin{equation}
D^{j} (K_{ij}- \gamma_{ij}K)=\gamma^{jm} \, D_{m} (K_{ij}- \gamma_{ij}K) 
\end{equation}

denotes spatial covariant differentiation using the Christoffel symbols $^{(3)}\Gamma^i_{jk}$ computed from the three-dimensional metric $\gamma_{ij}$.

Using the results for the fully covariant components $K_{ji}$ of the extrinsic curvature from Eq.~(\ref{eq:Kcovsubiexpj}), and the fact that in Nat\'{a}rio's spacetime $K=0$ by construction, it follows that:

\begin{equation}
\begin{pmatrix}
G_{1}^{\; 0}\\
G_{2}^{\; 0} \\
G_{3}^{\; 0}
\end{pmatrix}
=
\begin{pmatrix}
G^{0}_{\; 1}\\
G^{0}_{\; 2} \\
G^{0}_{\; 3}
\end{pmatrix}
=
\begin{pmatrix}
- v  \left( \frac{\partial^2 f}{\partial r^2}
+4 (1/r) \frac{\partial f}{\partial r}\right) \cos (\theta ) \\
\frac{1}{2} v  \left(4 \frac{\partial f}{\partial r}+r \left(r \frac{\partial^3 f}{\partial r^3}+6 \frac{\partial^2 f}{\partial r^2}
\right)\right) \sin (\theta ) \\
0
\end{pmatrix} .
\end{equation}
Tensor indices $(1,2,3)$ correspond to the spherical coordinates $(r,\theta,\phi)$ respectively. In contrast to the time-time curvature component $G^{00} = \varrho \kappa$, which is proportional to the energy density $\varrho$ and is a quadratic function of the variables, the time-space curvature components $G_{i}^{\; 0} = G^{0}_{\; i}=p_{i} \, \kappa$, which are proportional to the components of the momentum density vector $p_{i}$, are linear functions of the variables.  This is exemplified by a leading term of $-v^2 \left( \frac{r}{2} \frac{\partial^2 f}{\partial r^2}\right)^2 \sin^2(\theta)$ for $G^{00}$ and $-v \frac{\partial^2 f}{\partial r^2} \cos(\theta)$ for  $G_{1}^{\; 0} = G_{x}^{\; 0}$. This linear dependence results in an anti-symmetric $\cos (\theta )$ distribution along the $x$-axis direction of travel when the spacecraft moves at a constant velocity $v$.  Consequently, the time-space curvature component $G_{1}^{\; 0} = G_{x}^{\; 0}$ serves as an indicator of the direction of travel (towards $+x$ or $-x$) unlike the time-time component $G^{00} = \varrho \kappa$ or the scalar curvature $G$ which are symmetric functions fore and aft.     

Fig.~\ref{fig:G_1^0 v1.5 rho5} shows the distribution of the component of Einstein's tensor $G_{1}^{\; 0}$ associated with momentum-density observed by the Eulerian observer, within the radial range defined as $r \in [\rho - 3/\sigma, \rho + 3/\sigma]$, where $\rho$ represents the radius of the warp bubble and $1/\sigma$ denotes its thickness. The depicted `direction of travel' corresponds to the perspective of distant Eulerian observers. The spherical shape of the warp bubble is evident. The momentum-density is anti-symmetric in the direction of travel, as expected since momentum-density is proportional to velocity (unlike the energy-density observed by the Eulerian observer, which is proportional to the square of the velocity and therefore is symmetric fore and aft). Also, note that the momentum-density reaches maxima and minima in the fore and aft directions of travel and is zero at the location perpendicular to the direction of travel. This is the opposite of the energy-density distribution, which reaches its maximum absolute magnitude at the location perpendicular to the direction of travel.

The generation of the warp bubble, as proposed for example in Fig. 1 of Bobrick et al. \cite{Bobrick_2021}, involves utilizing `exotic' negative matter-energy density to curve spacetime within the warping region for superluminal travel. The distribution and magnitude of this negative energy-density in the warp drive determine the radius $\rho$ of the spherical warp bubble. White \cite{WhiteGRG, White101, White102} included a spaceship within the depicted bubble. His representation, particularly in the direction of travel, showed the warp bubble extending significantly beyond the spaceship in both the fore and aft directions, unconnected to the spaceship. 

The mechanism of how this negative energy-density is generated, maintained, and distributed remains an open question not discussed in much of the warp drive literature. One possible method is that the negative matter-energy-density could be associated with the fuselage of the warp drive. This would involve integrating the necessary energy-density distribution within the fuselage of the spacecraft itself \cite{Sarfatti}, ensuring that the warp bubble remains stable and centered around the spaceship. Another idea, is to create a distribution of negative energy-density around the spacecraft using advanced field manipulation technologies, such as those proposed in quantum field theory \cite{Pfenning, LoboVisser}.

Before Nat\'{a}rio's work, which imposed a zero value on the extrinsic curvature scalar $K$, a common understanding, originating from the emphasis on Alcubierre's seminal paper \cite{Alcubierre_1994}, was that warp drive solutions necessitated contracting space volume elements in front of and expanding them behind the warp-bubble. Despite Nat\'{a}rio's model featuring zero extrinsic curvature scalar $K$, both Nat\'{a}rio's and Alcubierre's warp drives share a critical characteristic: the anti-symmetry of momentum density in the direction of motion. Determined by the spatial gradient of the extrinsic curvature tensor, the momentum density, rather than the expansion or contraction of volume elements, stands as the essential physical quantity in both the Nat\'{a}rio and Alcubierre models at constant velocity.

\subsection{Newman-Penrose null tetrad and Weyl scalars}\label{Newman Penrose}

To date, the Petrov type of Nat\'{a}rio's spacetime has not been explicitly determined in the warp drive literature due to its historical focus on the Ricci curvature and the associated energy-momentum tensor, neglecting the contribution of the Weyl curvature. However, we demonstrate in this article that due to the sharp localization of the form function $f(r)$ near the warp-bubble radius, and particularly for Nat\'{a}rio's zero-expansion warp drive, due to the imposed constraint of no change in differential volume, the Weyl curvature invariant exhibits a higher magnitude than the Ricci curvature invariants, including the scalar curvature $R=-G$. This observation holds even when the curvature invariants are normalized to comparable units $[\text{m}^{-2}]$ by considering the signed square root of the Weyl curvature invariant.  Accordingly, due to the dominance of the Weyl curvature invariant in Nat\'{a}rio's spacetime, it is useful to determine its Petrov type. In other words, it is important to classify the algebraic properties of the Weyl tensor in Nat\'{a}rio's spacetime. This involves analyzing the tensor symmetries, eigenvalues, and eigenvectors, which provide information about the nature of spacetime curvature.  
This type of analysis finds widespread application across scientific disciplines. In optics and crystallography, researchers utilize eigenvector analysis to study various types of crystals. In structural dynamics it enables the prediction of natural frequencies and mode shapes of buildings and bridges, ensuring safety during earthquakes and dynamic loads. In quantum mechanics, analyzing the eigenvalues and eigenvectors of operators allows for predicting outcomes in quantum systems including quantum computing and cryptography. 
In protein structure prediction, eigenvalue techniques aid in studying protein folding pathways, protein-protein interactions, and drug binding. It also plays a fundamental role in the stability analysis of control systems in the design of feedback controllers for applications like aircraft autopilots and robotic manipulators.

As a final example, in machine learning, techniques such as Principal Component Analysis (PCA) and Singular Value Decomposition (SVD) leverage eigenvalue analysis to reduce data dimensionality, enabling applications like facial recognition.
The algebraic analysis of the  Weyl curvature tensor in four-dimensional spacetime requires specialized mathematical treatments involving complex matrices that go beyond the analysis of simple matrices or operators (which have real eigenvalues and orthogonal eigenvectors), as in many of the previous examples. 
The study of the Weyl tensor's algebraic structure benefits from techniques such as the Newman-Penrose null tetrad formalism, spinor decompositions, Jordan canonical forms (converting the Weyl tensor to a 3x3 complex $Q$ matrix and finding its Jordan normal form reveals its algebraic structure and Segre characteristics), holonomy and symmetry groups, and the application of criteria like the Bel conditions. Moreover, the `peeling property,' primarily observed in type $D$ spacetimes, describes the behavior of the gravitational field near null infinity in terms of a hierarchy of behaviors.
The Newman-Penrose null tetrad formalism provides a powerful way to decompose the Weyl tensor into scalar components (Weyl scalars) using a null tetrad basis.

We will use the Newman-Penrose (double-null, complex) formalism \cite{stephani2003,Griffiths,Chandrasekhar} to ascertain the Petrov spacetime classification. A tetrad (set of four vectors) is introduced at each point. The first two vectors, $l^{\mu}$ and $n^{\mu}$, are a pair of standard (real) null (lightlike) vectors such that $l^{\mu} n_{\mu} = -1$. We can consider $l^{\mu}$ as the outgoing null vector and $n^{\nu}$ as the ingoing null vector in spherical coordinates. A complex null vector is then constructed by combining a pair of real, orthogonal unit space-like vectors. The complex conjugate of this vector forms the fourth element of the tetrad. For signature and normalization, we utilize the $(-,+,+,+);\; l^{\mu} n_{\mu}=-1,\; m^{\mu} \overline{m}_{\mu}=1$ convention.

Mattingly et al.'s \cite{Mattingly}  assertion that ``The comoving null tetrad describes geodesics traveling parallel to the warp bubble'' requires clarification. A null tetrad comprises inherently lightlike vectors that align with the paths of light rays (null trajectories in spacetime). However, these null trajectories are not timelike paths, which are the trajectories that observers with mass, including test masses, must follow.  It is more precise to state that this double null (lightlike) tetrad is constructed through complex linear combinations of comoving timelike tetrad vectors. These timelike vectors describe geodesics that are comoving with the warp bubble.

A null tetrad consistent with the shift vector $\boldsymbol{\beta}$ components in the spherical coordinate system $(r,\theta,\phi)$, with $l^{1}$ as the outgoing radial null vector component, and $n^{1}$ as the ingoing radial null vector component is:

\begin{equation}
\begin{pmatrix}
l^0 \\
l^1 \\
l^2 \\
l^3
\end{pmatrix}
=
\frac{1}{\sqrt{2}}
\begin{pmatrix}
-1 \\
-1 + \beta^{1} \\
\beta^{2} \\
\beta^{3}
\end{pmatrix}
=
\frac{1}{\sqrt{2}}
\begin{pmatrix}
-1 \\
-1 - 2 \; v f \cos(\theta) \\
 v  \left(2  \frac{f}{r}+\frac{\partial f}{\partial r} \right) \sin(\theta) \\
0
\end{pmatrix} \; ,
\label{eq:null_vector_l}
\end{equation}

\begin{equation}
\begin{pmatrix}
n^0 \\
n^1 \\
n^2 \\
n^3
\end{pmatrix}
=
\frac{1}{\sqrt{2}}
\begin{pmatrix}
-1 \\
1 + \beta^{1} \\
\beta^{2} \\
\beta^{3}
\end{pmatrix}
=
\frac{1}{\sqrt{2}}
\begin{pmatrix}
-1 \\
1 - 2 \; v f \cos(\theta) \\
 v  \left(2  \frac{f}{r}+\frac{\partial f}{\partial r} \right) \sin(\theta)\\
0
\end{pmatrix} \; ,
\label{eq:null_vector_n}
\end{equation}

\begin{equation}
\begin{pmatrix}
m^0 \\
m^1 \\
m^2 \\
m^3
\end{pmatrix}
=
\frac{1}{\sqrt{2}}
\begin{pmatrix}
0\\
0\\
-\sqrt{\gamma^{22}} \\
i \;\sqrt{\gamma^{33}}
\end{pmatrix}
=
\frac{1}{\sqrt{2}}
\begin{pmatrix}
0\\
0 \\
- \frac{1}{r} \\
\frac{i}{r \sin(\theta)}
\end{pmatrix} \; ,
\label{eq:null_vector_m}
\end{equation}

\begin{equation}
\begin{pmatrix}
\overline{m}^0 \\
\overline{m}^1 \\
\overline{m}^2 \\
\overline{m}^3
\end{pmatrix}
=
\frac{1}{\sqrt{2}}
\begin{pmatrix}
0 \\
0 \\
-\sqrt{\gamma^{22} }\\
-i \; \sqrt{\gamma^{33}}
\end{pmatrix}
=
\frac{1}{\sqrt{2}}
\begin{pmatrix}
0 \\
0 \\
- \frac{1}{r} \\
- \frac{i}{r \sin(\theta)}
\end{pmatrix} .
\label{eq:null_vector_mbar}
\end{equation}

We verify that these tetrad vectors satisfy the following conditions:

Nullity:
\begin{equation}
l_{\mu} l^{\mu} = n_{\mu} n^{\mu} = m_{\mu} m^{\mu} = \overline{m}_{\mu} \overline{m}^{\mu} = 0 \; ,
\label{eq:null_vectors}
\end{equation}

Orthogonality:
\begin{equation}
l_{\mu} m^{\mu} = l_{\mu} \overline{m}^{\mu} = n_{\mu} m^{\mu} = n_{\mu} \overline{m}^{\mu} = 0 \; ,
\label{eq:perpendicularity}
\end{equation}

Cross-normalization:
\begin{equation}
l_{\mu} n^{\mu} = n_{\mu} l^{\mu} = -1 \; ,
\label{eq:normalization1}
\end{equation}
\begin{equation}
m_{\mu} \overline{m}^{\mu} = \overline{m}_{\mu} m^{\mu} = 1 \; ,
\label{eq:normalization2}
\end{equation}

Metric Reconstruction:
\begin{equation}
g_{\mu \nu} = -l_{\mu} n_{\nu} - n_{\mu} l_{\nu} + m_{\mu} \overline{m}_{\nu} + \overline{m}_{\mu} m_{\nu} \; ,
\label{eq:metric_covariant}
\end{equation}
\begin{equation}
g^{\mu \nu} = -l^{\mu} n^{\nu} - n^{\mu} l^{\nu} + m^{\mu} \overline{m}^{\nu} + \overline{m}^{\mu} m^{\nu} .
\label{eq:metric_contravariant}
\end{equation}

The five Newman-Penrose Weyl scalars $(\Psi_0,\Psi_1,\Psi_2,\Psi_3,\Psi_4)$ are derived from the Weyl tensor $C_{\alpha\beta\gamma\delta}$, which provides information about the curvature of spacetime, specifically how curvature changes anisotropically. The Weyl scalars for Nat\'{a}rio's spacetime are:

\begin{align}
\Psi_0 &= C_{\alpha\beta\gamma\delta} l^\alpha m^\beta l^\gamma m^\delta = -v^2 \left(\frac{1}{2r}\right) f \left( 2\frac{\partial f}{\partial r}+r\frac{\partial^2 f}{\partial r^2}\right) \sin^2(\theta), \\
\Psi_1 &= C_{\alpha\beta\gamma\delta} l^\alpha n^\beta l^\gamma m^\delta =v \left(\frac{1}{8 r}\right)  \Biggl(4 \frac{\partial f}{\partial r}-4 r \frac{\partial^2 f}{\partial r^2}-r^2 \frac{\partial^3 f}{\partial r^3} \nonumber \\
&\quad +2\; v \cos (\theta ) \left(r^2 f \frac{\partial^3 f}{\partial r^3}+\left(2 f-r \frac{\partial f}{\partial r}\right) \left(r \frac{\partial^2 f}{\partial r^2}+2 \frac{\partial f}{\partial r}\right)\right)\Biggr) \sin(\theta) , \\
\Psi_2 &= C_{\alpha\beta\gamma\delta} l^\alpha m^\beta \overline{m}^\gamma n^\delta=v^2 \left(\frac{1}{12}\right)  \left(-2 r^2 \sin ^2(\theta ) \left(\frac{\partial^2 f}{\partial r^2}\right)^2+(\cos (2 \theta )-13) \left(\frac{\partial f}{\partial r}\right)^2\right. \nonumber \\
&\quad \left.-\frac{\partial^2 f}{\partial r^2} \left(8 r \sin ^2(\theta ) \frac{\partial f}{\partial r}+f (9 \cos (2 \theta )+3)\right)\right), \\
\Psi_3 &= C_{\alpha\beta\gamma\delta} l^\alpha n^\beta \overline{m}^\gamma n^\delta=v \left( \frac{1}{8 r}\right)  \Biggl( 4\, r \, v\, \cos (\theta ) \left(\frac{\partial f}{\partial r}\right)^2\nonumber \\
&\quad+\frac{\partial f}{\partial r} \left(4+2 v \cos (\theta ) \left(-4 f+r^2 \frac{\partial^2 f}{\partial r^2}\right)\right)\nonumber \\
&\quad
-r \left(4 (1+v \cos (\theta )\, f)\frac{\partial^2 f}{\partial r^2}+r (1+2\, v \cos (\theta )\, f)\frac{\partial^3 f}{\partial r^3}\right) \Biggr) \sin(\theta), \\
\Psi_4 &= C_{\alpha\beta\gamma\delta} n^\alpha \overline{m}^\beta n^\gamma \overline{m}^\delta=\Psi_0.
\end{align}

Since all the Newman-Penrose Weyl scalars in Petrov type I spacetimes are non-vanishing, it is possible to rotate \cite{Re2003} the Newman-Penrose double-null tetrad in a Petrov type I spacetime into a frame where the Newman-Penrose Weyl scalars $\Psi_1$ and $\Psi_3$ are equal to zero. This rotated frame elucidates the combination of a static Coulomb-type effect associated with $\Psi_2$ and the transverse degrees of freedom $\Psi_0$ and $\Psi_4$, which are potentially associated with traveling transverse gravitational waves.

\subsection{Principal null directions encoded in the \textit{Q} Matrix}\label{Q matrix}

The principal null directions of the Weyl tensor correspond to the null eigenvectors of this curvature tensor and therefore provide information about its algebraic structure, which in turn determines the Petrov classification of spacetimes.  These null directions represent the directions along which gravitational radiation propagates in a given spacetime.
The $3\times3$ complex, traceless, symmetric matrix with purely spatial components,

\begin{equation}
Q^{\hat{a}}\,_{\hat{c}} \coloneqq -(E^{\hat{a}}\,_{\hat{c}} + i \, H^{\hat{a}}\,_{\hat{c}}) = Q^{\hat{c}}\,_{\hat{a}},
\end{equation}

encodes the principal null directions \cite{stephani2003}. $E^{\hat{a}}\,_{\hat{c}}$ is the `electric' part \cite{stephani2003, PlebanskiKrasinski, deFelice} of the Weyl tensor. It is obtained by contracting the Weyl tensor with a pair of arbitrary timelike unit vectors. For example, by contraction with the 4-velocity vector field of an Eulerian observer. It encodes the gravitational tidal forces, and therefore, it survives in the Newtonian non-relativistic limit, where it can be obtained from the second derivatives of the Newtonian gravitational potential. The `magnetic' part $H^{\hat{a}}\,_{\hat{c}}$ \cite{stephani2003, PlebanskiKrasinski, deFelice} of the Weyl tensor is obtained by contracting its Hodge dual \cite{deFelice} with a pair of timelike vectors. It arises solely due to the spacetime curvature and has no direct analog in Newtonian gravity. It is involved in rotational frame-dragging effects, where rotating masses tend to drag space-time along with their rotation, and it contributes to defining at spatial infinity the angular momentum of an asymptotically flat spacetime.

In cosmological perturbation theory, the magnetic part of the Weyl tensor is associated with the vector modes of the metric perturbations at the first post-Newtonian order. Importantly, the existence of magnetic Weyl components is required for the propagation of gravitational waves. A non-zero magnetic Weyl tensor is an essential requirement for spacetimes to be classified as Petrov type $N$ (representing a pure gravitational wave) or the fully general type I. Petrov types II, III, and $D$ can arise even if the magnetic part vanishes.  

The sign of $Q^{\hat{a}}\,_{\hat{c}} \coloneqq -(E^{\hat{a}}\,_{\hat{c}} + i \, H^{\hat{a}}\,_{\hat{c}})$, adopted by us and by \cite{stephani2003}, is arbitrary and does not affect the Petrov classification. For example, \cite{PlebanskiKrasinski, Pareja} adopt the opposite sign.

The Q matrix components for Nat\'{a}rio's spacetime are:

\begin{equation}
\begin{pmatrix}
Q^{\hat{1}}\,_{\hat{1}} \;& 0\; & Q^{\hat{1}}\,_{\hat{3}} \\
0 \;& Q^{\hat{2}}\,_{\hat{2}} \;& Q^{\hat{2}}\,_{\hat{3}} \\
Q^{\hat{3}}\,_{\hat{1}} \;& Q^{\hat{3}}\,_{\hat{2}} \;& Q^{\hat{3}}\,_{\hat{3}}
\end{pmatrix} \; ,
\label{eq: Q matrix}
\end{equation}

\begin{align}
Q^{\hat{1}}\,_{\hat{1}} &= \Psi_2 - \frac{\Psi_0 + \Psi_4}{2} \nonumber \\
&= \left( \frac{1}{12 \, r}\right) v^2 \left(-2 r^3 \sin ^2(\theta ) \left(\frac{\partial^2 f}{\partial r^2}\right)^2 + \frac{\partial f}{\partial r} \left(r (\cos (2 \theta )-13) \frac{\partial f}{\partial r} + 12 f \sin ^2(\theta )\right) \right. \nonumber \\
&\quad \left. - 4 r \frac{\partial^2 f}{\partial r^2} \left(2 r \sin ^2(\theta ) \frac{\partial f}{\partial r} + 3 f \cos (2 \theta )\right)\right) \; , \label{eq:Q11}
\end{align}

\begin{align}
Q^{\hat{2}}\,_{\hat{2}} &= \Psi_2 + \frac{\Psi_0 + \Psi_4}{2} \nonumber \\
&= - \left( \frac{1}{12 \, r}\right) v^2 \left(2 r^3 \sin ^2(\theta ) \left(\frac{\partial^2 f}{\partial r^2}\right)^2+\frac{\partial f}{\partial r} \left(-r (\cos (2 \theta )-13) \frac{\partial f}{\partial r}+12 f \sin ^2(\theta )\right)\right. \nonumber \\
&\quad \left.+4 r \frac{\partial^2 f}{\partial r^2} \left(2 r \sin ^2(\theta ) \frac{\partial f}{\partial r}+3 f \cos ^2(\theta )\right)\right) \; ,
\label{eq:Q22}
\end{align}

\begin{align}
Q^{\hat{3}}\,_{\hat{3}} &= -2 \, \Psi_{2}  \nonumber \\
&= \frac{1}{6} v^2 \left(2 r^2 \sin ^2(\theta ) \left(\frac{\partial^2 f}{\partial r^2}\right)^2-\left((\cos (2 \theta )-13) \left(\frac{\partial f}{\partial r}\right)^2\right)\right. \nonumber \\
&\quad \left.+\frac{\partial^2 f}{\partial r^2} \left(8 r \sin ^2(\theta ) \frac{\partial f}{\partial r}+f (9 \cos (2 \theta )+3)\right)\right) \; ,
\label{eq:Q33}
\end{align}

\begin{align}
Q^{\hat{1}}\,_{\hat{2}} &= Q^{\hat{2}}\,_{\hat{1}} =i \;\frac{ (\Psi_4-\Psi_0)}{2} =0 \; ,
\label{eq:Q12}
\end{align}

\begin{align}
Q^{\hat{1}}\,_{\hat{3}} &= Q^{\hat{3}}\,_{\hat{1}} \nonumber \\
&= \Psi_{1} -\Psi_{3} \nonumber \\
&= \left(\frac{1}{4 r} \right) v^2  \left(r^2 f \frac{\partial^3 f}{\partial r^3} + \left(2 f - r \frac{\partial f}{\partial r}\right) \left(r \frac{\partial^2 f}{\partial r^2} + 2 \frac{\partial f}{\partial r}\right)\right) \sin (2 \theta ) \; ,
\label{eq:Q13}
\end{align}

\begin{align}
Q^{\hat{2}}\,_{\hat{3}} &= Q^{\hat{3}}\,_{\hat{2}} \nonumber \\
&= i \;\left(\Psi_{1} +\Psi_{3}\right) \nonumber \\
&= - \left(\frac{1}{4 r} \right)i\; v  \left(r \left(r \frac{\partial^3 f}{\partial r^3} + 4 \frac{\partial^2 f}{\partial r^2}\right) - 4 \frac{\partial f}{\partial r}\right) \sin (\theta ) .
\label{eq:Q23}
\end{align}

Note that the components $Q^{\hat{a}}\,_{\hat{c}}$ are real and hence they are related to the electric part of the Weyl tensor, except for $Q^{\hat{1}}\,_{\hat{3}}=Q^{\hat{3}}\,_{\hat{1}}$ which are purely imaginary. Hence the $Q^{\hat{1}}\,_{\hat{3}}=Q^{\hat{3}}\,_{\hat{1}}$ components are related to the magnetic part of the Weyl tensor and, therefore meet a requirement for Natário's spacetime to be classified as Petrov type I.

\subsection{Petrov Classification of Natário's spacetime: type I}\label{Clarifying}

The assertion, made without proof, by Mattingly et al. \cite{Mattingly}  that: ``\dots Class $B_{1}$ spacetimes, which include all hyperbolic spacetimes, such as the general warp drive line element\dots'' is invalid.  The Nat\'{a}rio, and the Alcubierre warp drive line elements discussed in their article do not fall under Class $B_{1}$ Warped Product spacetimes.

\textit{Definition (Class $B$ Warped Product Spacetimes):} \textit{Class $B$ warped product spacetimes} \cite{Carot_1993, Carot2005, Santosuosso} are spacetimes expressible as the product of two distinct 2-dimensional spaces: one with a Lorentzian metric and the other with a Riemannian metric. The coupling between these spaces should be separable. 

The statement that``\dots Class $B_{1}$ spacetimes, which include all hyperbolic spacetimes, such as the general warp drive line element\dots'' may stem from a misunderstanding between `globally hyperbolic spacetimes' and `hyperbolic spacetimes.' Globally hyperbolic spacetimes are those for which scalar fields existing on them always have solutions, defining the causal nature of the spacetime. The statement by Carot et al. \cite{Carot2005} that ``spherically, plane and hyperbolically symmetric spacetimes are all special instances of $B_T$ warped spacetimes'' does not imply ``globally hyperbolic.'' Instead, it signifies that the spatial sections have a negative curvature resembling a hyperbola. Carot et al. \cite{Carot2005} does not state that all globally hyperbolic spacetimes fall under Class $B$, instead it states that all spacetimes with spatial sections shaped like hyperbolas are.
\par

The canonical form of a class $B_{1}$ warped product spacetime is:

\begin{equation}
\begin{aligned}
ds^2 &= -2\: f(u,v)\: du\: dv + r(u,v)^2\:g(\theta, \phi)^2\: (d\theta^2 + d\phi^2) \; ,\\
\end{aligned}
\label{B1Def}
\end{equation}\par
and the standard form for class $B_{2}$ warped product spacetime is:

\begin{equation}
\begin{aligned}
ds^2 &= f(u,v)^2\: (du^2+dv^2) - 2\:r(u,v)^2\:g(\theta, \phi)\: d\theta \: d\phi.\\
\end{aligned}
\label{B2Def}
\end{equation}\par

When juxtaposing Nat\'{a}rio's line element Eq.~(\ref{eq: Natario line element}) with the form of a class $B_{1}$ Eq.~(\ref{B1Def}) or a class $B_{2}$ Eq.~(\ref{B2Def}) warped product spacetime, it becomes evident that it does not conform to these Class $B$ spacetime structures. The problem is the coupling terms due to products of the functions $cos(\theta)$,  $sin(\theta)$, $f(r)$, and $\frac {\partial{f}} {\partial{r}}$ appearing in the time-time $g_{00}$ (Eq.~(\ref{eq: timetime metric})) and time-space $g_{0r}=g_{r0}, \; g_{0\theta}=g_{\theta 0}$ (Eq.~(\ref{eq: timespace metric})) components of the metric, that intertwine the spacetime coordinates introducing dependencies that prevent a straightforward decomposition of the Nat\'{a}rio spacetime metric into two independent 2-dimensional spaces, one Lorentzian and the other Riemannian.\par

Nat\'{a}rio's article \cite{Natário} did not specify the type of spacetime encompassed by Nat\'{a}rio's metric, and, to our knowledge, this aspect has not been explored in the existing literature. Our analysis demonstrates that Nat\'{a}rio's spacetime is of Petrov type I, which is the most general category within the Petrov classification. This type is characterized by the absence of algebraic symmetries and the presence of four distinct and real principal null directions. Furthermore, we establish that Nat\'{a}rio's spacetime does not conform to a class $B$ warped product spacetime. The classification of Nat\'{a}rio spacetime as Petrov type I is substantiated by the following considerations:

\begin{enumerate}
    \item In the case of Nat\'{a}rio's metric at a constant velocity \( v \), four out of the five Newman-Penrose scalars are distinct and non-zero: \( \Psi_{0} = \Psi_{4} \), \( \Psi_{1} \), \( \Psi_{2} \), and \( \Psi_{3} \) (refer to subsection \ref{Newman Penrose}). These scalars' uniqueness and non-zero nature suggest categorizing the spacetime as Petrov type I, according to the classification in \cite{stephani2003}. The realness of these scalars suggests the presence of the electric component of the Weyl tensor, without an accompanying magnetic component (associated with the twisting of null geodesic congruences) \cite{Hofmann2013}. This contrasts with the Alcubierre metric spacetime \cite{Rodal}, which is also of Petrov type I, but possesses five complex Newman-Penrose scalars, all unequal and non-zero (\( \Psi_{0} \), \( \Psi_{1} \), \( \Psi_{3} \), \( \Psi_{4} \) are complex, while \( \Psi_{2} \) is purely real for the  Alcubierre metric). The complex nature of the Newman-Penrose scalars \( \Psi_{0} \), \( \Psi_{1} \), \( \Psi_{3} \), \( \Psi_{4} \) in the Alcubierre metric is attributed to the involvement of both the electric and magnetic (twisting of null geodesic congruences) components of the Weyl tensor \cite{Rodal,Hofmann2013}.    It is important to note, as highlighted in \cite{baumgarte_shapiro_2010}, that the choice of the Newman-Penrose tetrad, which is not unique, can influence the form and physical interpretation of the Newman-Penrose Weyl scalars. Therefore, a more robust classification can be achieved by relying on the following criteria.

    \item The $3\times 3$ complex, traceless matrix of $Q^{\hat{a}}\,_{\hat{c}} = - (E^{\hat{a}}\,_{\hat{c}} + i \, H^{\hat{a}}\,_{\hat{c}}) $ (refer to subsection \ref{Q matrix}), which encodes the principal null directions \cite{stephani2003}, has three distinct, unequal eigenvalues $\lambda_{1} \neq \lambda_{2} \neq \lambda_{3}$. $E^{\hat{a}}\,_{\hat{c}}$ is the electric part, and $H^{\hat{a}}\,_{\hat{c}}$ the magnetic part of the Weyl tensor. The components $Q^{\hat{a}}\,_{\hat{c}}$ are real (electric part), except for $Q^{\hat{1}}\,_{\hat{3}}=Q^{\hat{3}}\,_{\hat{1}}$ which is purely imaginary (magnetic part). The Segre characteristic is $[111]$ since each of the three eigenvalues has a multiplicity of 1.  The eigenvalues of the Q matrix offer a robust method for classifying spacetimes, providing an alternative to relying solely on the Newman-Penrose scalars. In the case of Nat\'{a}rio's spacetime, the distinct and non-zero eigenvalues of the Q matrix affirm its classification as Petrov type I \cite{stephani2003}. This classification is analogous to that of Alcubierre's spacetime, which is also categorized as Petrov type I based on similar criteria. The mathematical expressions for the eigenvalues of large matrices, as encountered in this context, are exceedingly lengthy. Detailing them would necessitate several pages, owing to the sheer scale of the matrices and the calculations involved. We are willing to provide the detailed calculations as supplementary material.

    \item It is known \cite{stephani2003, Bini2023} that Petrov type I is the only spacetime that satisfies the inequality 

    \begin{equation}
     \widetilde{I}^{3} \neq  27 \: \widetilde{J}^{2}.   
    \end{equation}
    
    In this inequality, 

     \begin{equation}
     \widetilde{I}\coloneqq \frac{1}{32} Q^{\hat{a}}\,_{\hat{c}} \; Q^{\hat{c}}\,_{\hat{a}}
    \end{equation}

    is the quadratic complex curvature scalar invariant and 
    \begin{equation}
     \widetilde{J}\coloneqq \frac{1}{384} Q^{\hat{a}}\,_{\hat{c}} \; Q^{\hat{c}}\,_{\hat{f}} \; Q^{\hat{f}}\,_{\hat{a}}
    \end{equation}
    
     is the cubic complex curvature scalar invariant.  Further calculations reveal that for Nat\'{a}rio's spacetime, the condition \(\widetilde{I}^{3} \neq 27 \: \widetilde{J}^{2}\) holds, reinforcing that Nat\'{a}rio's spacetime is of Petrov type I \cite{stephani2003, Bini2023}, similar to the classification of Alcubierre's spacetime.

\end{enumerate}

Class $B$ warped product spacetimes can only be Petrov type $D$ or $O$ \cite{Carot_1993}. Therefore, since Nat\'{a}rio's spacetime is Petrov type I, it cannot be a Class $B$ warped product spacetime.

\subsection{Natário's radially-dependent warp-bubble form function \textit{f}}\label{form function}

Nat\'{a}rio defines the warp-bubble form function $f(r)$  as a function of the radial coordinate $r$ in the spherical coordinate system $(r,\theta,\phi)$, ensuring that Eulerian observers at the center of the bubble remain stationary, while those far away from the bubble are observed as moving with a relative speed $-v$. The sole conditions that Nat\'{a}rio \cite{Natário} imposed on the form function $f(r)$ to ensure a divergenceless field capable of generating a warp-bubble were the limits: $\lim_{r \to \infty} f(r) = \frac{1}{2}$ and $\lim_{r \to 0} f(r) = 0$.

Mattingly et al. \cite{Mattingly} state: ``Nat\'{a}rio’s chosen shape function is $n\left( r_s\right) = \frac{1}{2}\left(1 - \frac{1}{2}\left(1 - \tanh{\left(\sigma\left(r_s - \rho\right)\right)}\right)\right)$\dots''; however, Nat\'{a}rio \cite{Natário} did not choose that (or any other) $f(r)$ function.

We can express Nat\'{a}rio's form function $f(r)$ in terms of another function $g(r)$ that approaches zero at infinity instead of at the center of the warp-bubble. This expression satisfies Nat\'{a}rio's conditions: $\lim_{r \to \infty} f(r) = \frac{1}{2}$ and $\lim_{r \to 0} f(r) = 0$, if $\lim_{r \to \infty} g(r) = 0$ and $\lim_{r \to 0} g(r) = 1$:

\begin{equation}
f(r) = \frac{1}{2} \left(1 - g(r)\right) .
\label{f in terms of g}
\end{equation}

For example, defining $g(r)$ as Alcubierre's form function $g(r) = f_{Alc}$:

\begin{equation}
\begin{aligned}
f_{Alc}(r) &= \frac{\tanh{[\sigma (r + \rho)]} - \tanh{[\sigma (r - \rho)]}}{2 \tanh{[\sigma \rho]}} \; , \\
f(r) &= \frac{1}{2} \left(1 - f_{Alc}(r)\right) .\\
\end{aligned}
\label{fAlcubierreForm}
\end{equation}

This particular choice of form function $f(r)$ satisfies Nat\'{a}rio's limits: $\lim_{r \to \infty} f(r) = \frac{1}{2}$ and $\lim_{r \to 0} f(r) = 0$ exactly.

However, the form function $g(r) = f_{M}$ used by Mattingly et al. \cite{Mattingly}:

\begin{equation}
\begin{aligned}
f_{M}(r) &= \frac{\left(1 - \tanh{\left(\sigma\left(r - \rho\right)\right)}\right)}{2} \; , \\
f(r) &= \frac{1}{2} \left(1 - f_{M}(r)\right) \; , \\
\end{aligned}
\label{fMatt}
\end{equation}
meets only the $\lim_{r \to \infty} f(r) = \frac{1}{2}$ condition. The $\lim_{r \to 0} f(r) = 0$ condition is not exactly satisfied; instead, it yields $\lim_{r \to 0} f(r) = \frac{1}{4}\left(1 - \tanh\left(\rho \sigma\right)\right)$. Thus, Eq.~(\ref{fMatt}) is an approximation, valid when $\frac{1}{4}\left(1 - \tanh\left(\rho \sigma\right)\right) \approx 0$. 

In the following section, we present plots of the curvature invariants for the following warp-bubble parameters:

\begin{enumerate}
    \item $\rho = 5\,[\text{m}], \sigma = 4\,[\text{m}^{-1}]$, yielding $\lim_{r \to 0} f(r)=\frac{1}{4}(1 - \tanh(5 \times 4)) \approx  2.1241 \times 10^{-18}$ instead of zero.
    \item $\rho = 100\,[\text{m}], \sigma = 50000\,[\text{m}^{-1}]$, yielding $\lim_{r \to 0} f(r)=\frac{1}{4}(1 - \tanh(100 \times 50000)) \approx 1.8339 \times 10^{-348}$ instead of zero.
\end{enumerate}

Therefore, Eq.~(\ref{fMatt}) is an excellent approximation in these cases.

The most significant deviation from the ideal case ($\lim_{r \to 0} f(r) = 0$) occurs when using Eq.~(\ref{fMatt}) for scenarios where the warp-bubble `thickness' $1/\sigma$ is so large as to be comparable to its `radius' $\rho$. In the extreme and unrealistic case of $1/\sigma = \rho$, the limit $\lim_{r \to 0} f(r)$ yields $\frac{1}{4}(1 - \tanh(1)) \approx 0.05960$, diverging from the desired limit value of zero.

\section{\texorpdfstring{Visualizing curvature invariants in Nat\'{a}rio's spacetime}{Visualizing curvature invariants}}\label{Visualizing curvature invariants}

This section provides a detailed comparison and visualization of curvature invariants in Natário’s spacetime. It examines the radial and circumferential distributions and relative amplitudes. It compares them with those in Alcubierre's spacetime across different warp drive velocities and bubble dimensions, highlighting the differences and similarities between the two models. The aim is to comprehensively understand the behavior of these invariants in the warp-bubble zone and their implications for Natário's warp drive.

We highlight that the Weyl scalar is the curvature invariant with the highest amplitude. Previous warp drive studies have examined the Ricci curvature and the corresponding energy-momentum tensor, ignoring the crucial role of the Weyl curvature. We attribute the Weyl curvature's high amplitude to the sharp localization of the form function $f(r)$ near the warp-bubble radius. The form function approximates a top-hat function near the warp-bubble radius, leading to pronounced amplitudes of higher-order derivatives $\partial^n f / \partial r^n$ for orders $n \geq 2$. The high amplitude of these higher-order derivatives of $f(r)$ near the warp-bubble radius elucidates the critical, yet previously under-emphasized, local, and significant role of the Weyl curvature tensor in warp-bubble spacetimes.

Subsection \ref{Invariant comparison with Mattingly} critically examines and compares our computational results with those presented by Mattingly et al. \cite{Mattingly}. We address their computational and plotting challenges and highlight the discrepancies between our findings and the analysis in \cite{Mattingly}, including their significant underestimation of curvature invariant amplitudes by 21 orders of magnitude. Natário’s enforcement of zero-differential-volume-change $(K=0)$ makes his model even less practical than Alcubierre’s.

\subsection{\texorpdfstring{General behavior of curvature invariants}{General Behavior of Curvature Invariants}}

\par

All plots of curvature invariants were calculated and plotted using Wolfram \textit{Mathematica\textsuperscript{\textregistered}} \cite{Mathematica} based on Nat\'{a}rio's metric and its inverse (Eq.~(\ref{eq: spacetime metric}), and Eq.~(\ref{eq: spacetime contra metric}), respectively). To ensure the accuracy of our calculations, the results of all expressions used to calculate the curvature invariants were cross-verified against a range of known exact solutions. In Appendix~\ref{secA1}, we provide the closed-form analytical expressions for the curvature invariants $G$, $r_1$, $I$, and $r_2$. For the purpose of direct comparison, all plots are expressed in terms of a form with consistent curvature units, such as $G$, $\sign{\left(r_{1}\right)}\sqrt{\abs{r_{1}}}$, $\sign{\left(I\right)}\sqrt{\abs{I}}$, and $\sign{\left(r_{2}\right)}\sqrt[3]{\abs{r_{2}}}$. Notably, when rendered in this consistent curvature unit form, these invariants all depend on the square of the velocity, analogous to how energy density is proportional to velocity squared.

The form function defined by Eq.~(\ref{fMatt}) is employed for the plots of the curvature invariants discussed in this section. This choice is based on the warp-bubble parameters used in these examples, specifically $\rho = 5\,[\text{m}]$ and $\sigma = 4\,[\text{m}^{-1}]$, as well as $\rho = 100\,[\text{m}]$ and $\sigma = 50000\,[\text{m}^{-1}]$. As demonstrated in subsection \ref{form function}, these parameters provide an excellent numerical approximation for the scenarios considered. Additional calculations were performed using the exact form function defined by Eq.~(\ref{fAlcubierreForm}). While these calculations introduced greater algorithmic complexity and required more computational time, they did not yield any visible impact on the plots.

All plots are rendered within the radial range defined as $r \in [\rho - 3/\sigma, \rho + 3/\sigma]$, where $\rho$ represents the radius of the warp-bubble and $1/\sigma$ denotes its thickness. This range is selected based on the form function $f(r)$ behavior, which closely approximates a top-hat function and rapidly asymptotes to a flat function beyond this range. The flatness of $f(r)$ outside this range has been confirmed through separate computations. Choosing this specific range facilitates a clear three-dimensional visualization of both negative and positive values of the plotted function by eliminating the obstruction from the uninformative flat part of the function. Additionally, this range allows computational resources to be efficiently allocated for the precise numerical calculation of the amplitude and derivatives of the function, ensuring accuracy in regions where $f(r)$ undergoes significant changes.

The `direction of travel' depicted in all plots corresponds to the perspective of distant Eulerian observers. In Nat\'{a}rio's spacetime, the construction is such that Eulerian observers located at the center of the warp-bubble remain stationary while those outside the bubble move in the opposite direction to that shown in the plots.

\subsubsection{\texorpdfstring{Radial distribution of curvature invariants}{Radial distribution of curvature invariants}}

\begin{figure}[htbp]
    \centering
    \begin{subfigure}[b]{0.45\textwidth}
        \includegraphics[width=\textwidth]{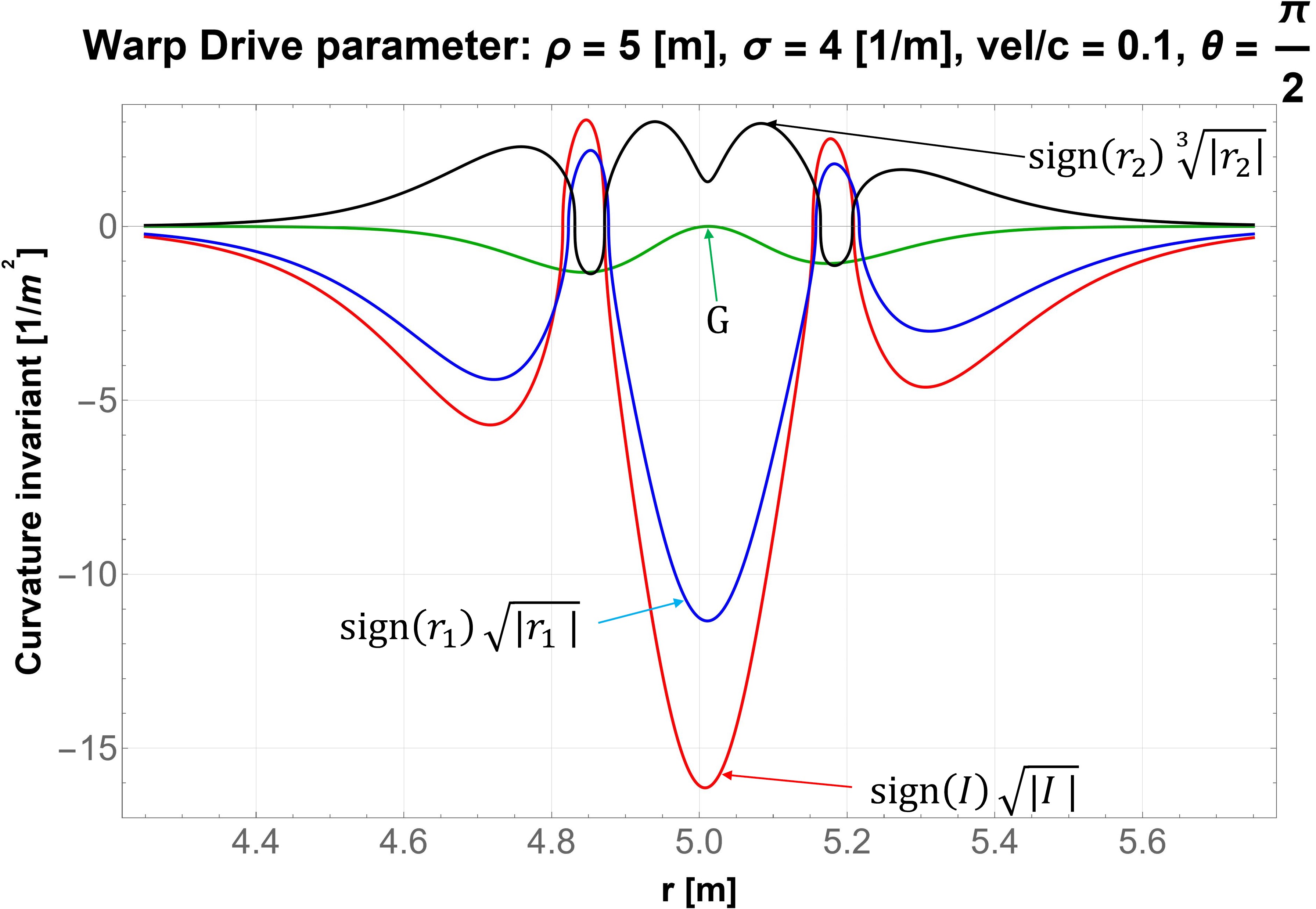}
        \caption{$\frac{v}{c}=0.1,\rho=5[\text{m}],\sigma=4[\text{m}^{-1}]$}
        \label{fig:pi2 v01 rho5}
    \end{subfigure}
    \hfill 
    \begin{subfigure}[b]{0.45\textwidth}
        \includegraphics[width=\textwidth]{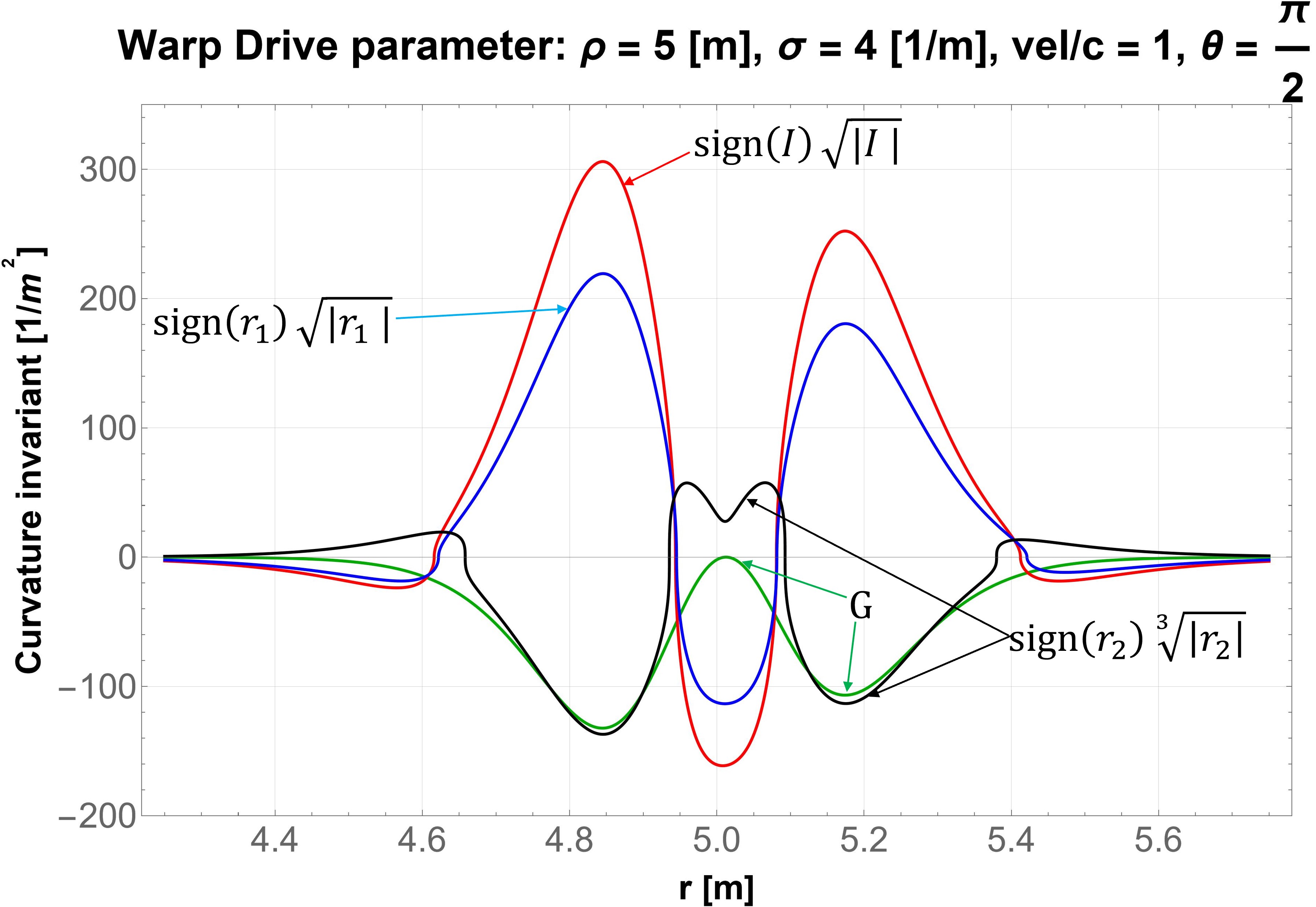}
        \caption{$\frac{v}{c}=1,\rho=5[\text{m}],\sigma=4[\text{m}^{-1}]$}
        \label{fig:pi2 v1 rho5}
    \end{subfigure}

    \begin{subfigure}[b]{0.45\textwidth}
        \includegraphics[width=\textwidth]{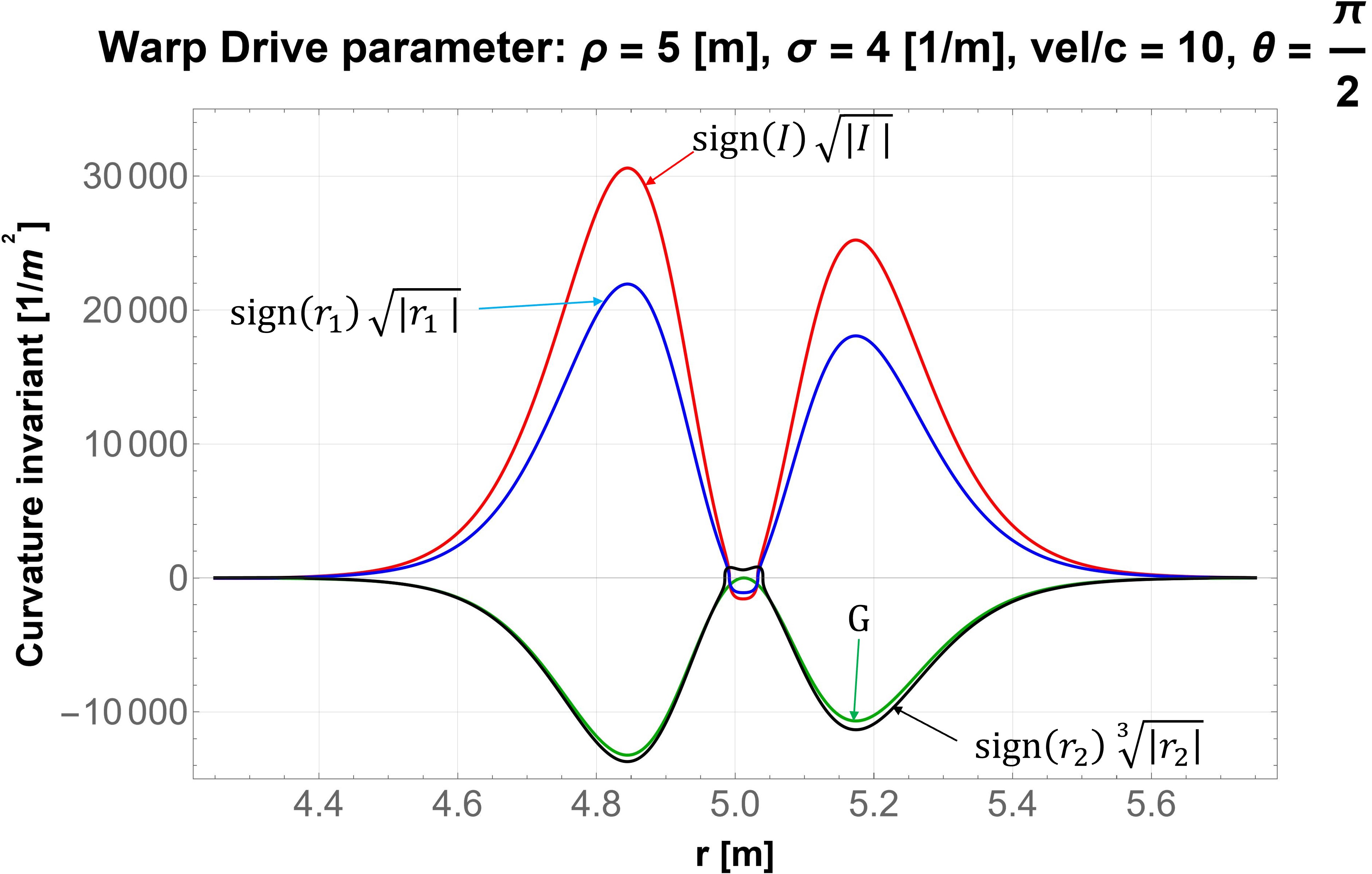}
        \caption{$\frac{v}{c}=10,\rho=5[\text{m}],\sigma=4[\text{m}^{-1}]$}
        \label{fig:pi2 v10 rho5}
    \end{subfigure}
    \hfill
    \begin{subfigure}[b]{0.45\textwidth}
        \includegraphics[width=\textwidth]{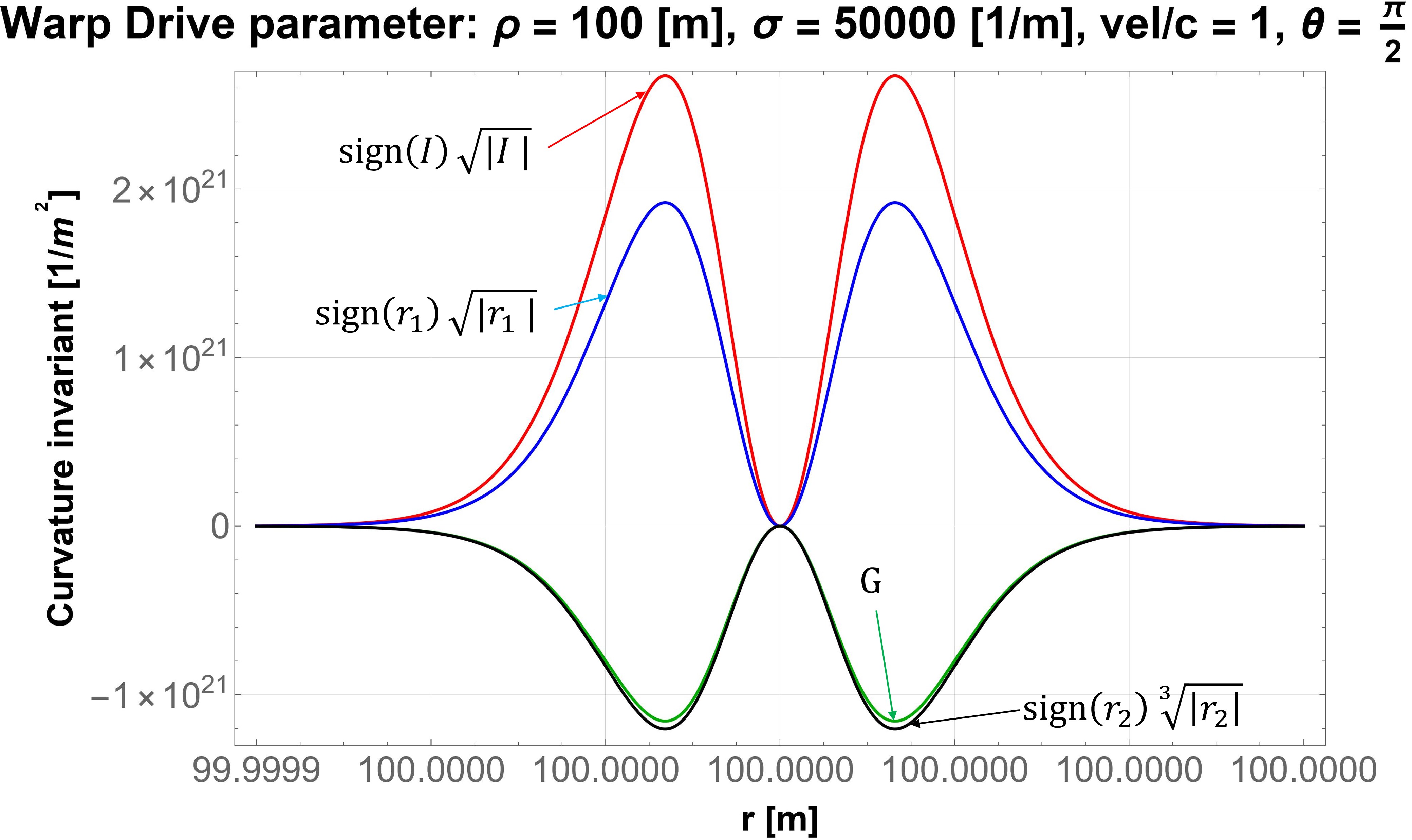}
        \caption{$\frac{v}{c}=1,\rho=100[\text{m}],\sigma=50000[\text{m}^{-1}]$}
        \label{fig:pi2 v1 rho100}
    \end{subfigure}
    \caption{Amplitude of the curvature invariants at the peak location $\theta=\frac{\pi}{2}$.}
    \label{fig:pi2}
\end{figure}

As elaborated in the subsequent subsections, the distribution of curvature invariants in the radial direction typically exhibits two peaks of similar amplitude under most conditions (see Fig.~\ref{fig:pi2}). These peaks are located at radii slightly shorter and longer than the radius of the warp-bubble. However, an exception occurs in the distribution of $\sign{\left(r_{1}\right)}\sqrt{\abs{r_{1}}}$ and $\sign{\left(I\right)}\sqrt{\abs{I}}$ at a subluminal speed of $v = 0.1c$, specifically for $\rho = 5 \, [\text{m}]$ and $\sigma = 4 \, [\text{m}^{-1}]$. In this scenario, the distribution exhibits three peaks (see Fig.~\ref{fig:pi2 v01 rho5}), with the central peak being the most pronounced in amplitude. Conversely, the cubic invariant measure $\sign{\left(r_{2}\right)}\sqrt[3]{\abs{r_{2}}}$ displays four peaks (see Fig.~\ref{fig:pi2 v01 rho5}), while Einstein's curvature scalar $G$ consistently shows only two peaks, similar to other cases.

\subsubsection{\texorpdfstring{Circumferential distribution of curvature invariants on a meridian}{Circumferential distribution on a meridian}}\label{Circumferential distribution meridian}

The distribution over a meridian (circle of longitude, Fig. \ref{Fig1_SpherCoord}) of all curvature invariants in Nat\'{a}rio spacetime primarily follows a $\sin^2(\theta) = \frac{1}{2}\left(1 - \cos(2\theta)\right)$ pattern. This distribution reaches its minimum at the intersections of the warp-bubble with the axis of travel at $\theta = 0$ and at $\theta = \pi$. It peaks at the intersections of the warp-bubble with the plane perpendicular to the axis of travel at $\theta = \pm \pi/2$. The distribution exhibits continuous rotational symmetry around the axis of travel since the components of the velocity vector and of the shift vector Eq.~(\ref{eq:BetaCtv},\ref{eq:BetaCov}) are independent of the azimuthal angle $\phi$ (Fig. \ref{Fig1_SpherCoord}).

\subsubsection{\texorpdfstring{Relative amplitude of curvature invariants}{Amplitude of curvature invariants}\label{sec:amplitude}}

\begin{figure}[htbp]
    \centering
    \includegraphics[width=\textwidth]{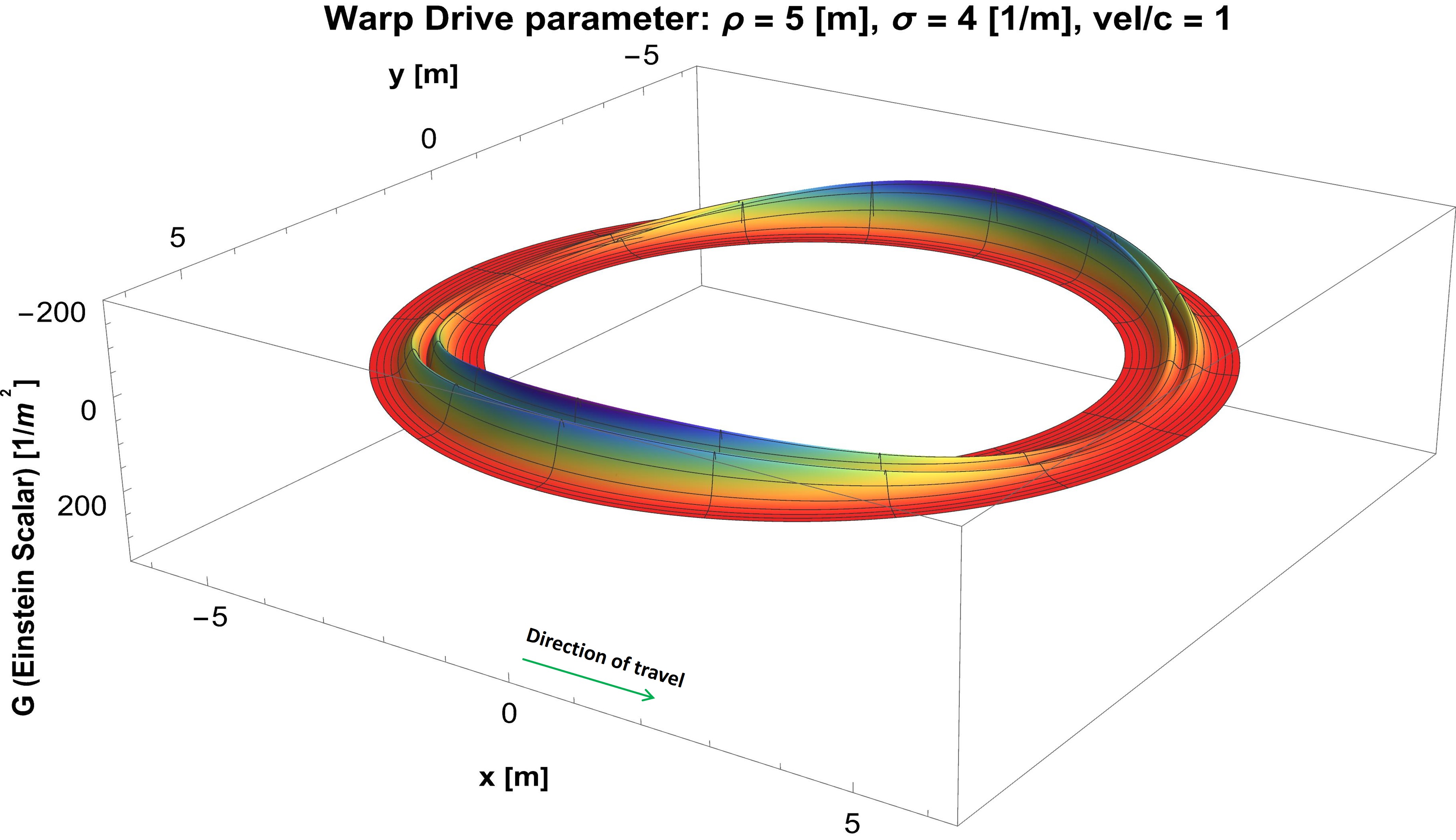}
    \caption{3-D plot (negative direction pointing upwards) of Einstein’s curvature scalar $G= G_{\alpha}\,^{\alpha}$ vs. the $x,y$ coordinates, for $\frac{v}{c}=1,\rho=5[\text{m}],\sigma=4[\text{m}^{-1}]$.}
    \label{fig:G v1 rho5}
\end{figure}

\begin{figure}[htbp]
    \centering
    \includegraphics[width=\textwidth]{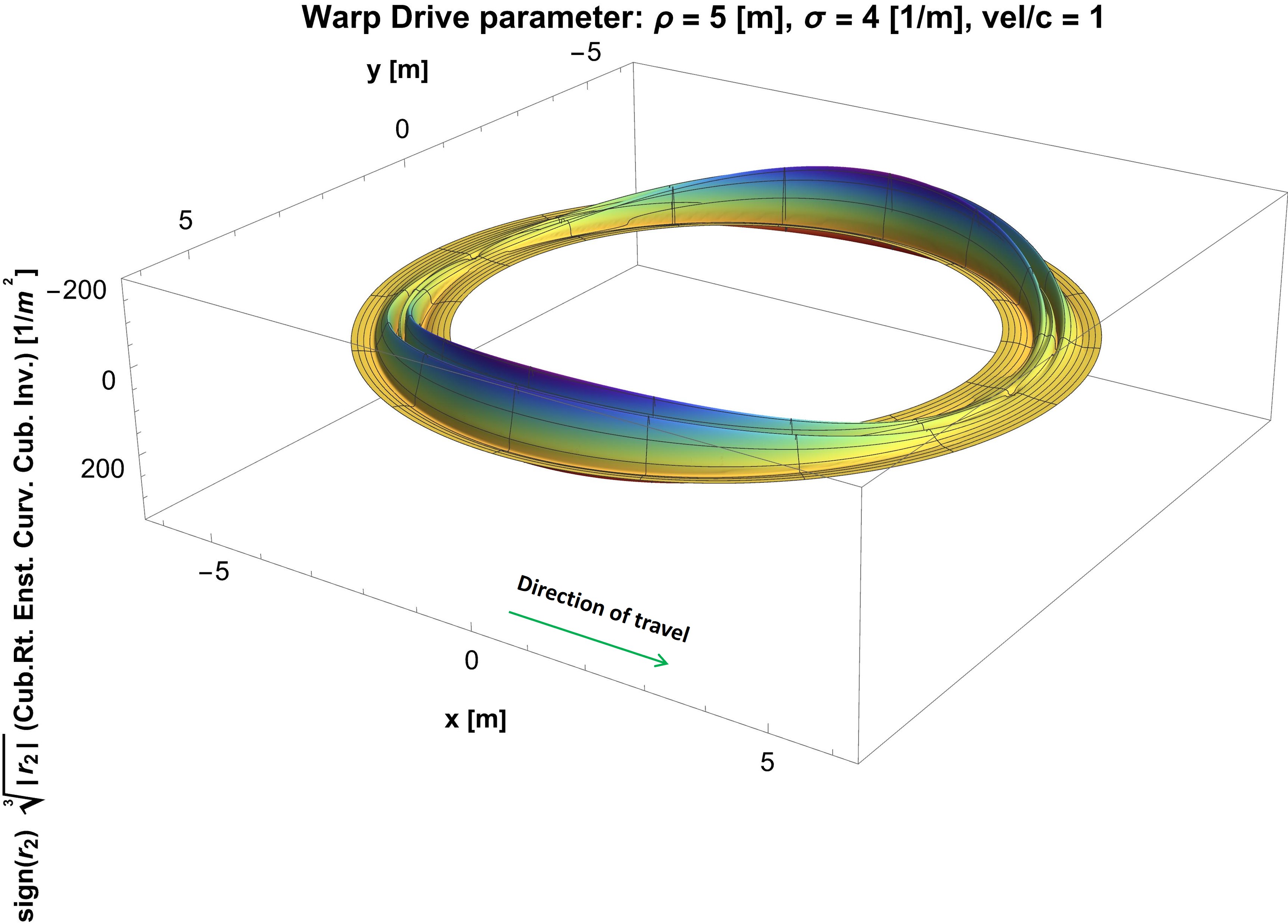}
    \caption{3-D plot (negative direction pointing upwards) of the signed cubic root $\sign{\left(r_{2}\right)}\sqrt[3]{\abs{r_{2}}}$ of the cubic scalar invariant
$r_{2}$ defined in Eqs.~(\ref{eq:r2_a},\ref{eq:r2_b}), vs. the $x,y$ coordinates, for $\frac{v}{c}=1,\rho=5[\text{m}],\sigma=4[\text{m}^{-1}]$.}
    \label{fig:r2 v1 rho5}
\end{figure}

\begin{figure}[htbp]
    \centering
    \includegraphics[width=\textwidth]{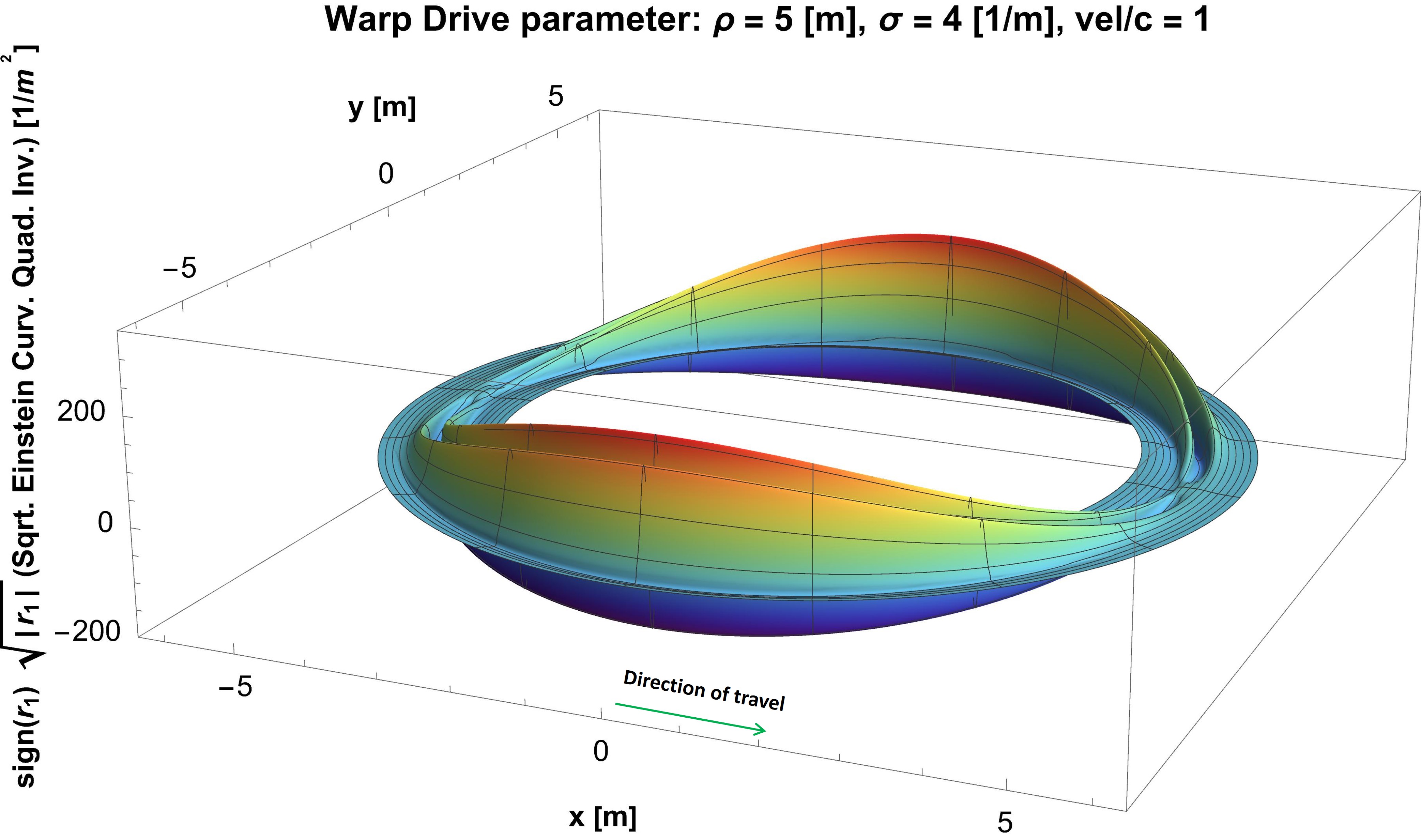}
    \caption{3-D plot of the signed square root $\sign{\left(r_{1}\right)}\sqrt{\abs{r_{1}}}$ of the quadratic scalar invariant
$r_{1}$ defined in Eqs.~(\ref{eq:r1_a},\ref{eq:r1_b}), vs. the $x,y$ coordinates, for $\frac{v}{c}=1,\rho=5[\text{m}],\sigma=4[\text{m}^{-1}]$.}
    \label{fig:r1 v1 rho5}
\end{figure}

\begin{figure}[htbp]
    \centering
    \includegraphics[width=\textwidth]{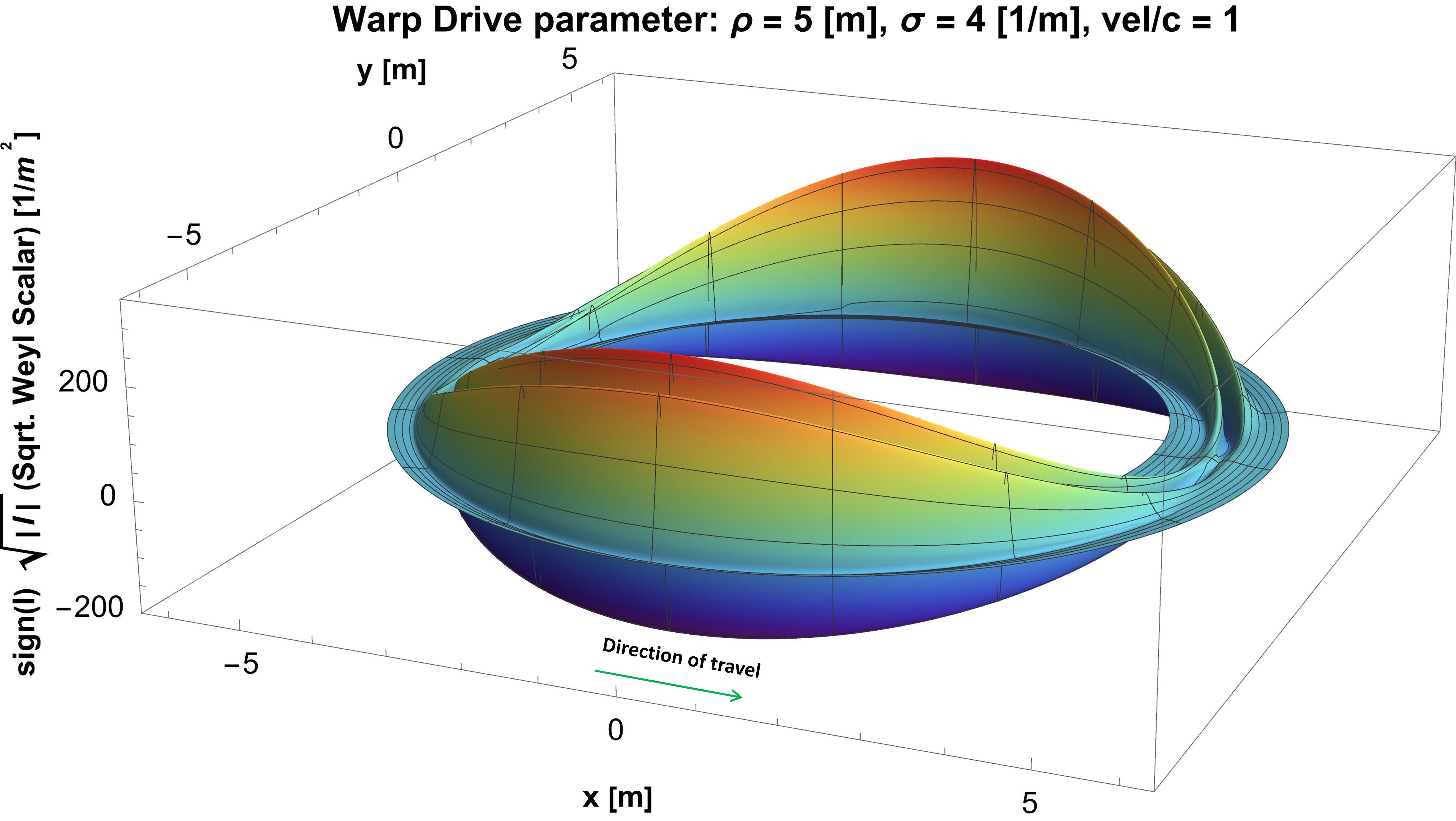}
    \caption{3-D plot of the signed square root $\sign{\left(I\right)}\sqrt{\abs{I}}$ of the Weyl scalar invariant $I\equiv C_{\alpha\beta\gamma\delta} C^{\alpha\beta\gamma\delta}$, defined in Eq.~($\ref{WeylScalar}$),  vs. the $x,y$ coordinates, for $\frac{v}{c}=1,\rho=5[\text{m}],\sigma=4[\text{m}^{-1}]$.}
    \label{fig:I v1 rho5}
\end{figure}

\begin{figure}[htbp]
    \centering
    \begin{subfigure}[b]{0.45\textwidth}
        \includegraphics[width=\textwidth]{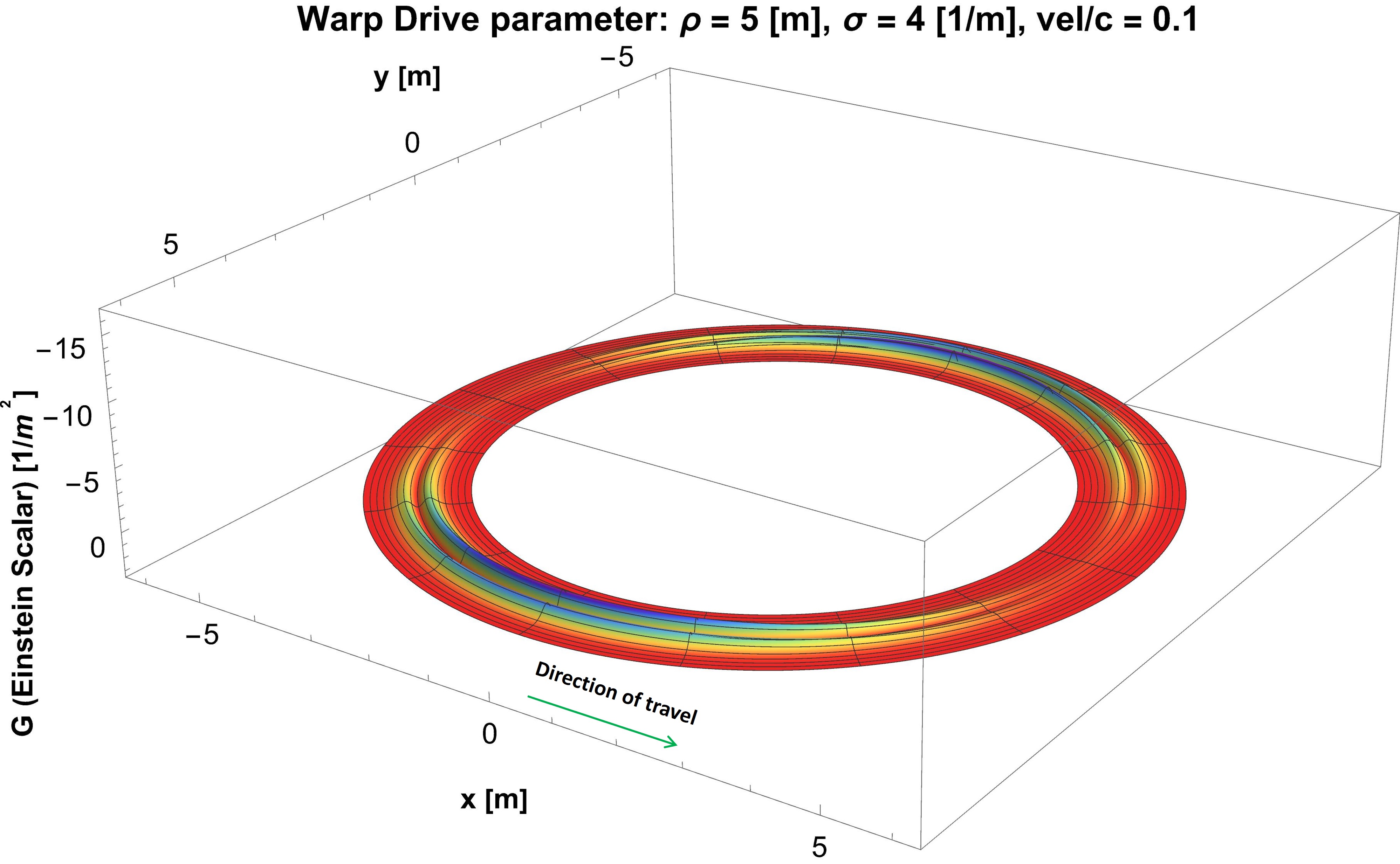}
        \caption{$G$ (negative direction pointing upwards)}
        \label{fig:G v01 rho5}
    \end{subfigure}
    \hfill 
    \begin{subfigure}[b]{0.45\textwidth}
        \includegraphics[width=\textwidth]{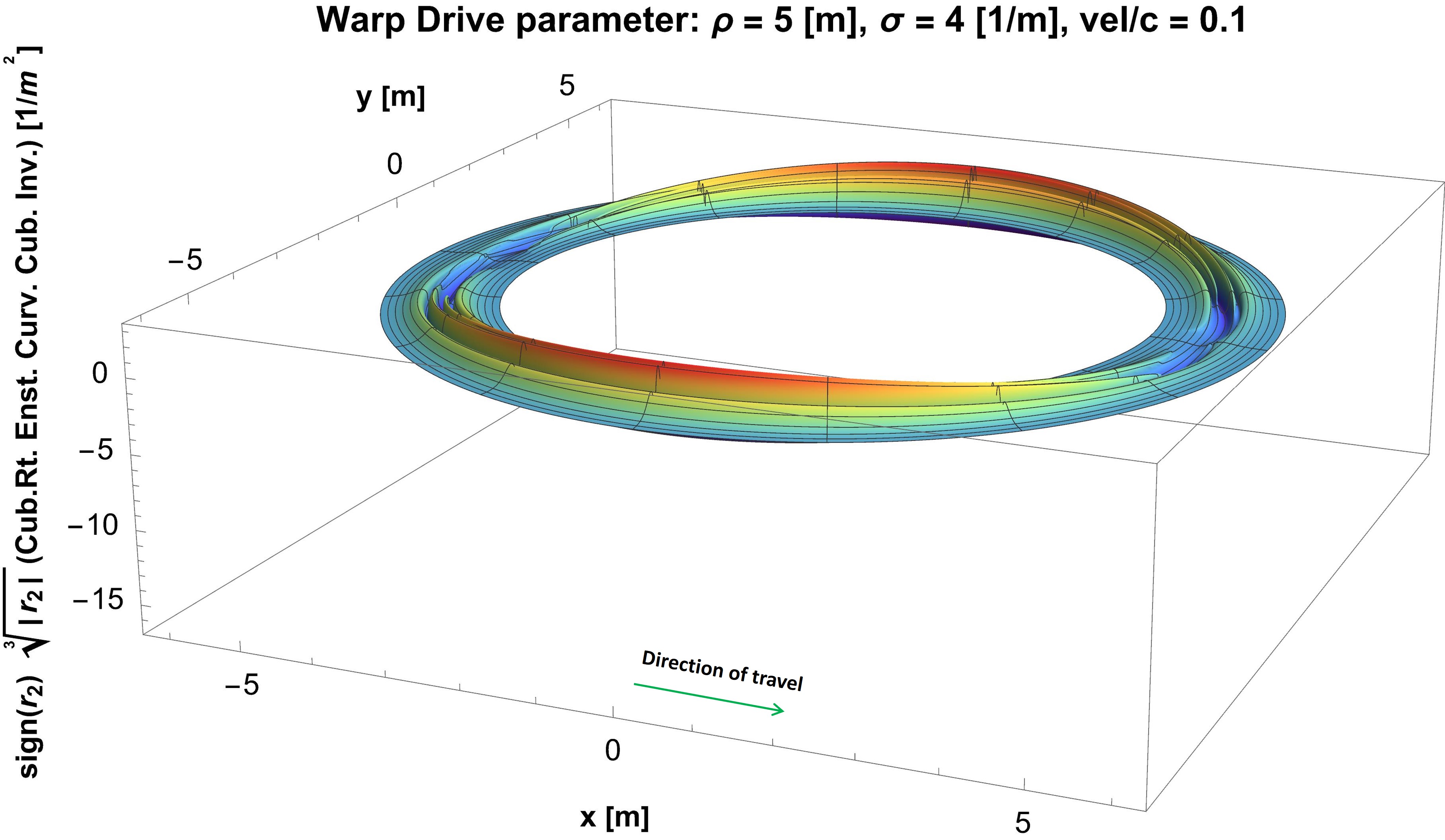}
        \caption{$\sign{\left(r_{2}\right)}\sqrt[3]{\abs{r_{2}}}$}
        \label{fig:r2 v01 rho5}
    \end{subfigure}

    \begin{subfigure}[b]{0.45\textwidth}
        \includegraphics[width=\textwidth]{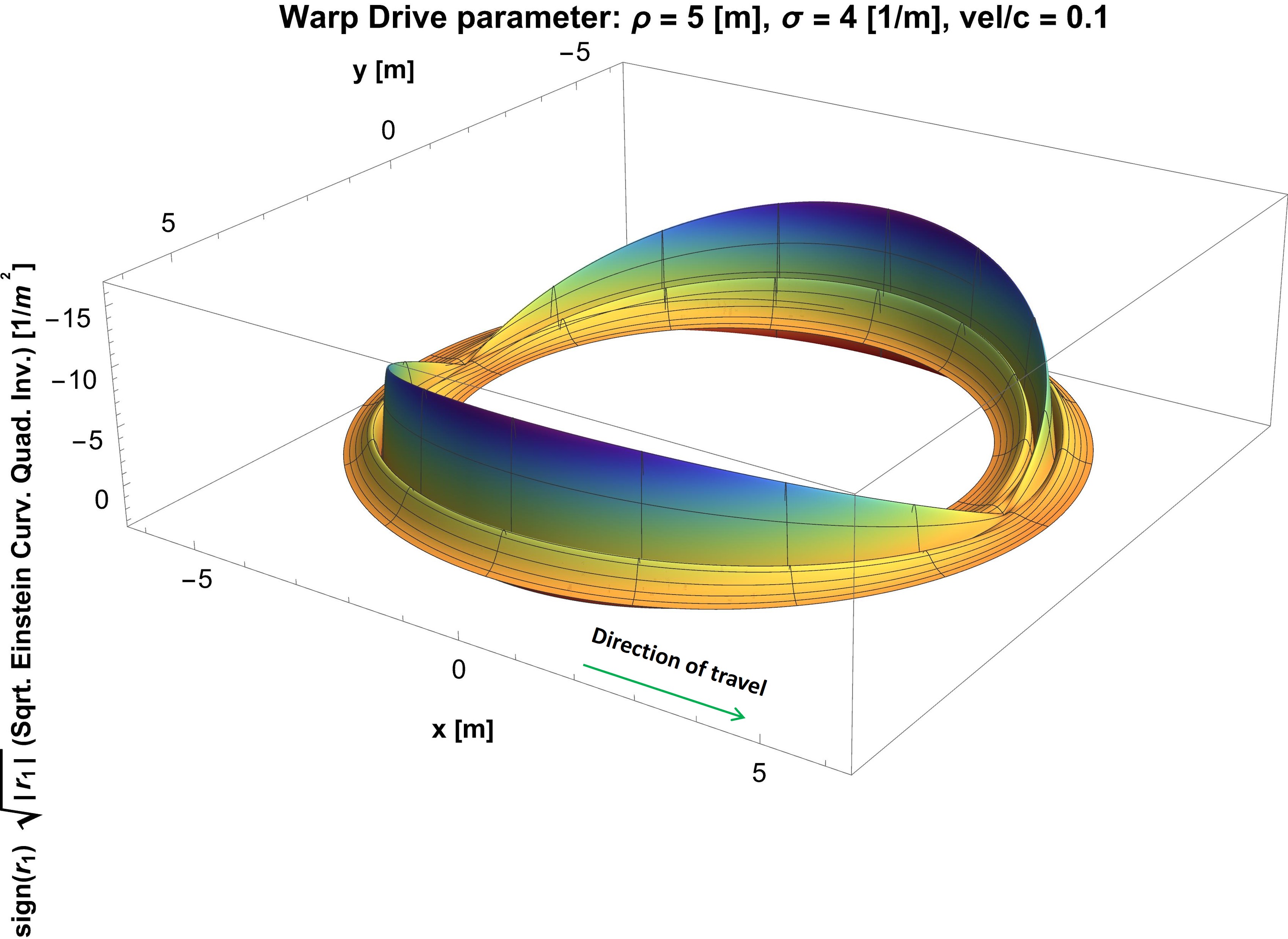}
        \caption{$\sign{\left(r_{1}\right)}\sqrt{\abs{r_{1}}}$ (negative direction pointing upwards)}
        \label{fig:r1 v01 rho5}
    \end{subfigure}
    \hfill
    \begin{subfigure}[b]{0.45\textwidth}
        \includegraphics[width=\textwidth]{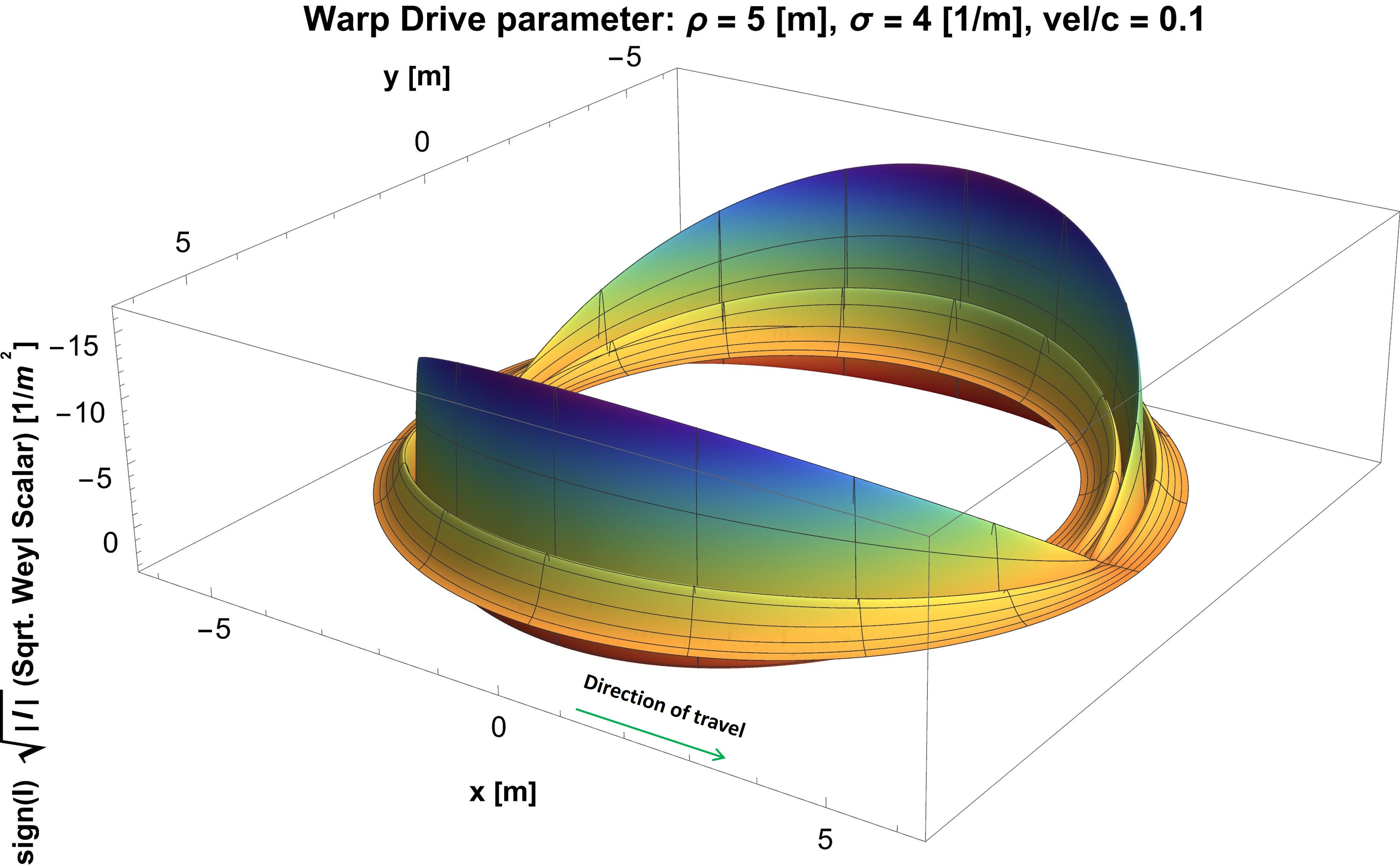}
        \caption{$\sign{\left(I\right)}\sqrt{\abs{I}}$ (negative direction pointing upwards)}
        \label{fig:I v01 rho5}
    \end{subfigure}
    \caption{Amplitude of the curvature invariants for $\frac{v}{c}=0.1,\rho=5[\text{m}],\sigma=4[\text{m}^{-1}]$.}
    \label{fig:v01 rho5}
\end{figure}

\begin{figure}[htbp]
    \centering
    \begin{subfigure}[b]{0.45\textwidth}
        \includegraphics[width=\textwidth]{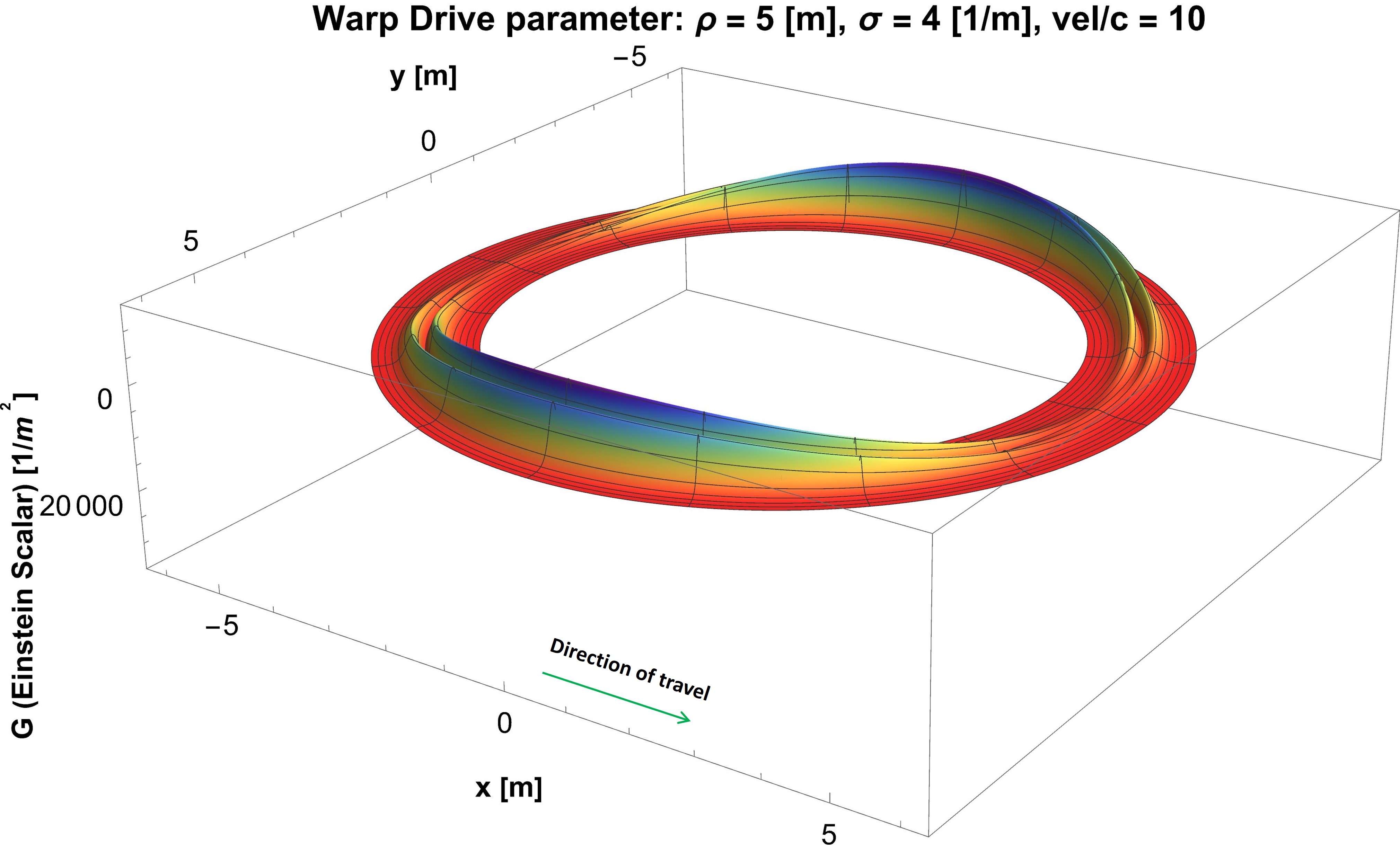}
        \caption{$G$ (negative direction pointing upwards)}
        \label{fig:G v10 rho5}
    \end{subfigure}
    \hfill 
    \begin{subfigure}[b]{0.45\textwidth}
        \includegraphics[width=\textwidth]{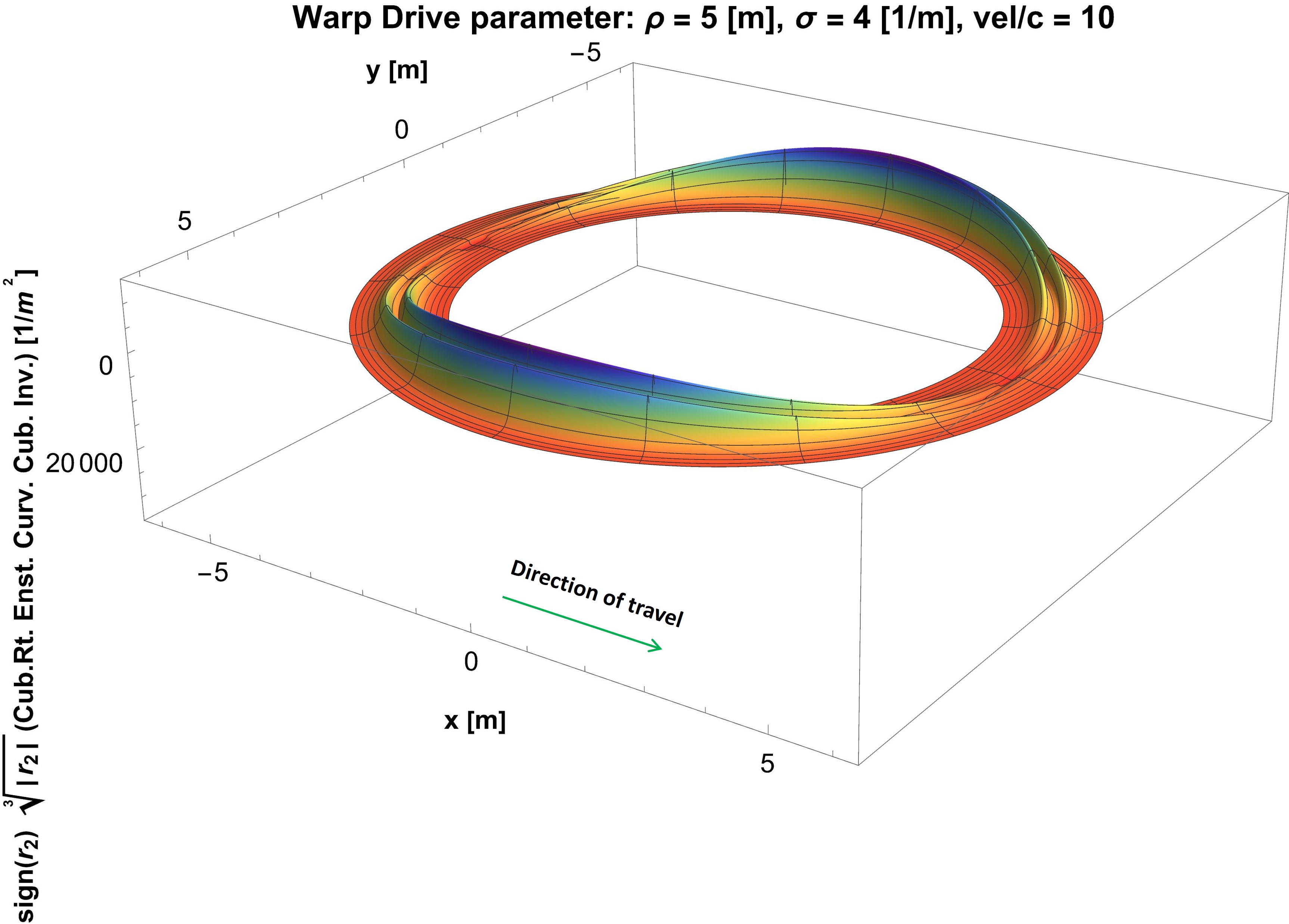}
        \caption{$\sign{\left(r_{2}\right)}\sqrt[3]{\abs{r_{2}}}$ (negative direction pointing upwards)}
        \label{fig:r2 v10 rho5}
    \end{subfigure}

    \begin{subfigure}[b]{0.45\textwidth}
        \includegraphics[width=\textwidth]{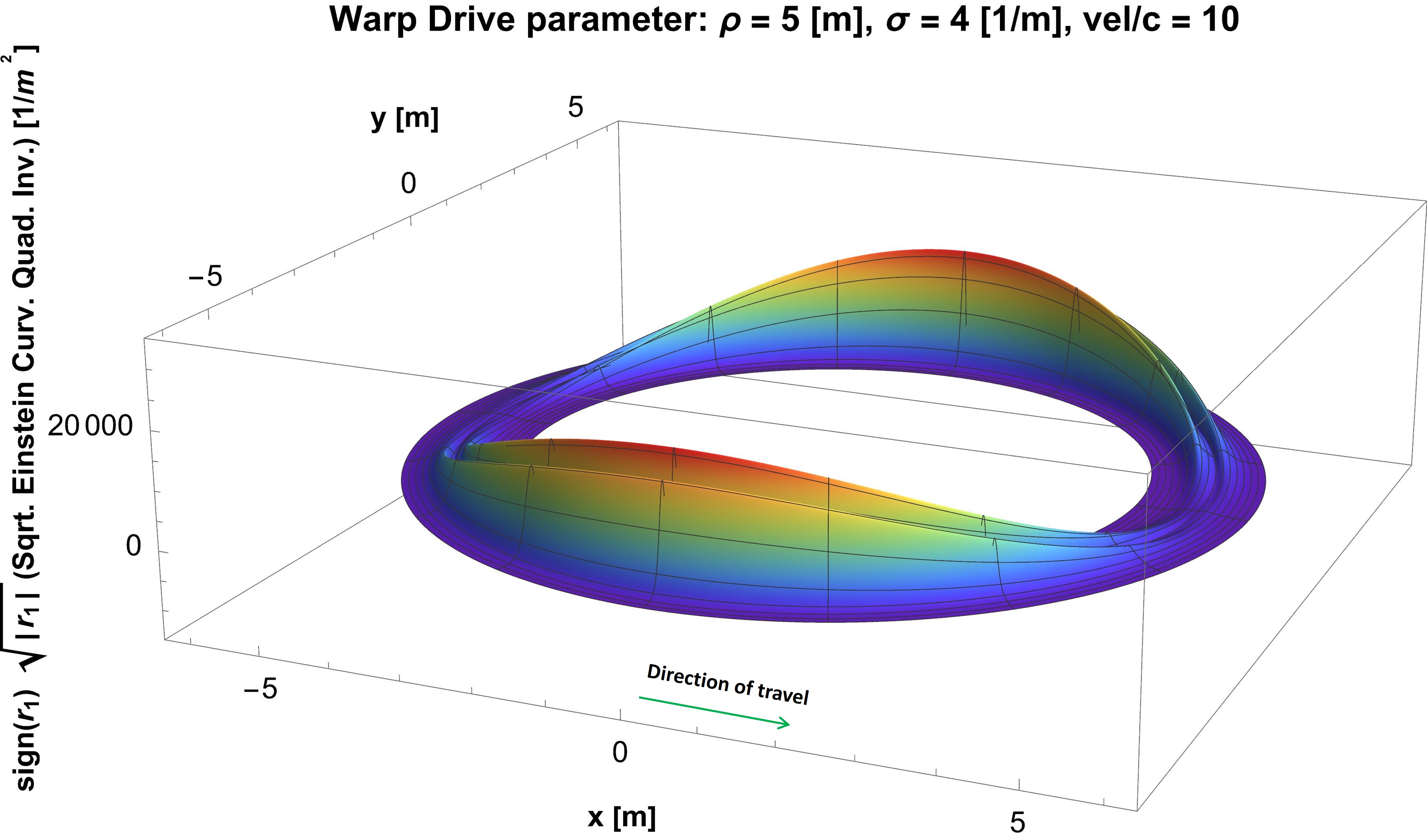}
        \caption{$\sign{\left(r_{1}\right)}\sqrt{\abs{r_{1}}}$}
        \label{fig:r1 v10 rho5}
    \end{subfigure}
    \hfill
    \begin{subfigure}[b]{0.45\textwidth}
        \includegraphics[width=\textwidth]{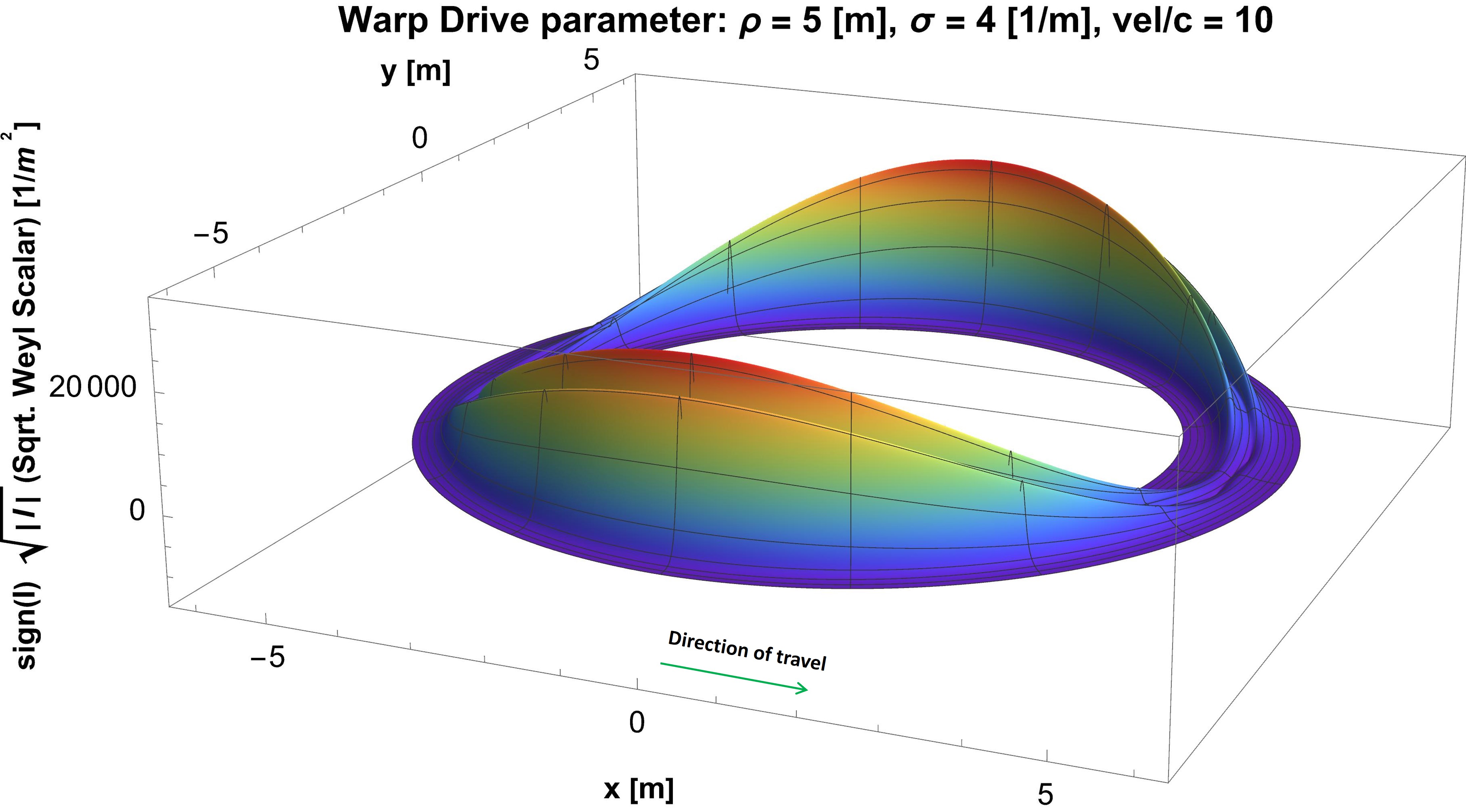}
        \caption{$\sign{\left(I\right)}\sqrt{\abs{I}}$}
        \label{fig:I v10 rho5}
    \end{subfigure}
    \caption{Amplitude of the curvature invariants for $\frac{v}{c}=10,\rho=5[\text{m}],\sigma=4[\text{m}^{-1}]$.}
    \label{fig:v10 rho5}
\end{figure}

In Nat\'{a}rio's spacetime, the amplitudes of curvature invariants span a range (Figs.~\ref{fig:G v1 rho5}-\ref{fig:v10 rho5}), with Einstein’s curvature scalar $G = G_{\alpha}\,^{\alpha}$ having the smallest amplitude. This is closely followed by the cubic invariant $\sign{\left(r_{2}\right)}\sqrt[3]{\abs{r_{2}}}$, and then, at a significantly higher amplitude, the quadratic invariant $\sign{\left(r_{1}\right)}\sqrt{\abs{r_{1}}}$. The Weyl invariant $\sign{\left(I\right)}\sqrt{\abs{I}}$ exhibits the largest amplitude among these invariants.

In the extensive literature on the Alcubierre and Nat\'{a}rio warp drives, it is noteworthy that the predominant focus has been on the Einstein and Ricci curvature measures, along with the corresponding stress-energy tensor components as a source, rather than on the Weyl curvature. Intriguingly, the Weyl curvature invariant $I$ exhibits a significantly higher amplitude, even when normalized to comparable units by considering its signed square root. Additionally, it is notable that the distribution of the Weyl curvature is located within the warp-bubble, coinciding with the location of Einstein's curvature. In this context, the Weyl curvature is not non-local as typically assumed; rather, it is just as local as the Einstein curvature.

Considering the deformation of spacetime, the square of the trace of the Einstein curvature, $G^2$, or equivalently, the square of the trace of the Ricci curvature, $R^2$, for a fluid under isotropic pressure, informs us about the curvature due to volume change in spacetime. The cubic invariant $r_{2}$ contains information about the curvature due to three-dimensional shear deformation of the purely spatial components, which is a more complex state of curvature deformation than the in-plane shear expressed by $r_{1}$.

The quadratic traceless invariants, $r_{1}$ and $I$, offer information about the curvature due to simple-shear changes in spacetime, which involve planar spatial rotation of the principal axes of stress. The invariant $r_{1}$ represents the curvature arising from the self-contraction of the traceless curvature tensor $\widehat{G}_{\alpha}^{\:\:\:{\beta}}\widehat{G}_{\beta}^{\:\:\:{\alpha}}$, sourced by the traceless stress-energy-momentum tensor $\widehat{T}_{\alpha \beta}$ as per the Einstein field equations. 

As discussed by Penrose \cite{Penrose} the Weyl tensor represents the free gravitational field while the Ricci tensor is associated with matter-energy sources. Through a Bianchi identity and the application of Einstein's field equations, the second derivatives of the Weyl curvature tensor are connected to the second derivatives of the stress-energy-momentum tensor, accompanied by nonlinear coupling terms. This is elaborated upon on pages 264-265 and 403 in Padmanabhan \cite{padmanabhan} and on pages 194-195 in de Felice and Clarke \cite{deFelice}. This relationship between the second derivatives of the stress-energy-momentum tensor and the second derivatives of the Weyl tensor highlights how high-amplitude second derivatives in matter-energy distributions can influence the structure and evolution of the Weyl curvature of spacetime.  This explains the Weyl curvature's local and significant role in warp drives' physics. The pronounced second derivatives of the warp-bubble, resulting from the sharpness of the form function $f(r)$—which closely approximates a top-hat function—are responsible for the amplitude of the Weyl scalar invariant $I\equiv C_{\alpha\beta\gamma\delta} C^{\alpha\beta\gamma\delta}$ for both the Alcubierre and Nat\'{a}rio spacetimes.

\begin{figure}[htbp]
    \centering
    \begin{subfigure}[b]{0.45\textwidth}
        \includegraphics[width=\textwidth]{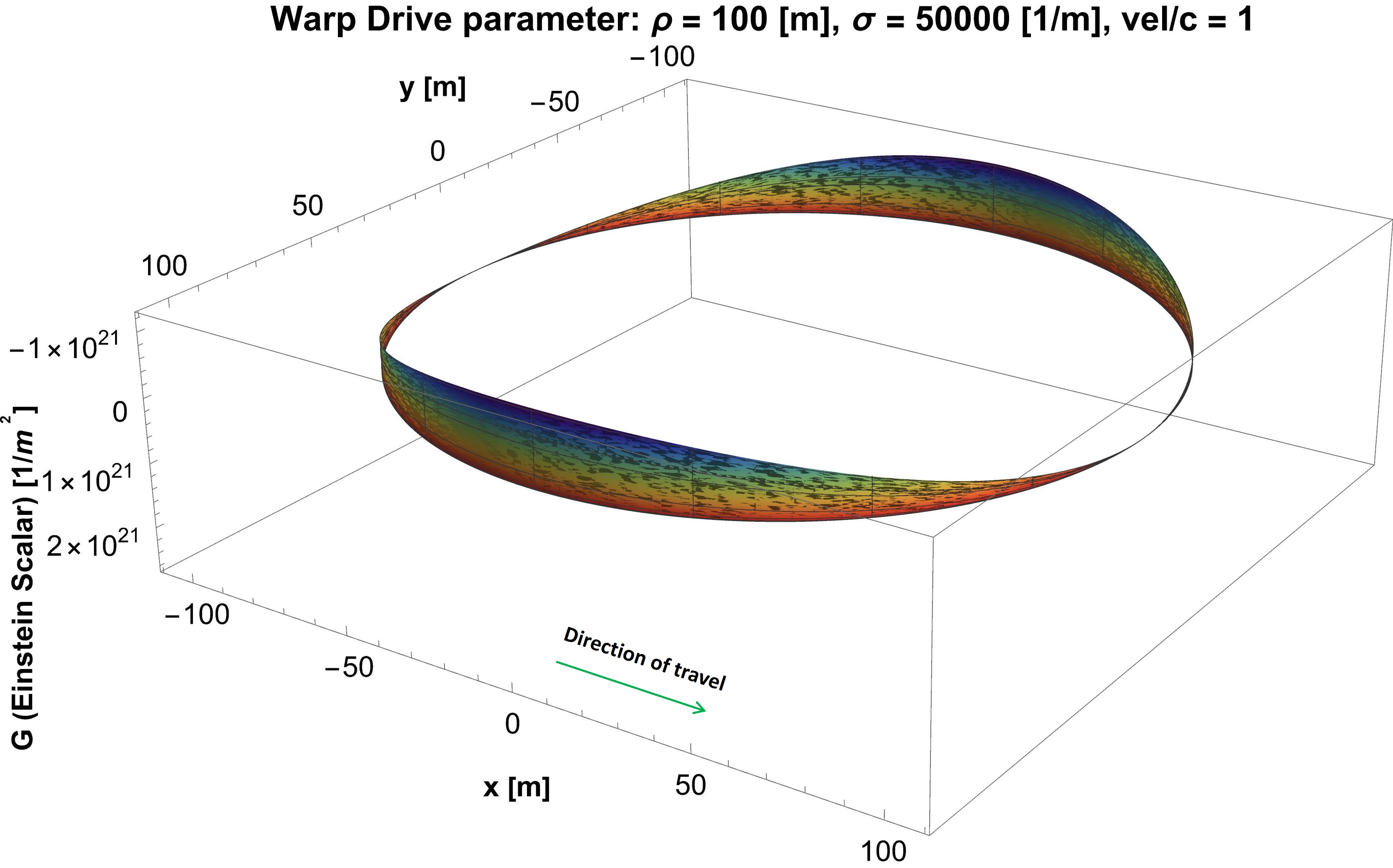}
        \caption{$G$ (negative direction pointing upwards)}
        \label{fig:G v1 rho100}
    \end{subfigure}
    \hfill 
    \begin{subfigure}[b]{0.45\textwidth}
        \includegraphics[width=\textwidth]{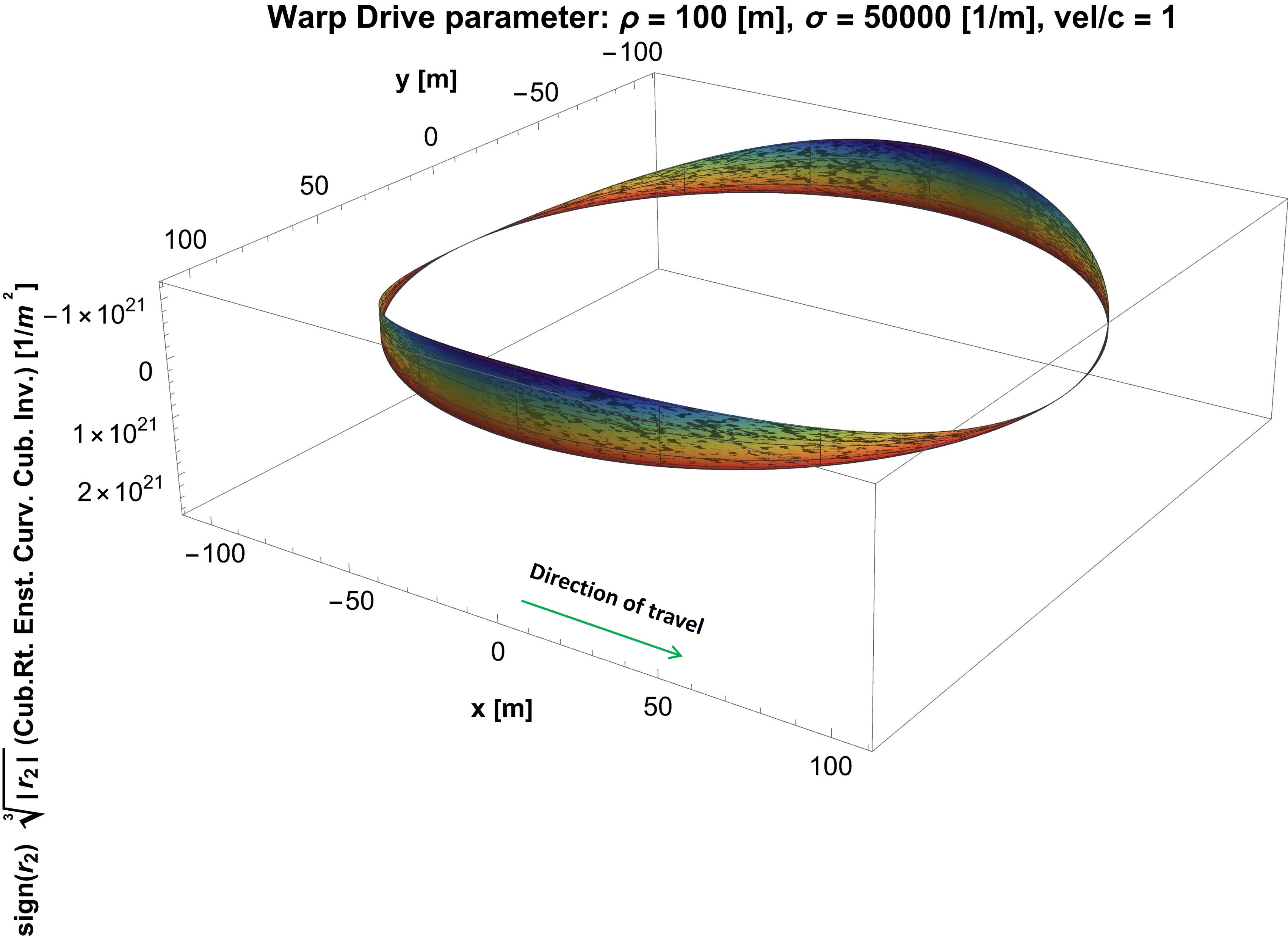}
        \caption{$\sign{\left(r_{2}\right)}\sqrt[3]{\abs{r_{2}}}$ (negative direction pointing upwards)}
        \label{fig:r2 v1 rho100}
    \end{subfigure}

    \begin{subfigure}[b]{0.45\textwidth}
        \includegraphics[width=\textwidth]{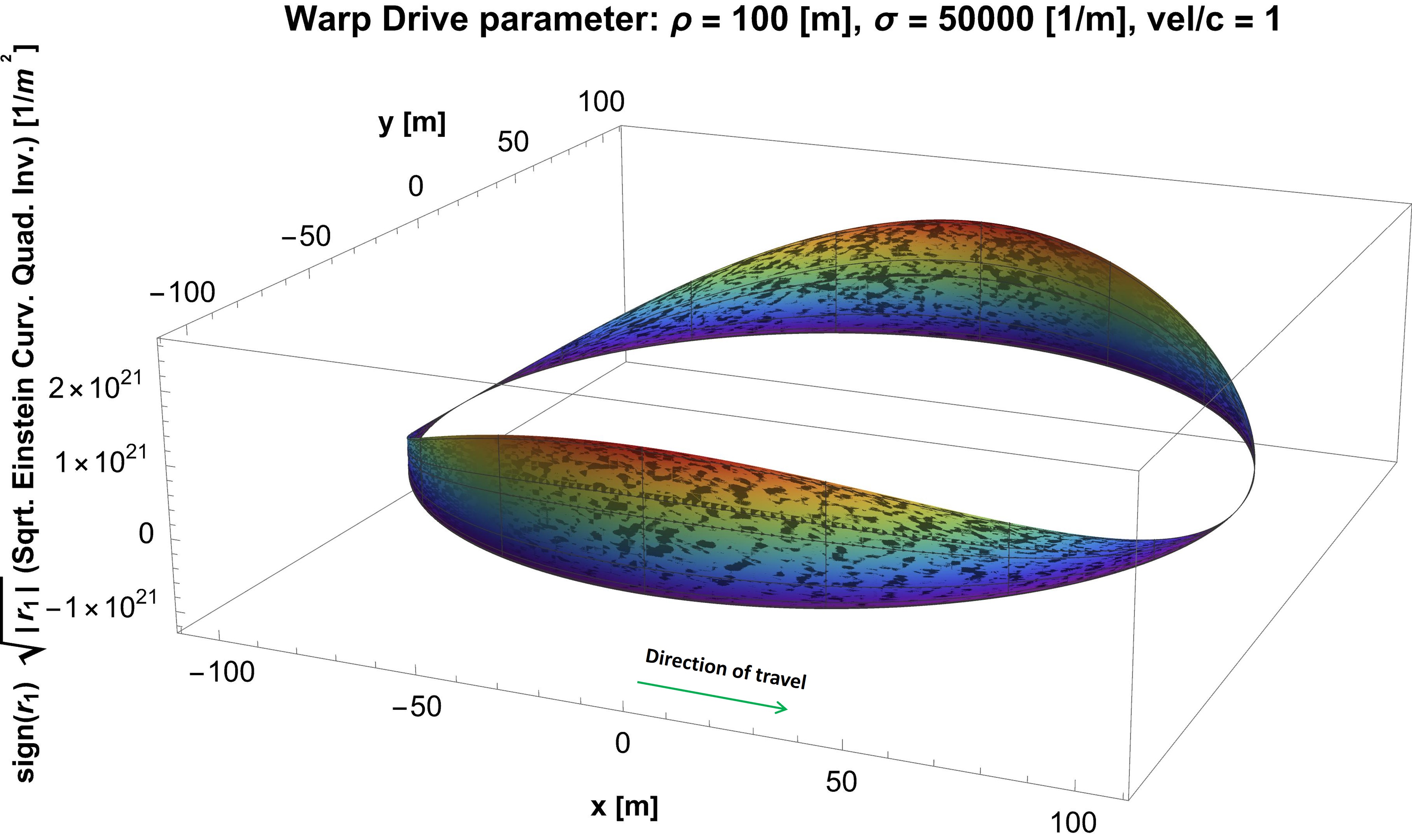}
        \caption{$\sign{\left(r_{1}\right)}\sqrt{\abs{r_{1}}}$}
        \label{fig:r1 v1 rho100}
    \end{subfigure}
    \hfill
    \begin{subfigure}[b]{0.45\textwidth}
        \includegraphics[width=\textwidth]{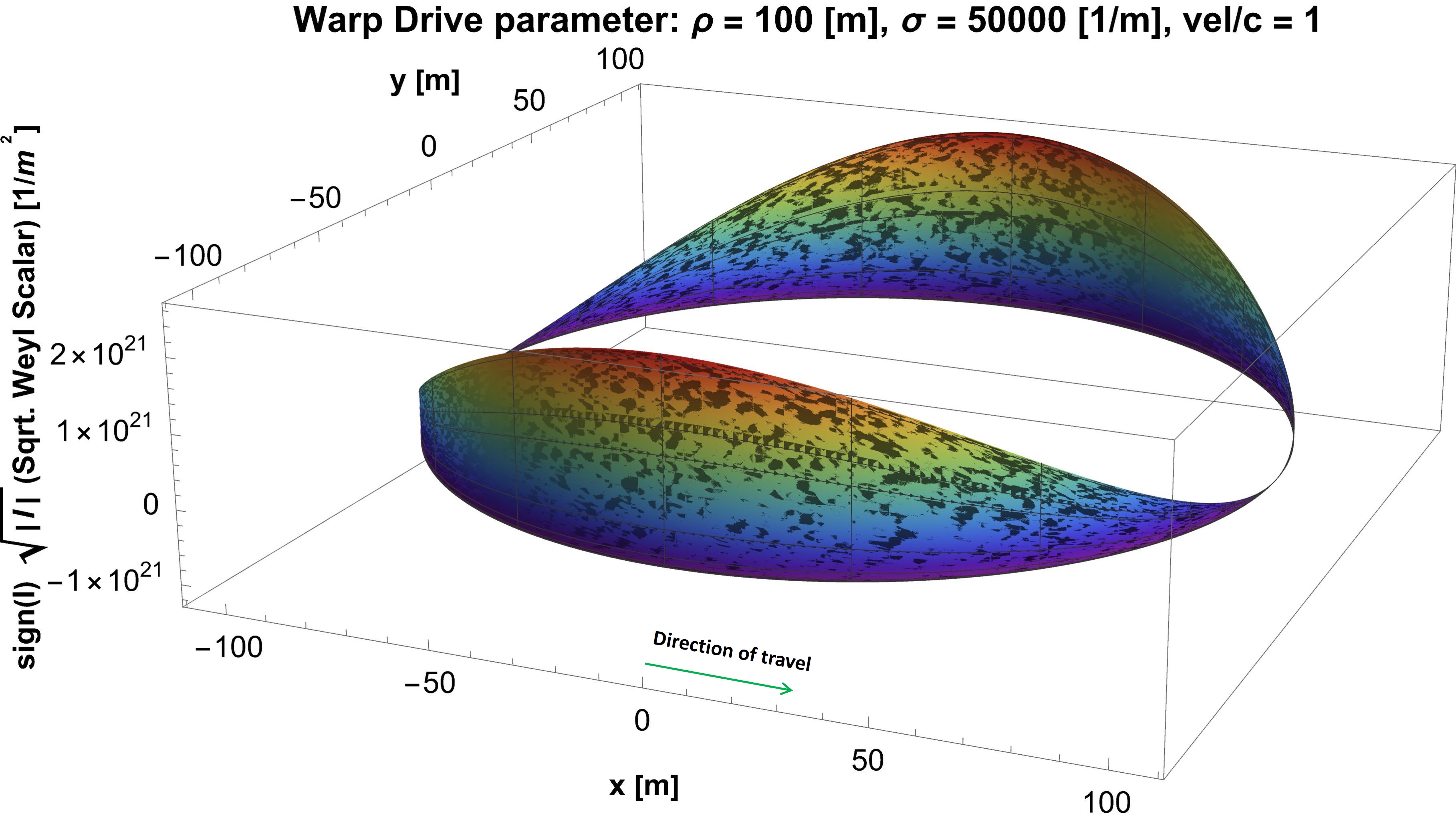}
        \caption{$\sign{\left(I\right)}\sqrt{\abs{I}}$}
        \label{fig:I v1 rho100}
    \end{subfigure}
    \caption{Amplitude of the curvature invariants for $\frac{v}{c}=1,\rho=100[\text{m}],\sigma=50000[\text{m}^{-1}]$.}
    \label{fig:v1 rho100}
\end{figure}

\subsection{\texorpdfstring{Comparison of curvature invariants in the Nat\'{a}rio and Alcubierre spacetimes for $ v = c $, $ \rho = 5 \, [\text{m}] $ and $ \sigma = 4 \, [\text{m}^{-1}] $}{Comparison of Curvature Invariants in Natario's and Alcubierre's Spacetimes}}

Figs.~\ref{fig:G v1 rho5}, \ref{fig:r2 v1 rho5}, \ref{fig:r1 v1 rho5}, and \ref{fig:I v1 rho5} provide a three-dimensional view of the distribution of curvature invariants in the $x,y$ plane, with the vertical axis representing the amplitude of each curvature invariant. The conditions for these plots are set at $v = c$, $\rho = 5 \, [\text{m}]$, and $\sigma = 4 \, [\text{m}^{-1}]$. These are the same conditions under which the Alcubierre spacetime was examined and plotted in \cite{Rodal}.

All plots are presented on the same vertical scale. However, Fig.~\ref{fig:G v1 rho5} for $G$ and Fig.~\ref{fig:r2 v1 rho5} for $\sign{\left(r_{2}\right)}\sqrt[3]{\abs{r_{2}}}$ are inverted, with the negative direction pointing upwards, to reflect that the peaks in these distributions have negative values. All four curvature invariant distributions exhibit two peaks of similar amplitude. These peaks are located at radii slightly shorter and longer than the radius of the warp-bubble.

The amplitudes of the curvature invariants range from the lowest corresponding to Einstein’s curvature scalar $G = G_{\alpha}\,^{\alpha}$, to the highest corresponding to the Weyl invariant $\sign{\left(I\right)}\sqrt{\abs{I}}$.

\subsubsection{\texorpdfstring{Comparison of radial distribution of curvature invariants in the Nat\'{a}rio and Alcubierre spacetimes}{Radial distribution of curvature invariants in Natario and Alcubierre}}

In Nat\'{a}rio's spacetime, Einstein's scalar $G$ (Fig.~\ref{fig:G v1 rho5}) consistently exhibits negative values and displays two negative peaks in the radial direction. These peaks are located at the intersection of the warp-bubble with the plane perpendicular to the axis of travel at $\theta = \pi/2$. The negative peaks occur at radii slightly smaller and larger than the warp-bubble's radius. In contrast, in Alcubierre's spacetime \cite{Rodal}, Einstein's scalar $G$ peaks in the perpendicular direction at the intersection of the warp-bubble with the direction of travel: the $x$-axis. In Alcubierre's model, $G$ shows a peak of negative amplitude at a radius smaller than the warp-bubble's radius and a peak of positive amplitude at a radius larger than the warp-bubble's radius.

In Nat\'{a}rio's spacetime, the curvature invariants $\sign{\left(r_{1}\right)}\sqrt{\abs{r_{1}}}$ (Fig.~\ref{fig:r1 v1 rho5}) and $\sign{\left(I\right)}\sqrt{\abs{I}}$ (Fig.~\ref{fig:I v1 rho5}) exhibit two positive peaks in the radial direction, at radii slightly smaller and larger than the warp-bubble radius. These invariants also display a single negative peak between the positive peaks, located at a radius slightly larger than the warp-bubble radius. In contrast, in Alcubierre's spacetime \cite{Rodal}, the curvature invariants $\sign{\left(r_{1}\right)}\sqrt{\abs{r_{1}}}$ and $\sign{\left(I\right)}\sqrt{\abs{I}}$ exhibit a more complex shape and orientation pattern with two positive and two negative peaks in the radial direction. 

\subsubsection{\texorpdfstring{Comparison of circumferential distribution of curvature invariants in the Nat\'{a}rio and Alcubierre spacetimes}{Comparison circumferential distribution on a meridian}}\label{Comparison of circumferential distribution}

Both the Nat\'{a}rio and Alcubierre distributions of curvature invariants exhibit continuous rotational symmetry around the axis of travel since the components of the velocity vector and of the shift vector Eq.~(\ref{eq:BetaCtv},\ref{eq:BetaCov}) are independent of the azimuthal angle $\phi$ (Fig. \ref{Fig1_SpherCoord}).

In Nat\'{a}rio's spacetime, the distribution of all curvature invariants over a meridian (circle of longitude, Fig. \ref{Fig1_SpherCoord}) primarily follows a $\sin^2(\theta)$ pattern. The distribution reaches its minimum at the intersections of the warp-bubble with the $x$-axis of travel and peaks at the intersections with the $y$-axis, perpendicular to the direction of travel. In contrast, as demonstrated in \cite{Rodal}, Einstein's scalar $G$ in Alcubierre's spacetime exhibits a $\cos^2(\theta)$ distribution with positive values at a radius slightly larger than the radius of the warp-bubble, and a $-\left(1 + 2 \cos^2\left(\theta\right)\right) = -\left(2 + \cos\left(2\theta\right)\right)$ distribution with larger negative values at a radius slightly smaller than the warp-bubble's radius. The maximum location in Alcubierre's spacetime coincides with the minimum location in Nat\'{a}rio's spacetime and vice versa. This inner radius distribution of $1 + 2 \cos^2(\theta)$ in Alcubierre's spacetime aligns with the prediction for Ricci's scalar based on analog gravity by Fischer and Visser \cite{Fischer_2003}, considering that $G = -R$.

In Alcubierre's spacetime, the distributions of $\sign{\left(r_{1}\right)}\sqrt{\abs{r_{1}}}$ and $\sign{\left(I\right)}\sqrt{\abs{I}}$ at outer radii larger than the warp-bubble's radius exhibit negative values and follow a $\sin^2(\theta)$ pattern, similar to the distribution observed in Nat\'{a}rio's spacetime. In contrast, the distributions of $\sign{\left(r_{1}\right)}\sqrt{\abs{r_{1}}}$ and $\sign{\left(I\right)}\sqrt{\abs{I}}$ in Alcubierre's spacetime at inner radii smaller than the warp-bubble's radius are positive and follow a pattern described by $a + b \cos^2(\theta)$, where $a$ and $b$ are constants.

\subsubsection{\texorpdfstring{Comparison of amplitude of curvature invariants in the Nat\'{a}rio and Alcubierre spacetimes}{Comparison of amplitude of curvature invariants in Nat\'{a}rio and Alcubierre}}\label{Comparison of amplitude of curvature invariants}

 The amplitudes of the curvature invariants in Nat\'{a}rio's spacetime are approximately 35 times greater than those in Alcubierre's spacetime under identical conditions of $v = c$, $\rho = 5\,[\text{m}]$, and $\sigma = 4\,[\text{m}^{-1}]$, as shown and discussed in \cite{Rodal}. This significant increase in amplitude is a direct consequence of Nat\'{a}rio's implementation of zero extrinsic curvature scalar ($K=0$), in contrast to Alcubierre's warp drive model, which does not enforce this constraint. In Alcubierre's model, a non-zero $K^2$ reduces the required amount of negative energy density, as evident from Eq.~(\ref{eq:Einsteinenergy}). Therefore, the claim by Mattingly et al. \cite{Mattingly} that ``The CM curvature invariants confirm that Nat\'{a}rio's warp drive is a more realistic alternative to Alcubierre’s'' is incorrect. In fact, Nat\'{a}rio's enforcement of zero extrinsic curvature scalar ($K=0$) makes the warp drive concept even less realistic than the already unrealistic Alcubierre warp drive due to the exorbitant amounts of negative energy density required.

\subsection{\texorpdfstring{Comparison of curvature invariants for Nat\'{a}rio's warp drive traveling at velocities from $0.1 c$ to $10 c$}{Comparison of curvature invariants at velocities}}

As shown in Appendix~\ref{secA1}, Eq.~(\ref{eq: G appendix}), for Nat\'{a}rio's isochoric flow model, Einstein's scalar $G = 2 \, \varrho \, \kappa$  is directly proportional to the energy density $\varrho$, thereby making it a quadratic function of velocity, as illustrated in Fig. \ref{fig:pi2}. Specifically, for relative velocities of $v/c = (0.1, 1.0, 10)$, the maximum absolute amplitudes of $G$ correspond to $(-1.31526, -131.526, -13152.6)$, respectively, aligning precisely with the expected velocity squared dependence.  Furthermore, Lobo and Visser \cite{LoboVisser} demonstrate that when the shift vector in Natário's spacetime is assumed to be both irrotational and divergenceless \cite{Fischer_2003}, the linearization of this specific scenario leads to the determination that \textit{all} components of the Einstein tensor are of the order $\mathcal{O}(v^2)$.

Weyl's invariant $\sign{\left(I\right)}\sqrt{\abs{I}}$ exhibits a velocity dependence that aligns with a quadratic relationship, particularly at higher velocities. For the same relative velocities of $v/c = (0.1, 1.0, 10)$, the maximum absolute amplitudes of $\sign{\left(I\right)}\sqrt{\abs{I}}$ are approximately $(2.78931, 303.914, 30415.4) [\text{m}^{-2}]$. The values for $v/c = 1.0$ and $v/c = 10$ closely match a velocity-squared dependence within a 0.1 percent margin. However, the value corresponding to the lower velocity $v/c = 0.1$ shows a more significant deviation. This deviation is attributed to the different, more complex pattern exhibited by the Weyl curvature invariant $\sign{\left(I\right)}\sqrt{\abs{I}}$ at this lower velocity, as depicted in Figs. \ref{fig:pi2 v01 rho5} and \ref{fig:I v01 rho5}.

Mattingly et al. \cite{Mattingly} constrain the amplitude of the curvature invariants on the vertical axis of their plots (specifically, their Fig. A7 for $r_{1}$, A8 for $r_{2}$, and A9 for $w_{2}$) to the range $[-1, 1]$ for velocities ranging from 0 to $v=100c$. This limitation is imposed without an explanation for such an arbitrary truncation. They recognize the inconsistency in the presentation of their results, which fail to demonstrate the dependence on velocity, stating: ``The prediction of an exponential increase in the magnitude of the invariants due to the velocity is not consistent with the invariants’ plots.'' This inconsistency, as will be further discussed in the following subsection, indicates that the plots of curvature invariants presented by Mattingly et al. \cite{Mattingly} are incorrect and misleading.

\subsection{\texorpdfstring{Comparison of curvature invariants for Nat\'{a}rio's warp drive with Mattingly et al.'s calculations for $ v = c $, $ \rho = 100 \, [\text{m}] $, and $ \sigma = 50000 \, [\text{m}^{-1}] $}{Comparison with Mattingly}}\label{Invariant comparison with Mattingly}

The curvature invariants are influenced by the radial derivatives of various orders of the form function $f(r)$. Specifically, when $\tanh(\rho \sigma) \approx 1$ (as is the case for the examples discussed here and in \cite{Mattingly} as well as in \cite{Rodal}, see subsection \ref{form function}), the dimensionless radial derivative is approximately $\left| r \frac{\partial f}{\partial r} \right|_{r \approx \rho} \approx \frac{\rho \sigma}{2}$, as detailed in \cite{AlcubierreLobo}. Consequently, the curvature invariants' amplitude increases with the dimensionless product's magnitude $\rho \sigma$, representing the ratio of the warp-bubble radius $\rho$ to its thickness $1/\sigma$. Furthermore, higher-order derivatives of order $n$ appearing in the expressions for the curvature invariants scale as $\left| r^{n} \frac{\partial^{n} f}{\partial r^{n}} \right|_{r \approx \rho} \approx \left(\frac{\rho \sigma}{2}\right)^{n}$. Upon normalization, the dominant term (the term containing the highest derivative) $-v^2 \left( \frac{r}{2} \frac{\partial^2 f}{\partial r^2}\right)^2 \sin^2(\theta)$ in the energy density equation (Eq.~\ref{eq:Einsteingausscodazzi-c}) therefore scales as $-v^2 \left( \frac{1}{64} (\rho \sigma)^4 \right) \sin^2(\theta)$.

In their analysis, Mattingly et al. \cite{Mattingly} utilized $\rho = 1\,[\text{m}]$ and $\sigma = 8\,[\text{m}^{-1}]$ for the Alcubierre plots across various velocities, yielding a $\rho \sigma$ value of 8. In contrast, for the Nat\'{a}rio spacetime, they selected $\rho = 100\,[\text{m}]$ and $\sigma = 50000\,[\text{m}^{-1}]$, resulting in a $\rho \sigma$ value of $5 \times 10^{6}$, which is $6.25 \times 10^{5}$ times larger. Therefore, it is expected that the energy will scale approximately as $-v^2 \left( \frac{1}{64} (6.25 \times 10^{5})^4 \right) \sin^2(\theta) \approx -v^2 \times (2.38 \times 10^{21}) \sin^2(\theta)$, which is 21 orders of magnitude greater.

Therefore, this significantly larger parameter choice for the Nat\'{a}rio spacetime implies a very unrealistic and exorbitantly high energy density requirement compared to the parameter choice for the Alcubierre drive. Mattingly et al. \cite{Mattingly} did not explain their unusual parameter selection, which also hinders a direct comparison between their Alcubierre and Nat\'{a}rio spacetime calculations.

Fig.~\ref{fig:v1 rho100} provides three-dimensional views of the distribution of curvature invariants in the $x,y$ plane, with the vertical axis representing the magnitude of each curvature invariant. The conditions for these plots are identical to those selected by Mattingly et al. \cite{Mattingly}, namely: $v = c$, $\rho = 100 \, [\text{m}]$, and $\sigma = 50000 \, [\text{m}^{-1}]$. These plots can be compared with Figs.~\ref{fig:G v1 rho5}, \ref{fig:r2 v1 rho5}, \ref{fig:r1 v1 rho5}, and \ref{fig:I v1 rho5}, which are set at the same speed $v = c$, but with $\rho = 5 \, [\text{m}]$, and $\sigma = 4 \, [\text{m}^{-1}]$. Based on the values of $\rho \sigma$, it is expected that the energy will scale approximately as $-v^2 \times (6.10 \times 10^{19}) \sin^2(\theta)$ between the two sets of conditions, shown in Fig.~\ref{fig:pi2 v1 rho5} and Fig.~\ref{fig:pi2 v1 rho100} for the curvature invariants.  

Using a simple extrapolation based on the dominant energy density term as a rough estimate for the magnitude of the curvature invariants, we consider Einstein's scalar $G$. For $v = c$, $\rho \sigma = 20$, $G$ peaks at $-131.53 \, [\text{m}^{-2}]$ as shown in Fig.~\ref{fig:pi2 v1 rho5}. Therefore, for $v = c$, $\rho \sigma = 5 \times 10^{6}$, extrapolating $G$ based on the energy density extrapolation, $G$ is expected to peak at approximately $(6.10 \times 10^{19}) \times (-131.53 \, [\text{m}^{-2}]) = -8.03 \times 10^{21} \, [\text{m}^{-2}]$. Fig.~\ref{fig:pi2 v1 rho100} shows that the actual peak of the curvature invariant $G$ occurs at $-1.16 \times 10^{21} \, [\text{m}^{-2}]$, which, while lower than the extrapolated value, is still within an order of magnitude of the rough estimate based on the dominant term in the energy density equation.

Fig.~\ref{fig:v1 rho100} and Fig.~\ref{fig:pi2 v1 rho100}, for the Nat\'{a}rio spacetime conditions $v = c$, $\rho = 100 \, [\text{m}]$, and $\sigma = 50000 \, [\text{m}^{-1}]$, demonstrate the following characteristics:

\begin{enumerate}
    \item \textbf{Amplitude of curvature invariants:} All curvature invariants exhibit amplitudes in the order of $10^{21} \, [\text{m}^{-2}]$. This observation aligns with a simple extrapolation based on the dominant term in the energy density equation.
    \item \textbf{Radial distribution of curvature invariants:} All curvature invariants display two peaks of equal magnitude in the radial direction. The large radius $\rho = 100 \, [\text{m}]$ and the extremely thin thickness $1/\sigma = 20 \times 10^{-6} \, [\text{m}]$ of the warp-bubble make these peaks invisible in the three-dimensional plots of Fig.~\ref{fig:v1 rho100}. However, they are clearly visible in Fig.~\ref{fig:pi2 v1 rho100}, which restricts the radial range for better visibility.
    \item \textbf{Circumferential distribution of curvature invariants:} All curvature invariants display a smooth $\sin^2(\theta)$ circumferential distribution over a meridian (circle of longitude, Fig. \ref{Fig1_SpherCoord}).  This distribution is consistent with a simple extrapolation based on the dominant term in the energy density equation. The numerical calculations for this distribution, performed using \textit{Mathematica\textsuperscript{\textregistered}} \cite{Mathematica} and based on the present tensorial formulation, encountered no computational issues.
    \item \textbf{Rotational symmetry around the axis of travel:} The warp-bubble geometry, as defined by Nat\'{a}rio's metric, exhibits spherical symmetry centered on its instantaneous central point since the components of the velocity vector and of the shift vector Eq.~(\ref{eq:BetaCtv},\ref{eq:BetaCov}) are independent of the azimuthal angle $\phi$ (Fig. \ref{Fig1_SpherCoord}). As the bubble moves along an axis, all curvature invariants show continuous rotational symmetry around this axis, a key feature of the metric and of the flow, under ideal conditions and in the absence of external disturbances.
    
\end{enumerate}

\subsubsection{Computational and plotting challenges in the analysis of Mattingly et al.}
\label{sec:discussion-mattingly-challenges}

Our analysis of Nat\'{a}rio's spacetime contrasts sharply with the computational and plotting challenges acknowledged by Mattingly et al. \cite{Mattingly}. They describe significant difficulties in their analysis, stating: 
\begin{quote}
``The shape of the $r_{1}$ invariant is that of a jagged disc at $r=\rho$. The disc has jagged edges in the negative direction, with sharp spikes at radial values $r=\rho$ and at polar angle values of $\theta= 0$ and $\theta=\pi$. The shape of the $r_{2}$ invariant is that of a jagged disc at $r=\rho$. Its edges vary between positive and negative values depending on the polar angle $\theta$. Similarly, the shape of the $w_{2}$ invariant is that of a jagged disc at $r=\rho$. In front of the harbor $(\theta> 0)$, the invariant has rapidly changing negative values between $-1$ and $0$. Behind the harbor $(\theta< 0)$, the invariant has rapidly changing positive values between $0$ and $1$. The jagged edges of the plots must mean that the $r_{1}$, $r_{2}$ and $w_{2}$ invariants oscillate rapidly between values of $-1$ and $1$ along the circumference of the warp-bubble. These oscillations must be occurring more rapidly than the program can plot.''
\end{quote}

It is important to address Mattingly et al.'s \cite{Mattingly} computational and plotting challenges for a comprehensive understanding of the correct solution:

\begin{enumerate}
    \item \textbf{Amplitude of curvature invariants:} Mattingly et al. \cite{Mattingly} presented all plots with the vertical axis, representing the magnitude of curvature invariants for the Nat\'{a}rio warp drive, restricted to the range $\in [-1,1]$. The implied unit of curvature is $[\text{m}^{-2}]$ since they employ meters as the unit of length in their plots. Consequently, they underestimate the actual range by 21 orders of magnitude, given the velocity and warp-bubble parameters used in their analysis. This can be shown by simple energy-based extrapolation (subsection \ref{Invariant comparison with Mattingly}). This underestimation of the actual energy density required could lead to misconceptions about the feasibility of the Nat\'{a}rio zero-expansion warp drive concept. This discrepancy might be attributed to a default behavior in the software used for plotting. With the default setting of \texttt{PlotRange $\rightarrow$ Automatic}, \textit{Mathematica\textsuperscript{\textregistered}} typically auto-adjusts and truncates the plot range. In their study of the Alcubierre spacetime, the same authors limited the plotted amplitude by 8 to 16 orders of magnitude (see \cite{Mattingly} and \cite{Rodal}). 
    
    \item \textbf{Jagged edges of the plotted variables:} Mattingly et al. \cite{Mattingly} write that the curvature invariants have jagged edges with sharp spikes at radial values $r=\rho$ and at polar angle values of $\theta= 0$ and $\theta=\pi$. This indicates severe numerical problems in their calculations that may stem from incorrect formulation, coding, or insufficient density of sampled data points.
    
    \item \textbf{Rapidly changing and oscillating values:} Mattingly et al. \cite{Mattingly} report that the curvature invariants exhibit rapidly changing values between their plotted limits of $-1$ and $1$ along the circumference of the warp-bubble. They also mention that these oscillations occur more rapidly than the program can plot. However, \textit{Mathematica\textsuperscript{\textregistered}} can plot rapidly changing oscillations accurately if the problem is correctly coded and enough points are set for plotting. These severe numerical issues in their calculations could be attributed to a potential deficiency in the density of sampled data points, errors in the computational algorithms employed, or inaccuracies in the mathematical model formulation. 
    
    \item \textbf{Lack of rotational symmetry around the axis of travel:} In the metric definitions of Nat\'{a}rio and Alcubierre, the geometry of the warp-bubble is characterized by spherical symmetry centered around the instantaneous central point of the warp-bubble. As this warp-bubble moves along a designated axis, all curvature invariants should exhibit continuous rotational symmetry around this axis of motion. This is a fundamental feature (regardless of the coordinate system) of the flow of spacetime as the spherical bubble travels along its axis of motion.  This is reflected by the components of the velocity vector and of the shift vector Eq.~(\ref{eq:BetaCtv},\ref{eq:BetaCov}) since they are independent of the azimuthal angle $\phi$ (Fig. \ref{Fig1_SpherCoord}).  Mattingly et al.'s \cite{Mattingly} figures A9, A10 and A11 show a lack of rotational symmetry (around the $x$-axis of travel) of the $w_{2}$ curvature invariant (e.g. $w_{2}<0$ for $y>0$ and $w_{2}>0$ for $y<0$), and hence a dependence on the azimuthal angle $\phi$.  This lack of rotational symmetry around the $x$-axis of travel in \cite{Mattingly} is inconsistent with the tetrad components and spherical coordinate system defined by Nat\'{a}rio \cite{Natário}, which \cite{Mattingly} ostensibly utilized in their analysis.

    \item  \textbf{Direction of travel:} Mattingly et al. \cite{Mattingly} state ``the harbor, which is defined as the flat portion of spacetime inside of the warp-bubble, as it travels.''  Therefore, when they \cite{Mattingly} mention ``in front of the harbor $(\theta> 0)$\dots'' they are likely referring to the forward direction relative to the ship's path. Conversely, when they \cite{Mattingly} write ``Behind the harbor $(\theta< 0)$\dots,'' it likely refers to the rearward direction in the context of the ship's movement. However, this interpretation implies that they \cite{Mattingly} considered the axis of travel to be perpendicular to the $x$-axis, which was defined by Nat\'{a}rio \cite{Natário} as the axis of travel (Fig.~\ref{Fig1_SpherCoord}).  In Nat\'{a}rio's \cite{Natário} coordinate system (Fig. \ref{Fig1_SpherCoord}), the front of the harbor is defined instead where $\cos \left({\theta}\right)> 0$ (corresponding to $- \pi/2 < \theta <  \pi/2$), whereas the space behind the harbor is characterized by $\cos \left({\theta}\right)< 0$.  This differs from \cite{Mattingly}'s statement that the region in front of the harbor is defined as $\theta> 0$ and the region behind the harbor by $\theta< 0$.  The region $- \pi/2 < \theta <  0$ is the front of the harbor according to Nat\'{a}rio \cite{Natário}, while it is the back of the harbor according to  \cite{Mattingly}.  Also, the region $ \pi/2 < \theta <  \pi$ is the back of the harbor according to Nat\'{a}rio \cite{Natário}, while it is the front of the harbor according to  \cite{Mattingly}.
    Therefore, the direction of travel stated in \cite{Mattingly} is inconsistent with the tetrad components and spherical coordinate system (Fig. \ref{Fig1_SpherCoord}) defined by Nat\'{a}rio \cite{Natário}, which \cite{Mattingly} purportedly applied in their analysis. 

   \item \textbf{Lack of quadratic dependence on velocity:} Mattingly et al.~\cite{Mattingly} acknowledge their failure to demonstrate the quadratic dependence on velocity, a key characteristic of energy density and curvature invariants when normalized to comparable units to $G$ (subsubsection \ref{sec:discussion-mattingly-challenges}). Furthermore Mattingly et al.~\cite{Mattingly} selected the warp-bubble parameters (radius) $\rho = 100\,[\text{m}]$, and (inverse thickness) $\sigma = 50000\,[\text{m}^{-1}]$, yielding a $\rho \sigma$ value of $5 \times 10^{6}$ for their Nat\'{a}rio spacetime calculations. This is $6.25 \times 10^{5}$ times larger than the $\rho \sigma$ value for the parameters they used for the Alcubierre spacetime. The energy for their Nat\'{a}rio calculations scales approximately 21 orders of magnitude greater than for their Alcubierre calculations.  They \cite{Mattingly} did not justify this extravagant parameter choice. Our calculated peak value of Einstein's scalar curvature is $G=-1.16 \times 10^{21} \, [\text{m}^{-2}]$ (see Fig.~\ref{fig:pi2 v1 rho100}) for Mattingly et al.'s parameters. Employing Eq.~\ref{eq: G appendix} with $\kappa=2.077 \times 10^{-43} \, [\text{N}^{-1}]$, this translates to an exorbitant value of the energy density of $\varrho=G /(2 \,\kappa)=-2.792 \times 10^{63} \, [\text{J m}^{-3}]$.  For context, its absolute value is approximately 29 orders of magnitude larger than the energy density of the densest neutron star, assuming an average density of $5.9 \times 10^{17} \, [\text{kg m}^{-3}]$, equivalent (using $E = mc^2$) to  $5.3 \times 10^{34} \, [\text{J m}^{-3}]. $

\end{enumerate}

\section{Summary of results}\label{SummaryResults}

This study delivers a precise visualization of scalar invariants within the Nat\'{a}rio warp drive spacetime. Through a comprehensive analysis, we have identified several key findings:

\begin{enumerate}

  \item \textbf{Nat\'{a}rio’s spacetime is Petrov type I; hence it is not a class \textit{B} warped product spacetime} (subsection \ref{Clarifying}). Our analysis proves that Nat\'{a}rio's spacetime \cite{Natário} falls within the Petrov type I classification, the most general category in the Petrov classification of spacetimes.  It is characterized by the lack of algebraic symmetries and the existence of four unique and real principal null directions.  The assertion, made without proof, by Mattingly et al. \cite{Mattingly}  that: ``\dots Class $B_{1}$ spacetimes, which include all hyperbolic spacetimes, such as the general warp drive line element\dots'' is invalid, because both the Nat\'{a}rio, and the Alcubierre (see \cite{Rodal}) warp drive line elements discussed in their article, are not Class $B_{1}$ Warped Product spacetimes. We find that Nat\'{a}rio’s spacetime does not fit the definition of a Class $B$ warped product spacetime. Within the classification of warped product spacetimes, Class $B$ types are restricted to Petrov types D or O. Consequently, as Nat\'{a}rio’s spacetime is identified as Petrov type I, it categorically excludes it from being a Class $B$ warped product spacetime. The Petrov type I classification is compatible with conditions under which gravitational waves may exist due to the lack of special algebraic symmetries in the warp-drive spacetime curvature. Since all the Newman-Penrose Weyl scalars in Petrov type I spacetimes are non-vanishing, it is possible to rotate \cite{Re2003} the Newman-Penrose double-null tetrad in a Petrov type I spacetime into a frame where the Newman-Penrose Weyl scalars $\Psi_1$ and $\Psi_3$ are equal to zero. This rotated frame elucidates the combination of a static monopole `Coulomb-type' term associated with $\Psi_2$ and the transverse degrees of freedom $\Psi_0$ and $\Psi_4$, which are potentially associated with traveling transverse gravitational waves.

  \item  \textbf{Weyl's scalar is the curvature invariant with the highest amplitude} (subsubsection \ref{sec:amplitude}). In Nat\'{a}rio's spacetime (with $K=0$ constraining expansion-contraction of Eulerian volume elements), curvature invariants range in amplitude, with the signed square root of the Weyl invariant $\text{sign}(I)\sqrt{\left|I\right|}$ exhibiting the largest and Einstein's scalar $G = G_{\alpha}^{\ \alpha}$ the smallest amplitudes. Conversely, in the Alcubierre spacetime \cite{Rodal} (where Eulerian volume element expansion-contraction is not constrained), Einstein's scalar $G$ presents the highest amplitude, closely followed by the Weyl invariant $\text{sign}(I)\sqrt{\left|I\right|}$, indicating the Weyl invariant's significant impact. In both spacetimes, the Weyl invariant $I \equiv C_{\alpha\beta\gamma\delta} C^{\alpha\beta\gamma\delta}$ consistently demonstrates a high amplitude even when normalized (to comparable $[\text{m}^{-2}]$ units to $G$) by considering its signed square root.  We attribute this to the sharp localization of the form function $f(r)$ near the warp-bubble radius (subsection \ref{form function}). As discussed in subsection \ref{sec:amplitude}, the second derivatives of the Weyl curvature tensor are connected to the second derivatives of the stress-energy-momentum tensor (accompanied by nonlinear coupling terms) through an equation formed from a Bianchi identity and Einstein's equations (\cite{padmanabhan}, pp. 264-265, 403; and  \cite{deFelice}, pp. 194-195). Near the warp-bubble radius, the form function approximates a top-hat function. This leads to a pronounced amplitude of the higher order derivatives $\frac{\partial^n f}{\partial r^n}$ for orders $n \geq 2$. The highest order derivative of $f(r)$ in the expressions for energy density as well as for Einstein's scalar $G$ is of the second order, $\frac{\partial^2 f}{\partial r^2}$, while in the expression for the Weyl scalar invariant $I$, it is of the third order, $\frac{\partial^3 f}{\partial r^3}$. Therefore, the high amplitude of these higher-order derivatives of $f(r)$ near the warp-bubble radius elucidates the critical yet previously under-emphasized, local, and significant role of the Weyl curvature tensor in warp-bubble spacetimes.

  \item  \textbf{Momentum density, rather than volume change, is the essential physical quantity in both the Nat\'{a}rio and Alcubierre models} (subsection \ref{Momentum density}).  Before Nat\'{a}rio's work, which imposed a zero value on the extrinsic curvature scalar $K$ (a necessary but not sufficient condition for maximal slicing in the 3+1 formalism), a common understanding, largely derived from Alcubierre's seminal work \cite{Alcubierre_1994}, was that warp drive solutions required contracting space volume elements in front of and expanding them behind the warp-bubble. Despite Nat\'{a}rio's model imposing this condition of zero scalar extrinsic curvature (analogous to isochoric fluid flow), both Nat\'{a}rio's and Alcubierre's warp drive models share a critical feature: the anti-symmetry of momentum density along the direction of motion. Determined by the spatial gradient of the extrinsic curvature tensor, it is the momentum density, rather than the expansion or contraction of volume elements, that stands as the essential physical quantity in both the Nat\'{a}rio and Alcubierre models at constant velocity (see subsection \ref{Momentum density} and Fig.~\ref{fig:G_1^0 v1.5 rho5}). The momentum density anti-symmetry answers White's question \cite{WhiteGRG}: ``\dots how does the ship know which way to go? The energy density curves local spacetime, but since it has no bias along the $x$-axis\dots .''

  \item  \textbf{Nat\'{a}rio’s warp drive is even more unrealistic than Alcubierre's warp drive} (subsection \ref{Comparison of amplitude of curvature invariants}).  The amplitudes of curvature invariants in Nat\'{a}rio’s spacetime are approximately 35 times greater than those in Alcubierre’s spacetime \cite{Rodal}, given identical warp-bubble parameters: $v = c$, $\rho = 5\, [\text{m}]$, and $\sigma = 4\, [\text{m}^{-1}]$. This finding contrasts with the claim by Mattingly et al. \cite{Mattingly} that: ``The CM curvature invariants confirm that Nat\'{a}rio’s warp drive is a more realistic alternative to Alcubierre’s.'' Our analysis indicates the contrary. Specifically, Nat\'{a}rio's imposition of a zero extrinsic curvature scalar ($K = 0$) exacerbates the negative energy density requirements, rendering Nat\'{a}rio's concept less viable than the Alcubierre warp drive, already burdened by its exorbitant negative energy density requirement. This conclusion follows from Eq.~\ref{eq:energy density}, which shows that the extrinsic curvature scalar contributes positively as $K^2$. Therefore, any non-zero value of $K$ mitigates the required amplitude of negative energy density.

  \item \textbf{Smooth $\sin^2(\theta)$ distribution of curvature invariants in Nat\'{a}rio’s spacetime} (subsections \ref{Circumferential distribution meridian}, \ref{Comparison of circumferential distribution}, and \ref{Invariant comparison with Mattingly}). Curvature invariants in Nat\'{a}rio’s spacetime display a smooth $\sin^2(\theta)$ circumferential distribution over a meridian (circle of longitude, Fig. \ref{Fig1_SpherCoord}). This distribution pattern arises because the dominant term in the curvature invariants, similar to the dominant term in the energy density, involves the highest order radial derivative of $f(r)$ multiplied by $\sin^2(\theta)$. The distribution attains its minimum at $\theta = 0$ and $\theta =\pm \pi$, corresponding to the intersections of the warp-bubble with the axis of travel. It reaches its peak at $\theta = \pm\frac{\pi}{2}$, at the intersection of the axis of travel with the plane perpendicular to this axis (Fig.~\ref{Fig1_SpherCoord}).  In contrast, in Alcubierre's spacetime \cite{Rodal}, Einstein's scalar $G$ shows a $\cos^2(\theta)$ distribution with positive values just outside the warp-bubble's radius, and a more pronounced negative distribution, $- (1 + 2 \cos^2(\theta)) = - (2 + \cos(2\theta))$, slightly inside the warp-bubble's radius. For all curvature invariants in Nat\'{a}rio’s spacetime, two similar-magnitude peaks are observed in the radial direction, except at lower velocities, such as $v=0.1c$, where three peaks are present.
  
  \item \textbf{Energy density extrapolation can be used for approximate estimation of curvature invariants in Nat\'{a}rio’s spacetime} (subsection \ref{Invariant comparison with Mattingly}). To facilitate direct comparison, curvature invariants are expressed in consistent curvature units $[\text{m}^{-2}]$, such as $G$, $\text{sign}\left(r_{1}\right)\sqrt{\left| r_{1} \right|}$, $\text{sign}\left(I\right)\sqrt{\left| I \right|}$, and $\text{sign}\left(r_{2}\right)\sqrt[3]{\left| r_{2} \right|}$. These invariants demonstrate a quadratic velocity dependence, as shown in Fig. \ref{fig:pi2}. This relationship mirrors the energy density's dependence on the square of velocity, with the dominant terms in the curvature invariants also exhibiting a similar dependence. For instance, the dominant term in the energy density equation $-v^2 \left( \frac{r}{2} \frac{\partial^2 f}{\partial r^2} \right)^2 \sin^2(\theta)$, when normalized, scales as $-v^2 \left( \frac{1}{64} (\rho \sigma)^4 \right) \sin^2(\theta)$ (Eq.~\ref{eq:Einsteingausscodazzi-c}). In subsection \ref{Invariant comparison with Mattingly} we estimate the magnitude of the curvature invariants. For $v = c$ and $\rho \sigma = 20$, $G$ peaks at $-131.53 \, [\text{m}^{-2}]$ (Fig.~\ref{fig:pi2 v1 rho5}). Extrapolating for $v = c$ and $\rho \sigma = 5 \times 10^{6}$, $G$ is expected to peak at approximately $(6.10 \times 10^{19}) \times (-131.53 \, [\text{m}^{-2}]) = -8.03 \times 10^{21} \, [\text{m}^{-2}]$. The actual peak value of $G$ at $-1.16 \times 10^{21} \, [\text{m}^{-2}]$ (Fig.~\ref{fig:pi2 v1 rho100}) is within an order of magnitude of this estimate, validating the extrapolation approach based on the energy density's dominant term.  Previously, Lobo and Visser \cite{LoboVisser} showed that in Natário's spacetime, if the shift vector is irrotational and divergence-free \cite{Fischer_2003}, linearization implies all Einstein tensor components are $\mathcal{O}(v^2)$.
    
  \item \textbf{Amplitude of curvature invariants:} subsection \ref{sec:discussion-mattingly-challenges} discusses in detail a number of errors in Mattingly et al. \cite{Mattingly}'s analysis. Most importantly, \cite{Mattingly} underestimated the actual amplitude of curvature invariants by 21 orders of magnitude, given the velocity and warp-bubble parameters used in their analysis of Nat\'{a}rio's zero-expansion warp drive. This underestimation of the actual energy density required could lead to misconceptions about the feasibility of the Nat\'{a}rio zero-expansion warp drive concept.

\end{enumerate}

\section{Conclusions}\label{Conclusions}

Our analysis proves that Nat\'{a}rio’s zero-expansion spacetime falls within the Petrov type I classification, which is characterized by the absence of algebraic symmetries and the existence of four unique and real principal null directions. This finding contradicts the assertion by \cite{Mattingly} that Nat\'{a}rio’s and Alcubierre’s warp drive line elements are Class $B_1$ warped product spacetimes. Petrov type I spacetimes do not fit the definition of Class $B$ warped product spacetimes, which are restricted to Petrov types $D$ or $O$. The Petrov type I classification allows for the potential propagation of wave-like perturbations in regions of the spacetime geometry. It is possible to rotate the Newman-Penrose double-null tetrad in a Petrov type I spacetime into a frame where the Newman-Penrose Weyl scalars $\Psi_1$ and $\Psi_3$ are equal to zero. This rotated frame elucidates the combination of a static monopole `Coulomb-type' term and transverse degrees of freedom which are potentially associated with traveling transverse gravitational waves.

To date, the Petrov type of Nat\'{a}rio’s zero-expansion spacetime has not been explicitly determined in the warp drive literature due to its historical focus on the Ricci curvature and the associated energy-momentum tensor, neglecting the contribution of the Weyl curvature. However, we demonstrate in this article that due to the sharp localization of the form function $f(r)$ near the warp-bubble radius, and particularly for Nat\'{a}rio’s zero-expansion warp drive, due to the imposed constraint of no change in differential volume, the Weyl curvature invariant exhibits a higher magnitude than the Ricci curvature invariants.

We found that in Nat\'{a}rio’s spacetime, which has no change of Eulerian differential volume elements, the signed square root of the Weyl invariant shows the largest amplitude, while Einstein’s scalar has the smallest. This is in contrast to Alcubierre’s spacetime, where Einstein’s scalar exhibits the highest amplitude, closely followed by the signed square root of the Weyl invariant. The pronounced amplitude of the Weyl invariant in both spacetimes, despite differences in the constraints on volume element expansion, highlights its significant impact near the warp-bubble radius due to the sharp localization of the warp-drive form function.

Momentum density emerges as a critical physical quantity in both the Nat\'{a}rio and Alcubierre models. Both models exhibit anti-symmetry of momentum density along the direction of motion. This anti-symmetry provides a clear physical mechanism for the directionality of the warp drive, addressing questions about how the ship `knows' which way to move at constant velocity.

Our comparative analysis indicates that Nat\'{a}rio’s warp drive is even less realistic than Alcubierre's. The amplitudes of curvature invariants in Nat\'{a}rio’s spacetime are 35 times greater than those in Alcubierre’s, given the same warp-bubble parameters. This increased amplitude exacerbates the already challenging negative energy density requirements, making Nat\'{a}rio’s concept less feasible. Importantly, \cite{Mattingly} underestimated the actual amplitude of curvature invariants by 21 orders of magnitude in their analysis of Nat\'{a}rio's warp drive. This significant underestimation could lead to misconceptions about the feasibility of the Nat\'{a}rio zero-expansion warp drive concept.

Curvature invariants in Nat\'{a}rio’s spacetime display a smooth \(\sin^2(\theta)\) circumferential distribution over a meridian, with peaks at the equator and minima at the poles. This distribution pattern contrasts with the \(\cos^2(\theta)\) distribution of Einstein's scalar in Alcubierre's spacetime, highlighting distinct geometric features of the two models.

Lastly, we demonstrated that energy density extrapolation can be used to estimate the magnitudes of curvature invariants in Nat\'{a}rio’s zero-expansion spacetime. The quadratic velocity dependence observed in the curvature invariants mirrors the energy density's dependence, allowing for accurate predictions of invariant amplitudes based on known parameters.

In conclusion, our findings challenge previous claims about the feasibility (in terms of required energy-density, and magnitude of curvature invariants) of Nat\'{a}rio’s zero-expansion warp drive and underscore the critical role of the Weyl curvature invariant in understanding the physical properties of proposed warp drive spacetimes. 

\section*{Acknowledgments}
We extend our sincere gratitude to the anonymous peer reviewers whose insightful comments and criticisms significantly contributed to enhancing the quality of this manuscript. Their dedication and expertise have been invaluable in improving our work.

\section*{Declarations}

\begin{itemize}
    \item Funding: No funding was received for conducting this study.
    \item Conflict of interest/Competing interests: The author has no competing interests to declare relevant to this article's content.
    \item Data Availability Statement: No traditional datasets were generated or analyzed during the current study. Computational analyses and visualizations were performed using \textit{Mathematica\textsuperscript{\textregistered}} \cite{Mathematica} commands such as \texttt{RevolutionPlot3D}. The \textit{Mathematica\textsuperscript{\textregistered}} commands are available from the author (J.R.) upon request.
\end{itemize}

\noindent


\begin{appendices}

\section{Curvature Invariants of the Nat\'{a}rio metric}\label{secA1}

In this appendix, we present the explicit formulas for the curvature invariants used in Section \ref{Visualizing curvature invariants} to calculate and visualize them in Natário's spacetime for different conditions. These curvature invariants are, namely, Einstein's curvature invariant $G \equiv G_{\alpha}^{\alpha}= - R$ as per Eqs. (\ref{eq:HatEins_a}, \ref{EinsRiccN4}), the quadratic Ricci invariant, $r_{1} \equiv \widehat{R}_{\alpha}^{\:\:\:\beta} \widehat{R}_{\beta}^{\:\:\:\alpha} = \widehat{G}_{\alpha}^{\:\:\:\beta} \widehat{G}_{\beta}^{\:\:\:\alpha}$ as per Eqs. (\ref{eq:r1_a}, \ref{eq:r1_b}), Weyl's curvature invariant, $I \equiv C_{\alpha\beta\gamma\delta} C^{\alpha\beta\gamma\delta}$ as per Eq. (\ref{WeylScalar}), and the cubic Ricci invariant,  $r_{2} \equiv \widehat{R}_{\alpha}^{\:\:\:\beta} \widehat{R}_{\beta}^{\:\:\:\gamma} = \widehat{G}_{\alpha}^{\:\:\:\beta} \widehat{G}_{\beta}^{\:\:\:\gamma} \widehat{G}_{\gamma}^{\:\:\:\alpha}$ as per Eqs. (\ref{eq:r2_a}, \ref{eq:r2_b}).

\begin{align}
G  =-2\, v^2 \left( 3 \left(\frac{\partial f}{\partial r}\right)^2 \,  \cos^2(\theta)+  \left( \frac{\partial f}{\partial r} + \frac{r}{2} \frac{\partial^2 f}{\partial r^2}\right)^2 \, \sin^2(\theta) \right) =2\, \varrho \; \kappa.
\label{eq: G appendix}
\end{align}

\begin{align}
r_1 &=\frac{1}{16 r^2} v^2 \Biggl[4 r^2 v^2 (-4 \cos (2 \theta )+3 \cos (4 \theta )+109) \frac{\partial f}{\partial r}^4 \notag \\
& \quad+r^2 \Biggl(8 r^2  \frac{\partial^3 f}{\partial r^3}^2 \sin ^2(\theta ) \left(4 v^2 f^2 \cos ^2(\theta )-1\right) \notag \\
& \quad+ \frac{\partial^2 f}{\partial r^2}^2 \Biggl(v^2 \Biggl(11 r^4 \sin ^4(\theta )  \frac{\partial^2 f}{\partial r^2}^2+16 r^2 f \sin ^2(\theta ) (5 \cos (2 \theta )+1)  \frac{\partial^2 f}{\partial r^2} \notag \\
& \quad+32 f^2 (4 \cos (2 \theta )+3 \cos (4 \theta )+5)\Biggr)+32 (4 \cos (2 \theta )-5)\Biggr) \notag \\
& \quad +32 r  \frac{\partial^3 f}{\partial r^3}  \frac{\partial^2 f}{\partial r^2} \left(v^2 f^2 \sin ^2(2 \theta )-3 \sin ^2(\theta )\right)\Biggr)\notag \\
& \quad+32 r v^2 \sin ^2(\theta ) \frac{\partial f}{\partial r}^3 \left(r^2 (4 \cos (2 \theta )+21)  \frac{\partial^2 f}{\partial r^2}-2 f (7 \cos (2 \theta )+9)\right)\notag \\
& \quad+8 r \frac{\partial f}{\partial r} \Biggl( \frac{\partial^2 f}{\partial r^2} \Biggl(v^2 \sin ^2(\theta ) \Biggl(11 r^4 \sin ^2(\theta )  \frac{\partial^2 f}{\partial r^2}^2+2 r^2 f (13 \cos (2 \theta )-5)  \frac{\partial^2 f}{\partial r^2}\notag \\
& \quad+8 f^2 (3 \cos (2 \theta )+5)\Biggr)+8 (\cos (2 \theta )-5)\Biggr)\notag \\
& \quad-4 r  \frac{\partial^3 f}{\partial r^3} \left(v^2 f \sin ^2(2 \theta ) \left(r^2  \frac{\partial^2 f}{\partial r^2}-2 f\right)+2 \sin ^2(\theta )\right)\Biggr)\notag \\
& \quad+8 \frac{\partial f}{\partial r}^2 \Biggl(v^2 \Biggl(\sin ^2(\theta ) \Biggl(r^4 (32-7 \cos (2 \theta ))  \frac{\partial^2 f}{\partial r^2}^2+4 r^2 f (5 \cos (2 \theta )-13)  \frac{\partial^2 f}{\partial r^2}\notag \\
& \quad+8 f^2 (3 \cos (2 \theta )+5)\Biggr)-8 r^3 f  \frac{\partial^3 f}{\partial r^3} \sin ^2(2 \theta )\Biggr)-8 (3 \cos (2 \theta )+5)\Biggr)\Biggr].
\end{align}

\begin{align}
I &= \frac{1}{6 r^2}\ v^2 \Biggl[-6 r^4  \frac{\partial^3 f}{\partial r^3}^2 \left(\sin ^2(\theta )-v^2 f^2 \sin ^2(2 \theta )\right)\notag \\
& \quad
+8 r^2  \frac{\partial^2 f}{\partial r^2}^2 \Biggl(v^2 \Biggl(r^4 \sin ^4(\theta )  \frac{\partial^2 f}{\partial r^2}^2+3 r^2 f \sin ^2(\theta ) (3 \cos (2 \theta )+1)  \frac{\partial^2 f}{\partial r^2}\notag \\
& \quad
+3 f^2 (4 \cos (2 \theta )+3 \cos (4 \theta )+5)\Biggr)-12 \sin ^2(\theta )\Biggr)\notag \\
& \quad
+r^2 v^2 (-52 \cos (2 \theta )-11 \cos (4 \theta )+351) \frac{\partial f}{\partial r}^4\notag \\
& \quad
+24 r^3  \frac{\partial^3 f}{\partial r^3}  \frac{\partial^2 f}{\partial r^2} \left(v^2 f^2 \sin ^2(2 \theta )-2 \sin ^2(\theta )\right)\notag \\
& \quad
+16 r v^2 \frac{\partial f}{\partial r}^3 \left(r^2 \sin ^2(\theta ) (\cos (2 \theta )+29)  \frac{\partial^2 f}{\partial r^2}-6 f \sin ^2(2 \theta )\right)\notag \\
& \quad
+4 r \frac{\partial f}{\partial r} \Biggl(4 \sin ^2(\theta )  \frac{\partial^2 f}{\partial r^2} \Biggl(v^2 \Biggl(4 r^4 \sin ^2(\theta )  \frac{\partial^2 f}{\partial r^2}^2+3 r^2 f (5 \cos (2 \theta )+1)  \frac{\partial^2 f}{\partial r^2}\notag \\
& \quad
+3 f^2 (3 \cos (2 \theta )+5)\Biggr)+12\Biggr)\notag \\
& \quad
+3 r  \frac{\partial^3 f}{\partial r^3} \left(v^2 f \sin ^2(2 \theta ) \left(4 f-r^2  \frac{\partial^2 f}{\partial r^2}\right)+4 \sin ^2(\theta )\right)\Biggr)\notag \\
& \quad
+6 \frac{\partial f}{\partial r}^2 \Biggl(2 \sin ^2(\theta ) \left(-5 r^4 v^2 (\cos (2 \theta )-3)  \frac{\partial^2 f}{\partial r^2}^2-8\right)\notag \\
& \quad
+r^2 v^2 f \left((76 \cos (2 \theta )+5 \cos (4 \theta )+15)  \frac{\partial^2 f}{\partial r^2}-4 r  \frac{\partial^3 f}{\partial r^3} \sin ^2(2 \theta )\right)\notag \\
& \quad
+8 v^2 f^2 \sin ^2(\theta ) (3 \cos (2 \theta )+5)\Biggr)\Biggr].
\end{align}

\vspace{4cm}

\begin{align}
r_2 &= \frac{1}{128 \
r^3}3 v^4 \Biggl[64 \ v^2 f^3 \left(r  \frac{\partial^2 f}{\partial 
r^2} \cos ^2(\theta )+\sin ^2(\theta ) \frac{\partial 
f}{\partial r}\right) \Biggl(\sin^2(2 \theta )  \
\frac{\partial^3 f}{\partial r^3}^2 r^4\notag \\
& \quad +2 (4 \cos (2 \theta )+5 
\cos (4 \theta )+7)  \frac{\partial^2 f}{\partial r^2}^2 r^2+4 \
\sin ^2(2 \theta ) \left(2 \frac{\partial f}{\partial r}+r  
\frac{\partial^2 f}{\partial r^2}\right)  \frac{\partial^3 
f}{\partial r^3} r^2\notag \\
& \quad +8 (\cos (2 \theta )+3) \sin ^2(\theta ) \
\frac{\partial f}{\partial r}  \frac{\partial^2 f}{\partial r^2} r+16 \sin ^2(2 \theta ) \frac{\partial f}{\partial r}^2\Biggr)\notag \\
& \quad -4 r v^2 \Biggl(2 \sin ^2(\theta )  \
\frac{\partial^2 f}{\partial r^2}^2  \Biggl(r^2  \frac{\partial^3 \
f}{\partial r^3}^2 \sin ^2(2 \theta )+4 r  \frac{\partial^2 \
f}{\partial r^2}  \frac{\partial^3 f}{\partial r^3} \sin ^2(2 \
\theta )\notag \\
& \quad -8 (4 \cos (2 \theta )+\cos (4 \theta )+1)  \
\frac{\partial^2 f}{\partial r^2}^2\Biggr) r^4\notag \\
& \quad 
+4 \frac{\partial \
f}{\partial r}  \frac{\partial^2 f}{\partial r^2} \Biggl(8 r^2 \cos ^2(\theta )  \frac{\partial^3 f}{\partial r^3}^2 \sin ^4(\theta )\notag \\
& \quad +(-20 \cos (2 \theta )+5 \cos (4 \theta )+47)  \
\frac{\partial^2 f}{\partial r^2}^2 \sin ^2(\theta )\notag \\
& \quad +2 r (\cos \
(2 \theta )+7) \sin ^2(2 \theta )  \frac{\partial^2 f}{\partial \
r^2}  \frac{\partial^3 f}{\partial r^3} \Biggr) \
r^3+\frac{\partial f}{\partial r}^2 \Biggl(96 r^2 \sin ^2(\theta \
)  \frac{\partial^3 f}{\partial r^3}^2 \cos ^4(\theta )\notag \\
& \quad
+(139 \
\cos (2 \theta )+226 \cos (4 \theta )+69 \cos (6 \theta )+718)  \
\frac{\partial^2 f}{\partial r^2}^2\notag \\
& \quad
+16 r (3 \cos (2 \theta )+11) \
\sin ^2(2 \theta )  \frac{\partial^2 f}{\partial r^2}  \
\frac{\partial^3 f}{\partial r^3}\Biggr) r^2\notag \\
& \quad +16 \frac{\partial f}{\partial r}^3 \Biggl((72 \cos (2 \theta )+19 (\cos (4 \theta \
)+7))  \frac{\partial^2 f}{\partial r^2} \sin ^2(\theta )
\notag \\
& \quad +2 r \
(\cos (2 \theta )+5) \sin ^2(2 \theta )  \frac{\partial^3 \
f}{\partial r^3}\Biggr) r\notag \\
& \quad -32 (-14 \cos (2 \theta )+\cos (4 \
\theta )-35) \sin ^2(\theta ) \frac{\partial f}{\partial \
r}^4 \Biggr) f^2\notag \\
& \quad +8 \Biggl(16 r^2 v^2 (38 \cos (2 \theta )+9 \cos \
(4 \theta )+49) \sin ^2(\theta ) \frac{\partial f}{\partial \
r}^5\notag \\
& \quad+4 r^3 v^2 \sin ^2(\theta ) \left(96 r  \frac{\partial^3 f}{\
\partial r^3} \cos ^4(\theta )+(48 \cos (2 \theta )-25 \cos (4 \
\theta )+201)  \frac{\partial^2 f}{\partial r^2}\right) \
\frac{\partial f}{\partial r}^4\notag \\
& \quad+2 \sin ^2(\theta ) \Biggl(r^4 \
v^2  \frac{\partial^2 f}{\partial r^2} \Biggl(16 r (\cos (2 \
\theta )+5)  \frac{\partial^3 f}{\partial r^3} \cos ^2(\theta \
)  \notag \\
& \quad +(-56 \cos (2 \theta )-25 \cos (4 \theta )+241)  \
\frac{\partial^2 f}{\partial r^2}\Biggr)-32 (7 \cos (2 \theta \
)+9)\Biggr) \frac{\partial f}{\partial r}^3\notag \\
& \quad +r \Biggl(v^2 (-148 \
\cos (2 \theta )-5 \cos (4 \theta )+137) \sin ^2(\theta )  \
\frac{\partial^2 f}{\partial r^2}^3 r^4\notag \\
& \quad +32 \sin ^2(\theta ) \
\left(3 v^2 \cos ^2(\theta ) \sin ^2(\theta )  \
\frac{\partial^2 f}{\partial r^2}^2 r^4-7 \cos (2 \theta \
)-9\right)  \frac{\partial^3 f}{\partial r^3} r\notag \\
& \quad +16 (-16 \cos (2 \
\theta )+\cos (4 \theta )-49)  \frac{\partial^2 f}{\partial \
r^2}\Biggr) \frac{\partial f}{\partial r}^2\notag \\
& \quad +4 r^2 \Biggl(-r^2 (7 \
\cos (2 \theta )+9)  \frac{\partial^3 f}{\partial r^3}^2 \sin \
^2(\theta )\notag \\
& \quad +4 r  \frac{\partial^2 f}{\partial r^2} \left(v^2 \
\cos ^2(\theta ) \sin ^2(\theta )  \frac{\partial^2 \
f}{\partial r^2}^2 r^4-13 \cos (2 \theta )-23\right)  \
\frac{\partial^3 f}{\partial r^3} \sin ^2(\theta )\notag \\
& \quad + \
\frac{\partial^2 f}{\partial r^2}^2 \left(19 r^4 v^2  \
\frac{\partial^2 f}{\partial r^2}^2 \sin ^6(\theta )+20 \cos (2 \
\theta )-17 \cos (4 \theta )-131\right)\Biggr) \frac{\partial \
f}{\partial r}\notag \\
& \quad
+2 r^3  \frac{\partial^2 f}{\partial r^2} \
\Biggl(r^4 v^2 (1-3 \cos (2 \theta )) \sin ^4(\theta )  \
\frac{\partial^2 f}{\partial r^2}^4+(56 \cos (2 \theta )-30 \
\cos (4 \theta )-58)  \frac{\partial^2 f}{\partial r^2}^2
\notag \\
& \quad +16 r (\
\cos (2 \theta )-2) \sin ^2(\theta )  \frac{\partial^3 \
f}{\partial r^3}  \frac{\partial^2 f}{\partial r^2}-4 r^2 \sin \
^2(\theta )  \frac{\partial^3 f}{\partial r^3}^2\Biggr)\Biggr) \
f\notag \\
& \quad +2 r \Biggl(48 r^2 v^2 \cos ^2(\theta ) (38 \cos (2 \theta )+3 \
\cos (4 \theta )-5) \frac{\partial f}{\partial r}^6 \notag \\
& \quad +8 r^3 v^2 \
(-44 \cos (2 \theta )+9 \cos (4 \theta )-133) \sin ^2(\theta )  \
\frac{\partial^2 f}{\partial r^2} \frac{\partial f}{\partial \
r}^5\notag \\
& \quad +2 \Biggl(r^4 v^2 (132 \cos (2 \theta )+41 \cos (4 \theta \
)-341) \sin ^2(\theta )  \frac{\partial^2 f}{\partial r^2}^2\notag \\
& \quad -16 \
(56 \cos (2 \theta )+3 \cos (4 \theta )+37)\Biggr) \
\frac{\partial f}{\partial r}^4\notag \\
& \quad +32 r \Biggl(-r^4 v^2 (\cos (2 \
\theta )+18)  \frac{\partial^2 f}{\partial r^2}^3 \sin \
^4(\theta )+2 r (3 \cos (2 \theta )+7)  \frac{\partial^3 \
f}{\partial r^3} \sin ^2(\theta )\notag \\
& \quad -4 (13 \cos (2 \theta )+\cos \
(4 \theta )-2)  \frac{\partial^2 f}{\partial r^2}\Biggr) \frac{\
\partial f}{\partial r}^3\notag \\
& \quad +4 r^2 \Biggl(r^4 v^2 (14 \cos (2 \theta \
)-33) \sin ^4(\theta )  \frac{\partial^2 f}{\partial r^2}^4\notag \\
& \quad +16 \
(-17 \cos (2 \theta )+2 \cos (4 \theta )+12)  \frac{\partial^2 \
f}{\partial r^2}^2+8 r (11-5 \cos (2 \theta )) \sin ^2(\theta )  \
\frac{\partial^3 f}{\partial r^3}  \frac{\partial^2 f}{\partial \
r^2}\notag \\
& \quad -2 r^2 (5 \cos (2 \theta )+1) \sin ^2(\theta )  \
\frac{\partial^3 f}{\partial r^3}^2\Biggr) \frac{\partial f}{\partial r}^2-4 r^3 \sin ^2(\theta )  \frac{\partial^2 f}{\partial r^2} \Biggl(9 r^4 v^2 \sin ^4(\theta )  \
\frac{\partial^2 f}{\partial r^2}^4\notag \\
& \quad +24 (3 \cos (2 \theta )-5)  \
\frac{\partial^2 f}{\partial r^2}^2+4 r (3 \cos (2 \theta )-11)  \
\frac{\partial^3 f}{\partial r^3}  \frac{\partial^2 f}{\partial \
r^2}-4 r^2 \sin ^2(\theta )  \frac{\partial^3 f}{\partial r^3}^2\
\Biggr) \frac{\partial f}{\partial r}\notag \\
& \quad +r^4 \sin ^2(\theta )  \
\frac{\partial^2 f}{\partial r^2}^2 \Biggl(-3 r^4 v^2 \sin \
^4(\theta )  \frac{\partial^2 f}{\partial r^2}^4+(48-96 \cos (2 \
\theta ))  \frac{\partial^2 f}{\partial r^2}^2 \notag \\
& \quad +48 r \sin^2(\theta )  \frac{\partial^3 f}{\partial r^3}  \
\frac{\partial^2 f}{\partial r^2}+4 r^2 \sin ^2(\theta )  \
\frac{\partial^3 f}{\partial r^3}^2\Biggr)\Biggr)\Biggr)\Biggr].
\end{align}

\vspace{1cm}




\end{appendices}


\bibliography{sn-bibliography}

\end{document}